\documentclass[11pt,a4paper]{article}
\usepackage{epsfig}
\usepackage[T1]{fontenc}    
\usepackage{graphics}
\usepackage{graphicx}
\usepackage{pstricks,pst-coil,pst-fill,pst-plot}
\usepackage[fleqn]{amsmath}    
\usepackage{amssymb}    
\usepackage{amsfonts}   
\usepackage{verbatim}   
\usepackage{mathrsfs}   
\usepackage{dsfont}
\usepackage{euscript}
\usepackage{yfonts}
\usepackage{txfonts}
\usepackage{marvosym}
\usepackage{vmargin}        

\setmarginsrb{1.8cm}{2cm}{1.8cm}{2cm}{1cm}{1cm}{1cm}{1.6cm}
\makeatletter
\@addtoreset{equation}{section}
\makeatother

\newcommand{\bra}[1]{\left\langle\,#1\,\right|}
\newcommand{\ket}[1]{\left|\,#1\, \right\rangle}

\let\ua=\uparrow
\let\da=\downarrow
\let\tend=\rightarrow

\newtheorem{theorem}{Theorem}[section]
\newtheorem{prop}{Proposition}[section]

\newtheorem{defin}{Definition}[section]

\newtheorem{lemme}{Lemma}[section]

\def\Proof{\medskip\noindent {\it Proof --- \ }}

\def\qed{\hfill\rule{2mm}{2mm}}

\long\def\symbolfootnote[#1]#2{\begingroup%
\def\thefootnote{\fnsymbol{footnote}}\footnote[#1]{#2}\endgroup}

\newcommand\beq{\begin{equation}}
\newcommand\enq{\end{equation}}
\newcommand\bem{\begin{multline}}
\newcommand\enm{\end{multline}}

\def\beqa{\begin{eqnarray}}
\def\eeqa{\end{eqnarray}}
\def\ba{\begin{array}}
\def\ea{\end{array}}

\def\det{\operatorname{det}}

\newcommand{\f}[2]{{\ensuremath{%
    \mathchoice%
    {\dfrac{#1}{#2}}
    {\dfrac{#1}{#2}}
    {\frac{#1}{#2}}
    {\frac{#1}{#2}}
}}}
\newcommand{\tf}[2]{\ensuremath{#1/#2}}
\newcommand{\pa}[1]{\ensuremath{\left(#1\right)}}
\newcommand{\paa}[1]{\ensuremath{\left\{#1\right\}}}
\newcommand{\pac}[1]{\ensuremath{\left[#1\right]}}
\newcommand{\paf}[2]{\ensuremath{\left(\f{#1}{#2}\right)}}
\newcommand{\pab}[2]{\ensuremath{\pa{\ba{c} #1 \vspace{2mm}\\ #2 \ea }}}

\def\a{\alpha}

\def\ga{\gamma}
\def\Ga{\Gamma}
\def\de{\delta}

\def\De{\Delta}
\def\eps{\epsilon}

\def\la{\lambda}

\def\sg{\sigma}

\def\Sg{\Sigma}

\def\Ups{\Upsilon}

\def\om{\omega}
\def\vp{\varphi}

\newcommand{\mc}[1]{\ensuremath{\mathcal{#1}}}
\newcommand{\mf}[1]{\ensuremath{\mathfrak{#1}}}
\newcommand{\msc}[1]{\ensuremath{\mathscr{#1}}}

\newcommand{\bs}[1]{\ensuremath{\boldsymbol{#1}}}

\newcommand{\ov}[1]{\ensuremath{\overline{#1}}}
\newcommand{\wt}[1]{\ensuremath{\widetilde{#1}}}
\newcommand{\wh}[1]{\ensuremath{\widehat{#1}}}

\newcommand{\Int}[2]{\ensuremath{\int\limits_{#1}^{#2}}}
\newcommand{\Oint}[2]{\ensuremath{\oint\limits_{#1}^{#2}}}
\newcommand{\Fint}[2]{\ensuremath{\fint\limits_{#1}^{#2}}}

\newcommand{\sul}[2]{\ensuremath{\sum\limits_{#1}^{#2}}}
\newcommand{\pl}[2]{\ensuremath{\prod\limits_{#1}^{#2}}}

\newcommand{\R}{\ensuremath{\mathbb{R}}}
\newcommand{\Cx}{\ensuremath{\mathbb{C}}}

\newcommand{\Dp}[1]{\ensuremath{\partial_{#1}}}

\newcommand{\limit}[2]{\ensuremath{\underset{#1 \tend #2}{\longrightarrow} }}
\newcommand{\J}[1]{\ensuremath{J_{\paa{#1}}}  }

\newcommand{\D}[1]{\ensuremath{D_{\paa{#1}}}  }

\newcommand{\ex}[1]{\ensuremath{\e{e}^{#1}}}

\newcommand{\braket}[2]{\ensuremath{\langle #1 \mid  #2 \rangle }}

\def\tr{\operatorname{tr}}

\newcommand{\ddet}[2]{\ensuremath{\det_{#1}\pac{#2}}}

\newcommand{\abs}[1]{\ensuremath{\left| #1 \right|}}

\newcommand{\norm}[1]{\ensuremath{\left\|#1\right\|}}

\newcommand{\dd}{\mathrm{d}}
\newcommand{\e}[1]{\ensuremath{\mathrm{#1}}}

\newcommand{\intff}[2]{\ensuremath{\left [ \, #1 \,; #2 \, \right ] }}
\newcommand{\intfo}[2]{\ensuremath{\left [ \, #1 \,; #2 \, \right [ }}

\newcommand{\intoo}[2]{\ensuremath{\left ] \, #1 \,; #2 \, \right [ }}

\newcommand{\intn}[2]{\ensuremath{[\![ \, #1 \,;\, #2 \,]\!]}}

\begin{document}

\begin{flushright}
DESY 10-227
\end{flushright}
\par \vskip .1in \noindent

\vspace{14pt}

\begin{center}
\begin{LARGE}
{\bf Riemann--Hilbert approach to the time-dependent generalized sine kernel.}
\end{LARGE}

\vspace{30pt}

\begin{large}

{\bf K.~K.~Kozlowski}\footnote[1]{DESY, Hamburg, Deutschland,
 karol.kajetan.kozlowski@desy.de},~~
\par

\end{large}

\vspace{40pt}

\centerline{\bf Abstract} \vspace{1cm}
\parbox{12cm}{\small We derive the leading asymptotic behavior and  build
a new series representation for the Fredholm determinant of integrable integral operators appearing in the
representation of the time and distance dependent correlation functions of integrable models
described by a six-vertex $R$-matrix.  This series representation
opens a systematic way for the computation of the long-time, long-distance asymptotic expansion  for the correlation functions of the aforementioned
integrable models \textit{away} from their free fermion point. Our method builds on a Riemann--Hilbert based analysis.}
\end{center}

\vspace{40pt}

\section{Introduction\label{INT}}

Highly structured determinants appear as a natural way for representing the correlation functions in integrable models
that are equivalent to the so-called free fermions. It was already shown by Kaufman and Onsager 
that certain two-point functions of the 2D-Ising model can be represented by Toeplitz determinants \cite{KafmanOnsagerFirstIntroDetRepCorrIsing2D}.
 Then Montroll, Potts and Ward 
\cite{MontrollPottsWardRowToRow2PtIsing2DToeplitzRepSpontMagProof} made this observation more systematic by expressing
the so-called row-to-row two-point function of this model in terms of a Toeplitz determinant.
It was  observed by Lieb, Mattis, Schultz \cite{LiebMattisSchultzXYSolutionByFreeFermionsCorrFunctsToo} that such Toeplitz determinant-based
representations also hold for the so-called XY model. 
Then, the systematic study of the correlation functions of the impenetrable Bose gas, the XY model or its isotrpoic version the XX model  
lead to the representation of various correlators in terms of Fredholm determinants (or their minors) of the so-called integrable 
operators \cite{ColomoIzerginKorepinTognettiEFPinXYtransversefield,ColomoIzerginKorepinTognettiTempCorrFctXX,KorepinSlavnovTimeDepCorrImpBoseGas,
LenardBoseImpGasredDensityMatrixFredminorsRepAndProofs,McCoyPArkShrockSpinTimeAutoCorrAsModSineKernel,
SchultzFredholmMinorForReduceDensityImpBosons}.
Such types of Fredholm determinants also appear in other branches of mathematical physics. For instance,  the
determinant of the so-called sine-kernel acting on an interval $J$ is directly related to the gap probability (probability that in the bulk scaling limit
a given matrix has no eigenvalues lying in $J$) in the Gaussian unitary ensemble \cite{GaudinMehtaDensityOfEigenvaluesRandomMatrices}.
Integrable integral operators \cite{DeiftIntegrableOperatorsDiscussion} are operators of the type $I+V$ where the integral kernel $V$ takes a very 
specific form. This fact allows for a relatively simple characterization of the resolvent kernel and often for a construction of a 
system of partial differential equations satisfied by the associated Fredholm determinant or its minors 
\cite{DeiftItsZhouSineKernelOnUnionOfIntervals,ItsIzerginKorepinTempCorrBoseGasIntSyst, 
ItsIzerginKorepinSlavnovDifferentialeqnsforCorrelationfunctions,JimMiwaMoriSatoSineKernelPVForBoseGaz,TracyWidomPDEforFredholms}.

In all of the aforementioned examples, the integrable integral operators  $I+V$ act
on some curve $\msc{C}$ with a kernel $V\pa{\la,\mu}$ depending, in an oscillatory way, on a parameter $x$.
In the previous examples a lot of interesting information can be drawn out of the asymptotic behavior of $\ddet{}{I+V}$ for large values of $x$.
For instance, when dealing with the correlation functions of integrable models, $x$ plays the role of a spacial and/or temporal separation
between the two operators entering in the correlation function. In such a case, computing the large-$x$ asymptotic expansion of the 
associated Fredholm determinants, allows one to test the predictions of conformal field theories. 
The form of the asymptotic behavior of the pure sine kernel determinant $\log \ddet{}{I+S}$ was strongly  argued in 
\cite{DescloizeauxMethaSineKernelFirstAsympotics,DysonSineKernelInverseScatteringAsymptoticExpansions}
and then proven, to some extend, using operator methods 
\cite{BudynBuslaevPureGammaSineKernelAsympt,EhrhardtConstantinPureSinekernelFredholmDeyt,WidomSinekernelOnSingleIntervals}.
Also, the discovery of non-linear differential equations of Painlev\'e V type for this determinant \cite{JimMiwaMoriSatoSineKernelPVForBoseGaz}
allowed to compute many terms in the large-$x$ asymptotic expansion of the associated correlation functions 
\cite{JimMiwaMoriSatoSineKernelPVForBoseGaz,McCoyPArkShrockSpinTimeSpaceAutoCorrAsModSineKernel,McCoyPArkShrockSpinTimeAutoCorrAsModSineKernel, 
McCoyTangSineKernelSubleadingFromPainleveV}

 However, a really systematic efficient approach to the asymptotic analysis of various quantities related 
to integrable integral operators $I+V$  has been made possible thanks to the results obtained in  
\cite{ItsIzerginKorepinSlavnovDifferentialeqnsforCorrelationfunctions}.
There it was shown that the analysis of such operators can be reduced to a resolution of an associated  Riemann--Hilbert problem (RHP).
The jump contour in this RHP coincides with the one on which the integral operator acts and the  jump
matrix is built out of the functions entering in the description of the kernel. In this way, one deals with a RHP depending on $x$
in an oscillatory way. The asymptotic analysis of their solutions is possible thanks to the non-linear steepest descent method of 
Deift-Zhou \cite{DeiftZhouSteepestDescentForOscillatoryRHPmKdVIntroMethod,DeiftZhouSteepestDescentForOscillatoryRHP}.
It is in this context, that the full characterization of the leading
asymptotic behavior of Fredholm determinants of kernels related to correlation functions in free-fermion equivalent models (the long-distance,
long-time/long-distance at zero and also non-zero temperature) has been carried out in the series of papers
 \cite{CheianovZvonarevZeroTempforFreeFermAndPureSine,DeiftZhouTimeautocorrInTempForXYatCritMag.Field,ItsIzerginKorepinTemperatureLongDistAsympBoseGas,
ItsIzerginKorepinNovokshenovTempeAutoCorrCriticalTransverseIsing,ItsIzerginKorepinSlavnovTempCorrFctSpinsXY,ItsIzerginKorepinVarguzinTimeSpaceAsymptImpBoseGaz}

This paper deals with an extension of these analysis to the case of a Fredholm determinant of an integrable integral operator whose 
integral kernel has a more involved structure then in the aforementioned cases. We call our kernel the time-dependent generalized sine kernel. The 
Fredholm determinant we analyze arises in the representation of the zero temperature long-distance/long-time asymptotic
behavior of two-point functions in a wide class of integrable models \textit{away} from their free fermion point.
In particular, its asymptotics expansion (and especially the new series representation that we obtain for it) plays a
crucial role in the computation of the long-time/long-distance asymptotic behavior of these two-point functions.

In a wide class of algebraic Bethe Ansatz solvable models, one is able to compute the so-called form factors (matrix elements of local operators)
and represent them as finite-size determinants \cite{KMTFormfactorsperiodicXXZ,OotaInverseProblemForFieldTheoriesIntegrability}  .
It has been shown in \cite{KorepinSlavnovTimeDepCorrImpBoseGas} that, for free fermion equivalent models,
it is possible to build on these representations so as to explicitly sum-up the form factor expansion
and compute the zero-temperature (and even the non-zero temperature) correlation functions of the model.
In the limit of infinite lattice sizes, a two-point function is then represented by a Fredholm determinant (or its minors) of an integrable
integral operator $I+V$ acting on some contour $\msc{C}$ determined by the properties of the model.
For time and space translation invariant models, the kernel $V$ depends on the distance separating the two operators as well
as on the difference of time evolution between them. One can show that for general free-fermion type models, the 
integral operator  $I+V$  associated with the form factor expansion of two-point functions
acts on a finite subinterval $\intff{-q}{q}$ of $\R$ and its kernel $V$  belongs to the class of kernels 
\beq
V\pa{\la,\mu} = 4  \f{ \sin \pac{\pi\nu\pa{\la}}  \sin \pac{\pi\nu\pa{\mu}} }{ 2i\pi \pa{\la-\mu}} 
\paa{E\pa{\la} e\pa{\mu}-E\pa{\mu} e\pa{\la} }  \; .
\label{definition noyau GSK FF}
\enq
There $\nu$ is some function encoding the fine structure of the excitations above the ground state whereas $e$, $E$ are oscillating factors. 
The function $E$ is expressed in terms of $e$
\beq
E\pa{\la} =i e\pa{\la} \paa{\Fint{ \msc{C}_{E} }{} \f{ \dd s }{2\pi } \f{ e^{-2}\pa{s} }{ s-\la }
 \; + \; \f{ e^{-2}\pa{\la} }{ 2 } \cot \pac{\pi \nu\pa{\la}}  } \;.
\label{definition fonction E+}
\enq
The functions $\nu$ and $e$ just as the integration curve $\msc{C}_{E}$ appearing in \eqref{definition fonction E+} depend on the specific model 
that one considers. We will give more precision about their
properties in the core of the paper. We stress that, in free-fermion equivalent models $\nu\pa{\la}$ is some constant and $e$ takes a simple form.
It was in such a context that the asymptotic analysis of $\ddet{}{I+V}$ has been carried out in the aformentioned papers. 

As will be shown in a series of subsequent publication 
\cite{KozKitMailTerNatteSeriesNLSECurrentCurrent,KozReducedDensityMatrixAsymptNLSE}, quite astoundingly, it is 
as well possible to build on the finite-size determinant representation for the form factors of local operators in integrable models out of their free 
fermion point so as to sum up the 
form factor series over the relevant sector of excited states.
The intermediate computations can be shown to boil down to effective generalized free fermionic models. As such, they involve, again, the
Fredholm determinants of operators $I+V$ with $V$ given by \eqref{definition noyau GSK FF}. However, then the functions $\nu$ and $e$ become much more
complex that at the free fermion point.
In some sense, the approach of \cite{KozKitMailTerNatteSeriesNLSECurrentCurrent,KozReducedDensityMatrixAsymptNLSE} shows that kernels 
\eqref{definition noyau GSK FF} appear as a natural basis of special functions
allowing one to represent the correlation functions of a wide class of interacting (\textit{ie} away from their free fermion point)
integrable models as certain linear combinations thereof. Therefore, the main purpose for our study of the time-dependent generalized sine kernel 
\eqref{definition noyau GSK FF}
is to obtain a convenient and effective representation for the associated Fredholm determinant allowing to re-sum the aforementioned linear combination in
some compact, explicit form, that moreover enjoys the property of giving an almost straightforward access to the asymptotic behavior of the
correlators.

This article contains two  main results. We first derive the leading asymptotic behavior of the Fredholm determinant of $I+V$
understood as acting on $L^{2}\pa{\intff{-q}{q}}$, with $q < +\infty$. This sets the ground for the second main result of the article.
Namely, we derive a new series representation -the Natte series- for the Fredholm determinant\symbolfootnote[2]{The origin of this name issues from
the so-called pig-tail (or braid) hairstyle that is called Natte in French. 
A braid is a specifically ordered reorganization of the loose hair-do style. 
Similarly, the Natte series reorganizes the Fredholm series in a very specific way, so that the resulting representation 
is perfectly fit for carrying out an asymptotic expansion.}.
This series is converging rather fast in the asymptotic $x\tend +\infty$ regime. Its main advantage is to provide a rather direct (without the need
to perform any additional analysis) approach to the asymptotic expansion of the determinant.
As already stressed out, this series representation plays a crucial
role in the computation of the long-time/long-distance asymptotic expansion of the two-point functions in integrable models corresponding to a six-vertex
$R$-matrix \cite{KozKitMailTerNatteSeriesNLSECurrentCurrent,KozReducedDensityMatrixAsymptNLSE}. 
Also, the very form of the asymptotic expansion stemming from the Natte series proves several conjectures relative to the structure of the 
asymptotic expansions for certain particularizations (for specific values of $\nu$, and $e$) of such Fredholm determinants
\cite{MullerShrockDynamicCorrFnctsTIandXXAsymptTimeAndFourier,TracyVaidyaRedDensityMatrixSpaceAsymptImpBosonsT=0}. 
Also, upon specialization, it yields the general structure of the asymptotic expansion of the fifth Painleve transcendent 
associated to the pure sine kernel \cite{DeiftItsZhouSineKernelOnUnionOfIntervals,JimMiwaMoriSatoSineKernelPVForBoseGaz}

This article is organized as follows. In section \ref{section Hypothese et resultats}, we outline the main assumptions
that we rely upon throughout the article and give a discussion of the class of functions $e$ that we deal with.
After introducing several notations, we present the two main results of the paper.
The remaining part is of technical nature.
In section \ref{section transfos RHP}, we present the RHP problem that is at the base of the asymptotic analysis of $\ddet{}{I+V}$ and the construction
of its Natte series. We also outline the chain of transformations corresponding to the implementation of the Deift-Zhou \cite{DeiftZhouSteepestDescentForOscillatoryRHP} steepest-descent method.
In section \ref{section parametrices}, we build the various local parametrices. This brings the original RHP into one that can be solved through
a series expansion of the associated singular integral equation \cite{ClanceyGohbergFactorizationMatrixFunctionSingIntEqnRHPreference}. 
The latter naturally provides the large-$x$ asymptotic expansion of the solution. 
We build on these results so as to derive the leading asymptotic expansion of the
Fredholm determinant in section  \ref{section DA determinant}. Finally, section \ref{section Series de Natte} is devoted to the construction
of Natte series for the Fredholm determinant of I+V. In particular, we establish the main properties of such series.
We then give a conclusion and discuss the further possible applications.
In appendix \ref{Appendix Properties CHF and Barnes}, we recall all the properties of the special functions that we use in this article.
In appendix \ref{Appendix Proofs asymptotic expansion for Pi and determinant}, we gather some proofs
relative to the structure of the large-$x$ asymptotic expansions of certain matrix valued Neumann series representing the solution
to a singular integral equation of interest to us. In appendix \ref{appendix Final bounds for PiN}, we establish some bounds
for certain matrices appearing in our analysis. 


\section{The main results and assumptions}
\label{section Hypothese et resultats}

In this article, we will focus on the case where the function $e$ takes the form
\beq
e^{-1}\pa{\la} = \ex{i \f{x u\pa{\la}}{2} + \f{g\pa{\la}}{2}} \; .
\label{definition fonction e}
\enq
$e\pa{\la}$ is quickly oscillating in the $x\tend +\infty$ limit and the function $g$ entering in the definition of $e\pa{\la}$
has been introduced so as to allow for some finite, $x$-independent oscillatory behavior of the function $e\pa{\la}$.
The principal value integral apperaing in the definition of $E$ \eqref{definition fonction E+}  is carried out along a curve $\msc{C}_{E}$ which corresponds to a slight deformation of the real axis
and is depicted in Fig.~\ref{contour pour transfo Cauchy et Hilbert}. Under the forthcoming hypothesis, such a contour allows to strengthen
the convergence of the integral defining $E$ at infinity (in the case of $\R$, the convergence would be the one of an oscillating non-absolutely
integrable power-law whereas it is exponentially fast along $\msc{C}_E$).


\subsection{The main assumptions}
\label{soussection hypotheses}

Throughout this paper, we make several assumptions on the function $u$, $g$ and $\nu$ entering in the description of
the integrable kernel \eqref{definition noyau GSK FF}.
\begin{itemize}
\item There exists an open neighborhood $U$ of $\R$ such that $u$ and $g$ are simultaneously holomorphic on $U$.
\item The function $g$ is bounded on $U$. 
\item The function $u$ is real valued on $\R$ and has a unique saddle-point in $U$ located at $\la_0 \in \R$. This saddle-point
is a zero of $u^{\prime}$ with multiplicity one, that is to say $ \exists \, ! \; \la_0 \in \R \; : \; u^{\prime}\!\pa{\la_0}=0$ and $u^{\prime\prime}\!\pa{\la_0}<0$.
We also assume that the saddle-point lies away from the boundaries: $\la_0\not=\pm q$.
\item $u$ is such that, given any $\eta>0$, $\ex{i\eta u\pa{\la}}$ decays exponentially fast in $\la$ when $\pm \Im\pa{\la} >\de>0$
for any fixed $\de>0$,  and $\Re\pa{\la} \underset{\la \in U}{\to}  \mp \infty$.
\item The function $\nu$ is holomorphic on $U$ and such that $\sin \pac{\pi \nu\pa{\la}}$ has no zeroes in some open neighborhood of $\intff{-q}{q}$
lying in $U$. 
\item The function $\nu$ has a "sufficiently" small real part at $ \pm q$, \textit{ie} $\abs{\Re\pac{\nu\pa{\pm q}} } <\tf{1}{2}$.

%
%
\end{itemize}

For technical reasons, one has to distinguish between two situations when the saddle point $\la_0$ is inside of $\intoo{-q}{q}$
or outside.  Following the tradition we refer to the first case ($-q<\la_0<q$) as the time-like regime and to the second one ($\abs{\la_0}>q$)
as the space-like regime. Actually, in this article we will only consider the case where $\la_0>q$.
Also, we do not treat the limiting case when $\la_0=\pm q$  as this would require a significant modification of our approach.


\subsection{The main result}

We now gather the main results of this paper into two theorems.

\begin{theorem}
\label{theorem DA determinant}
Let $V\pa{\la,\mu}$ be as in \eqref{definition noyau GSK FF} and $I+V$ act on $L^{2}\pa{\intff{-q}{q}}$.
Then, under the assumption stated in section \ref{soussection hypotheses}, the leading
$x\tend +\infty$  asymptotic behavior of $\ddet{}{I+V}$  reads:
\bem
\ddet{\intff{-q}{q}}{I+V} \ex{- \Int{-q}{q} \pac{i x u^{\prime}\pa{\la} + g^{\prime}\pa{\la}  } \nu\pa{\la} \dd \la } = B_x\pac{\nu,u}
\pa{1+\e{O}\paf{\log x}{x} }+ \f{b_1\pac{\nu,u,g} }{x^{\f{3}{2}}} B_x\pac{\nu,u}   \pa{1+\e{O}\paf{\log x}{x} }  \\
+ \ex{ix \pac{u\pa{q}-u\pa{-q}} + g\pa{q}-g\pa{-q} } B_x\pac{\nu+1,u} \pa{1+\e{O}\paf{\log x}{x} } + \ex{ix \pac{u\pa{-q}-u\pa{q}} + g\pa{-q}-g\pa{q} } B_x\pac{\nu-1,u} \pa{1+\e{O}\paf{\log x}{x} } \; .
\end{multline}
%
%
%

The functional $B_x\pac{\nu,u}$ takes the form
\beq
B_{x}\pac{\nu,u} =  \ex{C_1\pac{\nu}}
\f{G^2\!\pa{1+\nu\pa{q}} G^2\!\pa{1-\nu\pa{-q}}}
{ \pac{2qx \pa{u^{\prime}\!\pa{q}+i0^+} }^{\nu^2 \! \pa{q}} \pac{2qx u^{\prime}\!\pa{-q} }^{\nu^2 \! \pa{-q}}}
\pa{2\pi}^{\nu\pa{-q} - \nu\pa{q}}  \ex{i\f{\pi}{2} \pa{ \nu^2 \! \pa{q}-\nu^2 \! \pa{-q}} }  
%
%
%
\; .
\enq
It is expressed in terms of the Barnes $G$ function \cite{BarnesDoubleGaFctn2} and the auxiliary functional
\beq
C_1\pac{\nu}= \f{1}{2} \Int{-q}{q} \dd \la \dd \mu \f{\nu^{\prime}\pa{\la} \nu\pa{\mu}-\nu^{\prime}\pa{\mu} \nu\pa{\la} }{\la-\mu}
+\nu\pa{q} \Int{-q}{q} \f{\nu\pa{q} - \nu\pa{\la}}{q-\la} \dd \la  +\nu\pa{-q} \Int{-q}{q} \f{\nu\pa{-q} - \nu\pa{\la}}{q+\la}  \dd \la \; .
\label{definition fonctionnelle C1}
\enq
The functional $b_1\pac{\nu,u,g}$ takes different forms depending whether one is in the so-called space-like regime ($\la_0>q$) or in the
time-like regime ($\la_0 \in \intoo{-q}{q}$):
\beq
b_1\pac{\nu,u,g}= \f{  1  }{\sqrt{-2\pi u^{\prime \prime}\!\!\pa{\la_0}} }  \left\{ \ba{cc}
        \f{\nu\pa{-q}}{u^{\prime}\!\pa{-q} \pa{\la_0+q}^2} \f{\mc{S}_0}{\mc{S}_-}
        - \f{\nu\pa{q}}{u^{\prime}\!\pa{q} \pa{\la_0-q}^2} \f{\mc{S}_0}{\mc{S}_+}  \qquad & \e{time-like} \vspace{2mm}\\
         \f{\nu\pa{-q}}{u^{\prime}\!\pa{-q} \pa{\la_0+q}^2} \f{\mc{S}_-}{\mc{S}_0}
         - \f{\nu\pa{q}}{u^{\prime}\!\pa{q} \pa{\la_0-q}^2} \f{\mc{S}_+}{\mc{S}_0} \qquad& \e{space-like}  \ea\right.
\; .
\enq
There, we agree upon
\beqa
\mc{S}_+ &=& \pac{2q x u^{\prime}\!\pa{q} + i0^+}^{2\nu\pa{q}} e^2\!\pa{q}  \pa{\ex{-2i\pi \nu\pa{q}}-1} 
\f{ \Ga\pa{1-\nu\pa{q}} }{ \Ga\pa{1+\nu\pa{q}}    }
\exp\Bigg\{-2 \int_{-q}^{\,q} \f{\nu\pa{q}-\nu\pa{\mu} }{q-\mu} \dd\mu \Bigg\}
\; , \\
\mc{S}_- &=& \f{  \pa{\ex{-2i\pi \nu\pa{-q}}-1}   } { \pac{2q x u^{\prime}\pa{-q} }^{2\nu\pa{-q}} } e^2\pa{-q}
 \f{ \Ga\pa{1+\nu\pa{-q}} } { \Ga\pa{1-\nu\pa{-q}} }  \exp\Bigg\{2 \int_{-q}^{\,q} \f{\nu\pa{-q}-\nu\pa{\mu} }{q+\mu} \dd\mu \Bigg\}   \;,
\label{definition fonction S plus et moins}
\eeqa
\beq
\mc{S}_0= e^2\pa{\la_0} \ex{i\f{\pi}{4}} \paf{\la_0+q}{\la_0-q-i0^+}^{2\nu\pa{\la_0}} 
\exp\Bigg\{-2 \int_{-q}^{\,q} \f{\nu\pa{\la_0}-\nu\pa{\mu} }{\la_0-\mu} \dd\mu \Bigg\}
\times \left\{ \ba{cc}  \pa{\ex{-2i\pi \nu\pa{\la_0}}-1}^2  & \e{time-like} \vspace{2mm} \\ 
								1 						&  \e{space-like}   \ea \right.  
\; .
\label{definition fonction S zero et kappa}
\enq

\end{theorem}

The proof of this theorem will be given in section \ref{sous-section DA determinant preuve}. It heavily relies on the asymptotic
analysis of the RHP associated with $V$ that will be carried out in sections \ref{section transfos RHP} to \ref{section parametrices}.

Above, the $i0^+$ regularization is important only in the time-like regime as then $u^{\prime}\!\pa{q}<0$. It allows one for a non-ambiguous definition
of the power-laws appering above. In the space-like regime, the $i0^+$ regularization makes no difference .

A special limit of the kernel \eqref{definition noyau GSK FF} can be related to the generalized sine kernel studied in
\cite{KozKitMailSlaTerRHPapproachtoSuperSineKernel}. Indeed,  when the saddle point $\la_0$
is send to infinity, by deforming slightly the contours $\msc{C}_{E}$, the function $E$ can be seen to be proportional to $e^{-1}$, up to corrections that are
uniformly $\e{O}\pa{x^{-\infty}}$ on $\intff{-q}{q}$. In particular, one has that the $x \tend +\infty$ asymptotic expansion of the two Fredholm
determinants coincide in this limit. This can be seen directly by inspection of our formulae, at least in what concerns the leading asymptotics.

A specific case of our kernel $u\pa{\la}=\la-\tf{t\la^2}{x} $, $g=0$, $q=1$ and $\nu= cst$ has been studied in the litterature
in the context of its relation with the impenetrable fermion gas \cite{CheianovZvonarevZeroTempforFreeFermAndPureSine}. 
Upon such a specialization, our results agree with the coefficients of the asymptotic expansion obtained in that paper.

The second main result obtained in this paper is the Natte series representation for the Fredholm determinant.

\begin{theorem}
\label{theorem representation serie de Natte}

Under the assumptions stated in section \ref{soussection hypotheses}, the Fredholm determinant of the operator $I+V$ where the kernel V
is given by \eqref{definition noyau GSK FF} admits the below  absolutely convergent Natte series expansion. In other words, there exists
functionals $\mc{H}_n\pac{\nu, \ex{g}, u}$ such that
\beq
\ddet{}{I+V}\pac{ \nu,u,g }= \ddet{}{I+V}^{\pa{0}}\pac{\nu,u,g}
\paa{1 + \sul{n \geq 1}{} \mc{H}_n\pac{\nu, \ex{g}, u}  }\; .
\label{ecriture serie de Natte intro}
\enq
There
\beq
\ddet{}{I+V}^{\pa{0}}\pac{\nu,u,g} = B_x\pac{\nu,u} \cdot
\exp\Bigg\{ \int_{-q}^{\,q} \pac{ix u^{\prime}\!\pa{\la} + g^{\prime}\!\pa{\la} } \nu\pa{\la} \dd \la \Bigg\}
\; .
\enq
A more detailed  structure of the functionals $\mc{H}_N$ can be found in the core of the text, formulae \eqref{definition des fonctionnelles H_N}.
One has the following estimates for the functionals $\abs{ \mc{H}_n\pac{ \nu,\ex{g},  u } } \leq  \pac{m\pa{x}}^n$,
with $m\pa{x}=\e{O}\pa{x^{-w}}$ being $n$-independent and 
\beq
\hspace{-5mm} w=\f{3}{4} \e{min}\pa{\tf{1}{2}, 1-\wt{w}-2 \max_{\eps=\pm} \abs{ \Re\nu\pa{\eps q} }} \quad \e{with} \quad 
\wt{w} = 2 \sup\paa{ \abs{\Re \pac{ \nu\pa{\la}-\nu\pa{\eps q}}} \; : \; \abs{\la - \eps q}=\de \;, \; \eps=\pm } \;, 
\enq
where  $\de>0$ is taken small enough. Hence, the series is convergent for $x$ large enough.

The functionals $\mc{H}_n\pac{\nu,\ex{g}, u }$ take the form
\bem
\mc{H}_n\pac{\nu,\ex{g}, u } =  \mc{H}_{n}^{\pa{\infty}}\pac{\nu,\ex{g}, u}
\; + \;  \sul{ m=-\pac{\f{n}{2}} }{ \pac{\f{n}{2}} }
\f{ \ex{i x m \pac{u\pa{q}-u\pa{-q}}} }{ x^{2 m \pac{\nu\pa{q}+\nu\pa{-q}}} } \mc{H}^{\pa{m}}_n\pac{\nu, \ex{g},u} \\
+ \; \sul{ b=1 }{ \pac{\f{n}{2}} } \sul{p=0}{b} \sul{m= b - \pac{\f{n}{2}} }{\pac{\f{n}{2}}-b}
 \f{ \ex{i x m \pac{u\pa{q}-u\pa{-q}}} }{ x^{2 m \pac{\nu\pa{q}+\nu\pa{-q}}} }
 \cdot x^{\f{b}{2}} \f{ \ex{i x \bs{\eta} \pac{b u\pa{\la_0}-p u\pa{q}-\pa{b-p}u\pa{-q}  }   } }
{ x^{ 2\bs{\eta} \pa{b-p}\nu\pa{-q} - 2\bs{\eta} p  \nu\pa{q}} }
\cdot \mc{H}^{\pa{m,b,p}}_n\pac{\nu,\ex{g}, u} \;.
\label{ecriture serie Natte detaille pour chaque Hn Intro}
\end{multline}
Above, we agree upon $\bs{\eta}=1$ in the space-like regime and $\bs{\eta}=-1$ in the time-like regime. 
There $\mc{H}_{n}^{\pa{\infty}}\pac{\nu,\ex{g}, u}=\e{O}\pa{x^{-\infty}}$
and the functionals $\mc{H}^{\pa{\ell}}_n\pac{\nu, \ex{g},u}$ and $\mc{H}^{\pa{m,p,b}}_n\pac{\nu,\ex{g}, u}$ admit the asymptotc expansions
\beqa
\mc{H}_n^{\pa{m}}\pac{\nu, \ex{g},u} & \thicksim & \sul{r \geq 0}{} \mc{H}_{n; r}^{\pa{m}}\pac{\nu, \ex{g},u}  \qquad \e{with} \qquad
\mc{H}_{n; r}^{\pa{m}}\pac{\nu, \ex{g},u} = \e{O}\paf{ \pa{\log x }^{n+r -2m} }{ x^{n+r} } \; , \nonumber \\
\mc{H}^{\pa{m,b,p}}_n\pac{\nu,\ex{g}, u} & \thicksim & \sul{r \geq 0}{} \mc{H}_{n;r}^{\pa{m,b,p}}\pac{\nu,\ex{g}, u} \qquad \e{with} \qquad
\mc{H}_{n;r}^{\pa{m,b,p}}\pac{\nu,\ex{g}, u} = \e{O}\paf{ \pa{\log x }^{n+r -2\pa{m+b} } }{ x^{n+r} } \; .
\label{ecriture forme DA des fonctionnelle Hn}
\eeqa

\end{theorem}

This theorem, together with a more explicit expressions for the functionals $\mc{H}_n$, will be proven in section \ref{section Series de Natte}.
Here, we would however like to comment on the form of the asymptotic expansion. Indeed the above asymptotic expansion is not 
of the type usually encountered for higher transcendental functions. In fact, the large $x$-behavior of the functionals $\mc{H}_n\pa{\nu,\ex{g},u}$ and 
hence of the determinant $\ddet{}
{I+V}$  contains  a tower of different fractional powers of $x$, each appearing with its own oscillating pre-factor.
Once that one has fixed a given phase factor and its associated fractional power of $x$, then
the corresponding functional coefficients $\mc{H}^{\pa{m}}_n\pac{\nu,\ex{g}, u}$ or $\mc{H}^{\pa{m,b,p}}_n\pac{\nu,\ex{g}, u}$ admit an asymptotic 
expansion in the more-or-less standard sense. That is to say, each of their entries admits an asymptotic expansion
into a series whose $r^{\e{th}}$ term can be written as $\tf{P_{r+n}\pa{\log x}}{x^{n+r}}$ with $P_{r+n}$ being a polynomial of degree at most $r+n$.
One of the consequences of such a structure is that an oscillating term that appears in a sense "farther" (large values of $n$) in the asymptotic series 
might be dominant in respect to a non-oscillating term present in the "lower" orders of the Natte series.
This structure of the asymptotic expansion proves the conjectures raised in 
\cite{MullerShrockDynamicCorrFnctsTIandXXAsymptTimeAndFourier,TracyVaidyaRedDensityMatrixSpaceAsymptImpBosonsT=0}
for certain specializations of this kenrel. Also, upon specialization, it yields the general structure of the asymptotic expansion of the fifth 
Painlev\'{e} transcendent  associated to the pure sine kernel \cite{JimMiwaMoriSatoSineKernelPVForBoseGaz}. 

The series representation \eqref{ecriture serie de Natte intro} might appear abstract since there is no generic simple expression for the 
functionals $\mc{H}_n$, $n\geq 1$. 
However, the slightly more explicit (but also more cumbersome so that we did not present it a this point) 
characterization of the functionals $\mc{H}_n$, gives a thorough and explicit description of the way $\mc{H}_n$ acts on $\ex{g}$. 
This characterization, together with the overall form of the Natte series \eqref{ecriture serie de Natte intro}, 
is enough to build a multidimensional deformation of \eqref{ecriture serie de Natte intro} which describes 
a class of correlation functions appearing in integrable models \textit{away} from their free fermion points. 
The very fact that the series representation one starts with has good properties from the point of view of an
asymptotic analysis (for instance it immediately provides the leading asymptotics) leads to a multidimensional deformation
which has basically the same good properties in respect to the asymptotic analysis, in the sense that it admits 
an expansion of the type \eqref{ecriture serie de Natte intro}, \eqref{ecriture serie Natte detaille pour chaque Hn Intro}, 
\eqref{ecriture forme DA des fonctionnelle Hn}. 
As a consequence, the long-time/long-distance asymptotic behavior of two-point functions in an \textit{interacting} integrable model
can be simply \textit{read-off} by looking at the multidimensional series.

\subsection{Notations}
We now introduce several notations that we use throughout the article.

\begin{itemize}

\item $\mc{D}_{z_0,\de}=\paa{z \in \Cx \; : \; \abs{z-z_0}<\de}$ is the open disk of radius $\de$ centered at $z_0$.
$\Dp{}\mc{D}_{z_0,\de}$ stands for its canonically oriented boundary and $-\Dp{}\mc{D}_{z_0,\de}$
for the boundary equiped with the opposite orientation.

\item $\sg_3$, $\sg^{\pm}$ and $I_2$ stand for the below matrices
\beq
\sg_3=\pa{\ba{cc} 1 & 0\\ 0& -1 \ea}  \qquad \quad 
\sg^+=\pa{\ba{cc} 0 & 1\\ 0& 0 \ea}  \qquad \quad 
\sg^-=\pa{\ba{cc} 0 & 0\\ 1 & 0 \ea}  \qquad \quad 
I_2=\pa{\ba{cc} 1 & 0\\ 0 & 1 \ea} \;.
\enq

\item Given an oriented curve $\msc{C}$ in $\Cx$, $\Ga\pa{\msc{C}}$ stands for a small counterclockwise
loop around the curve $\msc{C}$. This loop is always chosen in such a way that the only potential singularities of the integrand inside of the loop
are located on $\msc{C}$. For instance, if $\msc{C}$
consists of one point $\la$, then $\Ga\pa{\msc{C}}$ can  be taken as $\Dp{} \mc{D}_{\la,\de}$, for some 
$\de>0$ and small enough.

\item When no confusion is possible, the variable dependence will be omitted, \textit{ie} $u\pa{\la}=u$, $g\pa{\la}=g$, \textit{etc}. 

\item $\log$ refers to the $\intoo{-\pi}{\pi}$ determination of the logarithm, and it is this determination that is used for defining powers.

\item Given a set $U$, $\overset{\circ}{U}$ refers to its interior and $\ov{U}$ to its closure.

\item $\mathbb{H}_+$, resp. $\mathbb{H}_-$, stands for the upper $\paa{z \in \Cx  : \; \Im\pa{ z} >0}$, resp. lower $\paa{z \in \Cx  : \; \Im\pa{ z} <0}$, half-planes.

\item  Given matrix valued functions $M\!\pa{\la}$, $N\!\pa{\la}$, the relation $M\!\pa{\la}= \e{O}\pa{N\!\pa{\la}}$ is to be understood
entry-wise $M_{k \ell}\pa{\la}= \e{O}\pa{N_{k\ell}\pa{\la}}$.

\item Given an oriented curve $\msc{C}$, one defines its $+$ (resp. $-$) side as the one lying to the left
(resp. right) when moving along the curve.  Above and in the following, given any function or matrix function $f$, $f_{\pm}\pa{\la}$ stands for the
non-tangential limit of $f\pa{z}$ when $z$ approches the point $\la \in \msc{C}$ from the $\pm$ side of the oriented curve $\msc{C}$.

\item Given a piecewise smooth curve $\msc{C}$ and matrix $M$ with entries in $L^{p}\pa{\msc{C}}$, $p=1,2, \infty$,
we use the canonical matrix norms ($\dagger$ stands for Hermitian conjugation):
\beq
\hspace{-4mm}\norm{M}_{ L^{\infty}\pa{\msc{C}} } = \max_{i,j} \norm{M_{ij}}_{ L^{\infty}\pa{\msc{C}} }  \; , \quad \;
\norm{M}_{ L^{2}\pa{\msc{C}} }  = \sqrt{ \norm{  \e{tr}\pac{ M^{\dagger} M }  }_{ L^{1}\pa{\msc{C}} }  }
\quad \e{and} \quad \norm{M}_{ L^{1}\pa{ \msc{C} } }  = \max_{ij} \norm{  M_{ij}  }_{ L^{1}\pa{\msc{C}} }
\;.
\enq

\item The distance between any two subsets $A,B$ of $\Cx$ will be denoted by $\e{d}\!\pa{A,B} \equiv \inf\paa{\abs{x-y} \; : \; x \in A \; , \; y \in B}$.

\end{itemize}

\subsection{Several remarks}
\label{Soussection remarques}
It now seems to be a good place so as to gather several remarks in respect to our assumptions.

\begin{figure}[h]
\begin{center}

\begin{pspicture}(6,4)

\psline[linestyle=dashed, dash=3pt 2pt]{->}(0,2)(6,2)


\psdots(2,2)(4,2)

\rput(2,1.8){$-q$}
\rput(4,1.8){$q$}

\pscurve(0,3.3)(0.7,3)(1,2.8)(1.2,2)(1.5,2)(2,2)(4,2)(4.5,2)(5,2)(5.5,1.4)(6,1)

\end{pspicture}

\caption{Contour $\msc{C}_{E}$ for the definition of $E$.\label{contour pour transfo Cauchy et Hilbert}}
\end{center}
\end{figure}
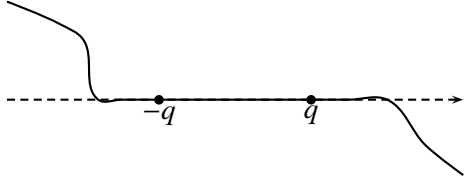

\begin{itemize}

\item The assumptions on the type of the saddle-point at $\la_0$
guarantee that there exists a local parametrization for $u\!\pa{\la}$ around $\la_0$,
$u\!\pa{\la}-u\!\pa{\la_0}=- \om^2\!\pa{\la}$ with $\om\!\pa{\la} =\pa{\la-\la_0} h\pa{\la}$, where $h\pa{\la_0}\not= 0$ and $h$
is holomorphic on $\mc{D}_{\la_0,\de}$ for some $\de>0$.

\item As it will become apparent from our asymptotic analysis, given functions $u,\nu, g$ satisfying to all the hypothesis,
one has that $\ddet{}{I+V} \not=0$ for $x$ large enough.

\item The assumption on the number of saddle-points and their order can be relaxed in principle.
RHP with multiple saddle-points have been considered in \cite{VarzuginOscillatoryRHPTypePDE}. This work was later
extended to the case of less regular functions and higer order saddle-points in \cite{DoRHPApproachGeneralizationVarzuginOscRHP}.

\item  The restriction on the real part of $\nu$ in the vicinity of $\pm q$ is of technical nature. It allows us to avoid the
analysis related to the so-called ambiguous Fisher--Hartwig symbols. The method for dealing with
such kinds of problems in the framework of Riemann--Hilbert problems
has been proposed in
\cite{DeiftItsKrasovskyAsymptoticsofToeplitsHankelWithFHSymbols,DeiftItsKrasovskyProofOfGeneralizedFHConjectureAndMoreAnnouncmentResults}. The cases where
$\Re\pa{\nu\pa{\pm q}} \geq \tf{1}{2}$ could in principle
be treated along these techniques, but we chose not to venture into these technicalities.

\item We have depicted the contour $\msc{C}_{E}$ appearing in
principal value integral in \eqref{definition fonction E+} on Fig.~\ref{contour pour transfo Cauchy et Hilbert}.
This contour $\msc{C}_{E}$ is chosen in such a way that the integral is converging exponentially fast at infinity.
This avoids us unnecessary complications and corresponds to most, if not all, situations that can arize in interacting integrable models.

\item  In the case of kernels involved in the representation of the two-point functions in integrable models, the function $u$
takes the form  $u\pa{\la}=p\pa{\la} -\tf{t \varepsilon\pa{\la} }{x} $.
$p$ corresponds to the momentum of excitations whereas $\varepsilon$ corresponds to their energy.
The parameter $t$ plays the role of the time-shift between the two operators and $x$ that of their  distance of separation.
In general, one is interested in the large-distance/long-time behavior of the two-point function in the case where the ratio $\tf{t}{x}$
is fixed. In such a limit, for many models of interest, the function $u$ has a unique saddle-point on $\R$.
This physical interpretation can be seen as a motivation for certain of our assumptions.

\item It is not a problem to carry out the same analysis in the case where the contour $\msc{C}_E$ given in 
Fig.\ref{contour pour transfo Cauchy et Hilbert} is replaced by $\msc{C}_E^{\pa{w}} = \msc{C}_E\cap\paa{ z \in \Cx \; : \; \abs{\Re\pa{z}}\leq w}$,
with $w \in \R^+$ such that $q$, $-q$  and $\la_0$ belong to $\intoo{-w}{w}$. Up to minor modifications due to such a truncation
of the remote part of the contour, the results remain unchanged. 

\end{itemize}

%
%
%


\section{The initial RHP and some transformation}
\label{section transfos RHP}

\subsection{The RHP for $\chi$}

The kernel of any integrable integral operator can be recast in a form allowing one to give a convenient characterization of the kernel $R\pa{\la,\mu}$
of the resolvent operator $I-R$ to $I+V$.

Namely, in the case of the kernel $V$ given in \eqref{definition noyau GSK FF} one sets
\begin{equation}
\ket{E^R\pa{\la}}=\f{  2 \sin\pac{\pi \nu\pa{\la} } }{i\pi}
                            \begin{pmatrix}
                                E\pa{\la}\\
                                e\pa{\la}
                            \end{pmatrix}, \qquad
\bra{E^L\pa{\la}}= \sin\pac{\pi \nu\pa{\la} }
                           \begin{pmatrix}
                               -e\pa{\la}\, , \,
                                E\pa{\la}
                            \end{pmatrix},
\end{equation}
\noindent so that the kernel $V$ is expressed as the scalar product:
\begin{equation}
V\pa{\la,\mu}=\f{\braket{E^L\pa{\la}}{E^R\pa{\mu}}}{\la-\mu} \;.
\end{equation}
The resolvent $I-R$ of $I-V$ exists if $\ddet{}{I+V}\not=0$. In that case, one defines $\ket{F^R\pa{\la}}$ as the unique solution
to the integral equation:
\begin{equation}
\ket{F^R\pa{\mu}}+\Int{-q}{q}  V\pa{\mu,\la} \ket{F^{R}\pa{\la}} \dd
\la= \ket{E^{R}\pa{\mu}}\;, \qquad 
\ket{F^{R}\pa{\la}} = \f{  2 \sin\pac{\pi \nu\pa{\la} } }{i\pi}
                            \begin{pmatrix}
                                F_1\pa{\la}\\
                                F_2\pa{\la}
                            \end{pmatrix} 
\label{definition fction fplus-moins}
\end{equation}
where the integration is to  be understood entry-wise. $\bra{F^L\pa{\la}}$ corresponds to the solution of the integral equation
where $\ket{E^R\pa{\la}}$ has been replaced with $\bra{E^L\pa{\la}}$. It was shown in 
\cite{ItsIzerginKorepinSlavnovDifferentialeqnsforCorrelationfunctions} that the resolvent kernel can be represented as:
\beq
R(\la,\mu)= \f{\braket{F^L\pa{\la}}{F^R\pa{\mu}}}{\la-\mu}\; .
\label{Reconstruction du Resolvent}
\enq
It is well know since the results established in
\cite{DeiftItsZhouSineKernelOnUnionOfIntervals,ItsIzerginKorepinTempCorrBoseGasIntSyst,ItsIzerginKorepinSlavnovDifferentialeqnsforCorrelationfunctions}
that the study of many properties
(construction of the resolvent, calculation of the Fredholm determinant, construction of a system of partial differential equations for the determinant)
of the so-called integrable integral operators
$I+V$ can be  deduced from the solution of a certain RHP. In the case of the kernel of interest, this RHP reads
\begin{itemize}
\item $\chi$ is analytic on  $\Cx \setminus\intff{-q}{q}$ and has continuous boundary values on $\intoo{-q}{q}$\; ;
\item $\chi\pa{\la} = \e{O}\left(\ba{cc} 1 & 1 \\
                                 1 & 1 \ea   \right)
                                 \log\abs{\la^2-q^2}
                                 \quad \text{for}\ \la \tend \pm q$\; ;
\item $\chi\pa{\la} = I_2  + \la^{-1}\e{O}\left(\ba{cc} 1 & 1 \\
                                 1 & 1 \ea   \right) $ \; uniformly in $\la \tend \infty$\; ;
\item $\chi_{+}\pa{\la} G_{\chi}\pa{\la}=\chi_-\pa{\la}  \quad \text{for} \  \la \in \intoo{-q}{q}$ \;.
\end{itemize}
We remind that $\chi_{\pm}$ stands for the $\pm$ boundary value of $\chi$ from the $\pm$-side of its jump curve.

The jump matrix $G_{\chi}\pa{\la}$ appearing in the formulation of this RHP reads
\beq
G_{\chi}\pa{\la}=I_2+ 4\sin^2\pac{ \pi\nu\pa{\la} } \pa{ \ba{c} E \pa{\la} \\ e\pa{\la} \ea} \pa{-e\pa{\la} ,\, E\pa{\la} } \; .
\enq
The above RHP admits a solution as long as $\ddet{}{I+V}\not=0$. Indeed, it has been shown in \cite{ItsIzerginKorepinSlavnovDifferentialeqnsforCorrelationfunctions} that
the matrix
\beq
\chi(\la)=I_2-\Int{-q}{q}
\f{\ket{F^R(\mu)}\bra{E^L(\mu)}}{\mu-\la}\, \dd \mu \;, \qquad
\chi^{-1}(\la)=I_2+\Int{-q}{q}
\f{\ket{E^R(\mu)}\bra{F^L(\mu)}}{\mu-\la}\, \dd \mu \; .
\label{forme explicite de chi en terms f plus moins}
\enq
solves the above RHP.
This solution is in fact unique, as can be  seen by standard arguments
\cite{BealsCoifmanScatteringInFirstOrderSystemsEquivalenceRHPPRoofRHPUniqueness}. It follows \cite{ItsIzerginKorepinSlavnovDifferentialeqnsforCorrelationfunctions}
readily from \eqref{forme explicite de chi en terms f plus moins} that the solution $\chi\pa{\la}$ allows one to construct $\ket{F^{R}\pa{\la}}$
and $\bra{F^{L}\pa{\mu}}$:
\beq
\ket{F^{R}\pa{\la}}= \chi\pa{\la}\ket{E^{R}\pa{\la}} \; , \qquad  \qquad \bra{F^{L}\pa{\la}}=
\bra{E^{L}\pa{\la}} \chi^{-1}\pa{\la} \;.
\label{reconstruction F en fonction E}
\enq

\subsection{Relation between $\chi$ and $\ddet{}{I+V}$}

One can express partial derivatives of $\ddet{}{I+V}$
in respect to the various parameters entering in the definition of $V\pa{\la,\mu}$ in terms of the solution
$\chi\pa{\la}$ to the above RHP. We will derive a set of such identities below.
 These will play an important role in our analysis.
\begin{prop}
Let $\eta \geq 0$ and $\Ga\pa{\msc{C}_{E}}$ be a loop in $U$ enlacing counterclockwisely $\msc{C}_{E}$ and such that 
it goes to infinity in the regions where $\ex{i\eta u\pa{z}}$, $\eta>0$ is decaying exponentially fast.
Then,
\beqa
\Dp{x} \log \ddet{}{I+V}  &=&  -i \f{\Dp{}}{\Dp{}{\eta}} \paa{ \;   \Oint{ \Ga\pa{ \msc{C}_{E} } }{} \f{ \dd z }{ 4\pi }  \ex{i\eta u\pa{z}}
\e{tr}\pac{ \pa{\Dp{z}\chi}\pa{z}  \pa{ \sg_3+ 2C\pac{e^{-2}}\pa{z} \sg^+ } \chi^{-1}\pa{z}  }   }_{\eta=0^+} \hspace{-4mm} ,
 \label{formule derviee x log det}\\
\Dp{\la_0} \log \ddet{}{I+V}  &=&  x
\paa{ \;  \Oint{\Ga\pa{ \msc{C}_{E} } }{} \f{ \dd z }{ 4\pi }  \pac{\Dp{\la_0} u \pa{z} } \ex{i\eta u\pa{z}}
\e{tr}\pac{ \pa{\Dp{z}\chi}\pa{z}  \pa{ \sg_3+ 2C\pac{e^{-2}}\pa{z} \sg^+ } \chi^{-1}\pa{z} }  }_{ \eta=0^+}  \hspace{-5mm} .
\label{formule derviee lambda 0 log det}
\eeqa
There, $C\pac{f}$  stands for the Cauchy transform on $\msc{C}_{E}$ and $C_{\pm}\pac{G}$ for its $\pm$ boundary values on $\msc{C}_{E}$.
One has more explicitly
\beq
C\pac{G}\pa{\la}=\Int{ \msc{C}_{E} }{} \! \f{ G\pa{\mu} }{ \mu-\la } \, \f{\dd \mu}{2i\pi}   \; ,
\quad \e{and} \qquad C_+\pac{G}\pa{\la}-C_-\pac{G}\pa{\la}=G\pa{\la} \; , \quad \e{for} \quad \la\in \msc{C}_{E}.
\label{definiton transforme de Cauchy}
\enq
\end{prop}

\Proof
The proof goes along similar lines to \cite{KozKitMailSlaTerRHPapproachtoSuperSineKernel}. It is straightforward that
\beq
\Dp{x}\log\ddet{}{I+V} = \Int{-q}{q}  \pac{ \Dp{x} V \cdot \pa{I-R} }\pa{\la,\la} \dd \la \;.
\label{equation donnant Dp x log det trivial}
\enq
In order to transform \eqref{equation donnant Dp x log det trivial} into \eqref{formule derviee x log det}, one should start by writing a convenient
representation for $\Dp{x}V\pa{\la,\mu}$.
One has that
\beq
\Dp{x} e\pa{\la} = -\f{i}{2} u\pa{\la}  e\pa{\la}  \quad , \qquad
\Dp{x} E\pa{\la} = \f{i}{2} u\pa{\la} E\pa{\la}   -
 e\pa{\la} \Int{ \msc{C}_{E} }{}  \f{\dd s}{2\pi}  \f{u\pa{s}-u\pa{\la}  }{ s-\la }  e^{-2}\pa{s} \;.
\enq
The last integral can be recast in a more convenient form
\bem
\Int{ \msc{C}_{E} }{}  \f{\dd \mu}{2\pi}  \f{u\pa{\mu}-u\pa{\la}  }{ \mu-\la }  e^{-2}\pa{\mu} =
  \Int{ \msc{C}_{E} }{}  \f{\dd \mu}{2\pi}   \Oint{\Ga\pa{\paa{\la,\mu}}}{} \hspace{-2mm}\f{\dd z}{2i\pi} \, \f{u\pa{z} }{ \pa{z-\la}\pa{z-\mu} }
  e^{-2}\pa{\mu}     \\
= -i \f{\Dp{}}{\Dp{}{\eta}}  \paa{ \; \Int{ \msc{C}_{E} }{}  \f{\dd \mu}{2\pi}   \Oint{\Ga\pa{\paa{\la,\mu}}}{}\hspace{-2mm} \f{\dd z}{2i\pi} \f{\ex{i\eta u\pa{z}} }{ \pa{z-\la}\pa{z-\mu} }  e^{-2}\pa{\mu} }_{\eta=0^+}
= -i \f{\Dp{}}{\Dp{}{\eta}}  \paa{ \;  \Int{\msc{C}_{E}}{}  \f{\dd \mu}{2\pi}   \Oint{\Ga\pa{ \msc{C}_{E} }}{} \hspace{-2mm}\f{\dd z}{2i\pi}
\f{\ex{i\eta u\pa{z}} }{ \pa{z-\la}\pa{z-\mu} }  e^{-2}\pa{\mu} }_{\eta=0^+}
  \\
= -i \f{\Dp{}}{\Dp{}{\eta}}  \paa{ \;  \Oint{\Ga\pa{ \msc{C}_{E} }}{} \hspace{-2mm} \f{\dd z}{2i\pi} \f{\ex{i\eta u\pa{z}} }{ \pa{z-\la} }
\; \Int{ \msc{C}_{E} }{}  \f{\dd \mu}{2\pi}  \f{ e^{-2}\pa{\mu} }{z -\mu}}_{\eta=0^+}
 =  i \f{\Dp{}}{\Dp{}{\eta}} \paa{ \;   \Oint{\Ga\pa{ \msc{C}_{E} }}{} \f{\dd z}{2\pi} \f{\ex{i\eta u\pa{z}} C\pac{e^{-2}}\pa{z} }{ \pa{z-\la} }  }_{\eta=0^+}  \hspace{-4mm}.
\label{ecriture suite de transfos pour obtenir eta derivee}
\end{multline}

We first have replaced the ratio of differences by a contour integral on $\Ga\pa{\paa{\la,\mu}}$. Here $\Ga\pa{\paa{\la,\mu}}$
consists of two small loops around the points $\la$ and
$\mu$. In order to manipulate convergent integrals, we then wrote the integral as an $\eta$-derivative.
The derivative symbol could then be taken out of the integral. Next we deformed the contour of integration from a compact one $\Ga\pa{\paa{\la,\mu}}$
into $\Ga\pa{\msc{C}_{E}}$.
Such a replacement is allowed as $\msc{C}_{E}$ is chosen precisely in such a
way so as to make $\ex{i \eta u\pa{\la}}$, $\eta>0$, decay exponentially on a small neighborhood of $\msc{C}_{E}$
where one can draw $\Ga\pa{\msc{C}_{E}}$.
Such a choice of contours allows us to satisfy to the hypothesis of Fubini's theorem and hence permute the orders of integration.
Also, we stress that one should compute  the $\eta$-derivative only once that all integrals have been computed. Indeed,
for generic choices of functions $u$, permuting the $\eta$-derivation with the $\la$-integration
in the last line of \eqref{ecriture suite de transfos pour obtenir eta derivee}, leads to an $a priori$ divergent integral.

Once that this differential identity is established, one readily convinces oneself that
\beq
\Dp{x} V\pa{\la,\mu} = i  \f{\Dp{}}{\Dp{}{\eta}}   \paa{\; \Oint{ \Ga\pa{\msc{C}_{E}} }{} \!\!
\f{\ex{i\eta u\pa{z}} }{ \pa{z-\la}\pa{z-\mu}  }
\bra{E^L\pa{\la}} \pa{\sg_3 +2C\pac{e^{-2}}\pa{z} \sg^+ } \ket{ E^R\pa{\mu} } \; \f{\dd z}{4\pi}  }_{\eta=0^+}  \hspace{-3mm}.
\enq
Denoting $S\pa{z}=\sg_3 +2C\pac{e^{-2}}\pa{z} \sg^+$, using the representation \eqref{Reconstruction du Resolvent} of the resolvent $R$   and the
fact that
\newline $\braket{F^L(\la)}{F^R(\mu)}=\tr \pac{ 
\ket{F^R(\mu)}\bra{F^L(\la)}}$, we get
\begin{multline}
\Dp{x}\log \ddet{}{I+V}
= i\f{\Dp{}}{\Dp{}{\eta}}  \paa{ \; \Oint{ \Ga\pa{ \msc{C}_{E} } }{} \hspace{-1mm} \f{\dd z}{4\pi}\, \ex{i\eta u\pa{z}}
\Int{-q}{q} \dd \la \, \f{\bra{E^{L}\pa{\la}} S\pa{z} \ket{E^R\pa{\la}}}{\pa{z-\la}^2} }_{\eta=0^+}  \\
 \hspace{1.5cm} -i \f{\Dp{}}{\Dp{}{\eta}}    \e{tr}\paa{\; \Oint{ \Ga\pa{ \msc{C}_{E} } }{} \hspace{-1mm}
 \f{\dd z}{4\pi}\, \ex{i\eta u\pa{z}}   \Int{-q}{q} \dd \la \dd \mu\,
\ket{F^R\pa{\la}}\bra{E^L\pa{\la}} \right.
 \\
\times \left. \pa{ \f{1}{\la-z}-\f{1}{\la-\mu} } S\pa{z}\f{\ket{E^R\pa{\mu}}\bra{F^L\pa{\mu}} }{\pa{\mu-z}^2}
 \vphantom{ \Oint{ \Ga\pa{ \msc{C}_{E} } }{}  } }_{\eta=0^+} \hspace{-4mm}.
\end{multline}
Using the integral expressions \eqref{forme explicite de chi en
terms f plus moins} for $\chi$ and $\chi^{-1}$, we obtain
\begin{align}
\Dp{x}\log \ddet{}{I+V}
&= i  \f{\Dp{}}{\Dp{}{\eta}}    \left\{  \; \Oint{\Ga\pa{ \msc{C}_{E}} }{} \hspace{-1mm} \f{\dd z}{4\pi}\, \ex{i\eta u\pa{z}}  \Int{-q}{q} \dd \la \,
\f{\bra{E^{L}\pa{\la}} S\pa{z} \ket{E^R\pa{\la}}}{\pa{z-\la}^2}   \right.
            \nonumber\\
&\qquad
   \left.  -i \f{\Dp{}}{\Dp{}{\eta}}   \e{tr}\pac{ \;   \Oint{\Ga\pa{ \msc{C}_{E} } }{} \hspace{-1mm} \f{\dd z}{4\pi}\, \ex{i\eta u\pa{z}} \Int{-q}{q} \dd \mu
\pa{\chi(\mu)-\chi(z)} S\pa{z} \f{\ket{E^R(\mu)}\bra{F^L(\mu)} }{\pa{\mu-z}^2} }    \right\}_{\eta=0^+}
            \nonumber\\
 &=
-i  \f{\Dp{}}{\Dp{}{\eta}}  \paa{ \;  \Oint{\Ga\pa{ \msc{C}_{E} } }{} \hspace{-1mm} \f{\dd z}{4\pi} \, \ex{i\eta u\pa{z}} \e{tr}\paa{ \Dp{z}\chi\pa{z} S\pa{z}
\chi^{-1}\pa{z}} }_{\eta=0^+} \hspace{-4mm}, \hspace{3cm}
\end{align}
where we used \eqref{reconstruction F en fonction E}.
The proof of identity \eqref{formule derviee lambda 0 log det} goes along very similar lines.

\qed


\section{The first set of transformations on the RHP}
\label{Section first set of transformations on RHP}

We now perform several transformations on the original RHP. We first simplify the form of the oscillating functions $E$ appearing in the formulation
of the RHP. This step in carried in the spirit of \cite{ItsIzerginKorepinVarguzinTimeSpaceAsymptImpBoseGaz}.
Then, we map this new RHP into one whose jump matrix can be written as the identity plus some purely off-diagonal matrix.
Finally, we apply the
non-linear steepest descent method by deforming the contour so as to obtain jump matrices that are a $\e{O}\pa{x^{-\infty}}$
uniformly away from $\pm q$ and $\la_0$.
These last steps are a standard implementation of the Deift-Zhou steepest descent method 
\cite{DeiftZhouSteepestDescentForOscillatoryRHPmKdVIntroMethod,DeiftZhouSteepestDescentForOscillatoryRHP}.

\subsection{Simplification of the function $E$}

In order to replace the complicated function $E$ by $e^{-1}$, we perform the substitution
\beq
\chi\pa{\la}=\wt{\chi}\pa{\la}\pa{I_2+\sg^+ C\pac{e^{-2}}\pa{\la} }
\label{definition chi tilde en termes de chi}
\enq
where $C$ is the rational Cauchy transform with support on $\msc{C}_{E}$ defined in \eqref{definiton transforme de Cauchy}.

It is readily checked that this new matrix $\wt{\chi}$ is the unique solution to the RHP

\begin{itemize}
\item $\wt{\chi}$ is analytic on  $\Cx \setminus\msc{C}_{E} $ and has continuous boundary values on $\msc{C}_{E}\setminus \paa{\pm q}$\;;
\item $\wt{\chi}\pa{\la} = I_2+ \la^{-1} \e{O}\left(\ba{cc} 1 & 1 \\
                                 1 & 1 \ea   \right) $, uniformly in  $\la \tend \infty$\;;
\item $\wt{\chi}\pa{\la} = \e{O}\left(\ba{cc} 1 & 1 \\
                                 1 & 1 \ea   \right)
                                 \log\abs{\la^2-q^2}
                                 \quad \text{for}\ \la \tend \pm q$\;;
\item $\wt{\chi}_{+}\pa{\la} G_{\wt{\chi}}\pa{\la}=\wt{\chi}_-\pa{\la}  \quad \text{for} \  \la \in \msc{C}_{E} \;\;$ .
\end{itemize}
The jump matrix for $\wt{\chi}$ takes two different forms

\beq
 G_{\wt{\chi}} \pa{\la}  = \pa{ \ba{cc}
                            \ex{-2i\pi\nu\pa{\la}}  & 0 \\
                            \ex{2i\pi\nu\pa{\la}} e^{2}\!\pa{\la}\pa{\ex{-2i\pi\nu\pa{\la}} -1}^2  &  \ex{2i\pi\nu\pa{\la}}
                            \ea } \;  \qquad
		\e{for} \; \la \in \intoo{-q}{q}
\enq
and
\beq
 G_{\wt{\chi}} \pa{\la}   = \pa{ \ba{cc}
                            1  &  e^{-2}\pa{\la}\\
                            0  &  1
                            \ea } \;   \qquad
					 \e{for}  \; \la \in \msc{C}_{E} \setminus \intff{-q}{q} \; .
\enq

The existence and uniqueness of solutions for the RHP for $\wt{\chi}$ ensures that there is a one-to-one correspondence between $\chi$ and $\wt{\chi}$.

\subsection{Uniformization of the jump matrices}

We now carry out the second substitution that will yield a RHP with upper or lower diagonal jump matrices whose diagonal is the identity.
For this purpose, we define
\beq
\a\pa{\la} = \kappa\pa{\la} \paf{\la+q}{\la-q}^{\nu\pa{\la}} \; , \qquad \e{where} \qquad  \log\kappa\pa{\la} = - \Int{-q}{q} \f{\nu\pa{\la}-\nu\pa{\mu}}{ \la-\mu} \dd \mu \; .
\enq
The function $\a\pa{\la}$ is holomorphic on $\Cx \setminus \intff{-q}{q}$, $\a\pa{\la} \limit{\la}{\infty} 1$ and satisfies to the jump condition
\beq
\a_+\pa{\la} \ex{2i\pi \nu \pa{\la}}=\a_-\pa{\la} \; \quad \e{for} \qquad  \la \in \intoo{-q}{q}  \;.
\label{ecriture condition saut alpha}
\enq

Then we set
\beq
\Xi\pa{\la}= \wt{\chi}\pa{\la} \a^{\sg_3}\pa{\la} \; .
\enq
The matrix $\Xi\pa{\la}$ is the unique solution to the RHP:
\begin{itemize}
\item $\Xi$ is analytic on $\mathbb{C}\setminus\msc{C}_{E}$ and has continuous boundary values on $\msc{C}_{E} \setminus \paa{\pm q}$ ;
\item $\Xi(\la) =  \e{O}\left(\ba{cc}
                                1 & 1 \\
                                1 & 1 \ea\right) \pa{\la-q}^{-\sg_3 \nu\pa{q}}
\pa{\la+q}^{\sg_3 \nu\pa{-q} }  \log\abs{\la^2-q^2}$ \; for \; $\la \tend \pm q$ ;
\item $\Xi (\la) = I_2 + \la^{-1} \e{O}\left(\ba{cc}
                                1 & 1 \\
                                1 & 1 \ea\right) \; $ uniformly in $\la \tend \infty$ ;
\item $\Xi_{+}(\la)\, G_{\Xi}(\la)=\Xi_-(\la)   \quad\text{for}\; \la \in \msc{C}_{E} \; .$
\end{itemize}

The new jump matrix $G_{\Xi}\pa{\la}$ reads
\beq
G_{\Xi}=\pa{ \ba{cc}
                            1  &   \a^{-2} e^{-2} \\
                             0 &  1
                            \ea } \quad \la \in \msc{C}_{E} \setminus \intff{-q}{q} \quad \e{and} \quad
G_{\Xi}=\pa{ \ba{cc}
                            1  &  0 \\
                             \ex{2i\pi\nu} \a_+\a_- e^{2}\pa{\ex{-2i\pi\nu} -1}^2 &  1
                            \ea } \quad \la \in \intoo{-q}{q} \;.
\nonumber
\enq

Again, there is a one-to-one correspondence between $\Xi$ and $\chi$.

\subsection{Deformation of the contour}

We now perform the third substitution that will result in a change of the shape of jump contour. Due to the fact that $e^{\pm 1}\!\pa{\la}$
are exponentially small in $x$ in appropriate regions of the complex plane, we will end up with a RHP for an unknown matrix $\Ups$
whose jump matrices are $I_2+\e{O}\pa{x^{-\infty}}$ and this for $\la$ uniformly away from the points $\pm q$ and $\la_0$.


\subsubsection{The time-like regime}

\begin{figure}[h]
\begin{center}

\begin{pspicture}(12,7)


\pscurve[linestyle=dashed, dash=3pt 2pt] (0,4.4)(0.5,4.3)(1.5,4)(2,3.5)(2.5,3.5)(3,3.5)(6,3.5)(9,3.5)(9.5,3.5)(10,3.3)(11,3)(11.5,2.7)

\rput(1,3.9){$\msc{C}_{E}$}


\psdots(3,3.5)(6,3.5)(9,3.5)
\rput(2.7,3.3){$-q$}
\rput(5.7,3.7){$\la_0$}
\rput(9.2,3.7){$q$}


\pscurve(0.2,6)(1,5.5)(2,5.7)(2.2,5)(2.5,4.5)(3,3.5)(3.3,3.2)(3.4,3)(3.5,2.5)
(3.7,2)(4,1.5)(4.5,1)(5,1.2)(5.5,1.6)(5.7,2.2)(5.9,3)(6,3.5)(6.2,4)(6.5,4.4)(7,6)(7.2,5.5)
(8.5,5.8)(8,4.9)(9,3.5)(9.5,2)(10,1.5)(11.5,2)


\psline[linewidth=3pt]{->}(2.75,4)(2.8,3.85)





\rput(0.2,5.5){$\Ga^{\pa{L}}_{\ua}$}
\rput(3,2){$\Ga^{\pa{L}}_{\da}$}

\rput(6.5,6){$\Ga^{\pa{R}}_{\ua}$}

\rput(8.9,2.5){$\Ga^{\pa{R}}_{\da}$}

\rput(1,4.8){$\Ups=\Xi M$}

\rput(4,5){$\Ups=\Xi$}
\rput(1,2.3){$\Ups=\Xi$}

\rput(4.7,2.65){$\Ups=\Xi \pac{N^{\pa{L}}} ^{-1}$}

\rput(7.5,4){$\Ups=\Xi N^{\pa{R}}$}

\rput(10.5,2.5){$\Ups=\Xi  M^{-1}$}

\rput(7.5,2){$ \Ups=\Xi$}

\end{pspicture}

\caption{Contour $\Sg_{\Ups}=\Ga^{\pa{L}}_{\ua}\cup \Ga^{\pa{L}}_{\da}\cup\Ga^{\pa{R}}_{\ua}\cup \Ga^{\pa{R}}_{\da}$ appearing in the RHP for $\Ups$  (time-like regime) .\label{contour pour le RHP de Upsilon time-like}}
\end{center}
\end{figure}
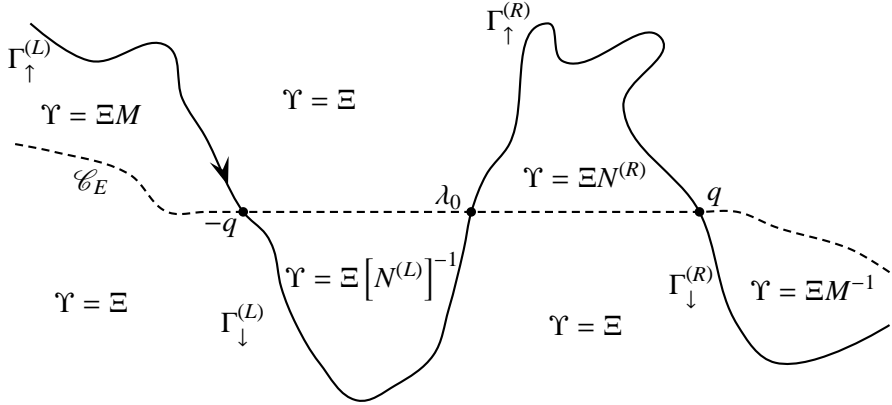

We first introduce three auxiliary matrices $M\!\pa{\la}$ and $N^{\pa{L/R}}\!\pa{\la}$
\beqa
M\!\pa{\la} &=& \hspace{8mm}\pa{ \ba{cc}
                            1  &  \a^{-2}\pa{\la} e^{-2}\!\pa{\la} \\
                             0 &  1
                            \ea }  \hspace{8mm}\;= \; I_2 + P\pa{\la} \sg^+
\label{definition matrice M time like} \\ 
N^{\pa{L}}\!\pa{\la}&=&\pa{ \ba{cc}
                            1  &  0 \\
                              \a^{2}_-\!\pa{\la} e^{2}\!\pa{\la}\pa{\ex{-2i\pi\nu\pa{\la}} -1}^2 &  1
                            \ea } \;= \; I_2 + Q^{\pa{L}}\!\pa{\la} \sg^-     \nonumber \\ 
N^{\pa{R}}\!\pa{\la}&=&\pa{ \ba{cc}
                            1  &  0 \\
                    \a^{2}_+\!\pa{\la} \ex{4i\pi\nu\pa{\la}} e^{2}\!\pa{\la}\pa{\ex{-2i\pi\nu\pa{\la}} -1}^2 &  1
                            \ea } \;= \; I_2 + Q^{\pa{R}}\!\pa{\la} \sg^- \; .
\label{definition matrices N left et right}
\eeqa
Note that although the matrices $N^{\pa{R/L}}\!\pa{\la}$ have different expressions, they coincide on $U$ due to the jump conditions for $\a\pa{\la}$.
It is clear from its very definition that $N^{\pa{L}}\!\pa{\la}$, resp. $N^{\pa{R}}\!\pa{\la}$,  has an analytic continuation to some
neighborhood of  $\intff{-q}{q}$ in the lower, resp. upper, half-plane.
Also, the matrix $M\pa{\la}$ has an analytic continuation to $U\setminus \intff{-q}{q}$  starting from $\msc{C}_{E}\setminus \intff{-q}{q}$.

The functions $P$ and $Q^{\pa{L}}$ and $Q^{\pa{R}}$ have the local parameterizations around $\pm q$
\beq
P\pa{\la}=\ex{i\zeta_{-q}} \pac{\zeta_{-q}}^{-2\nu\pa{\la}} \f{\ex{2i\pi\nu\pa{\la}}-1}{C^{\pa{L}}\!\pa{\la}} \quad
\e{for} \; \la \in \Dp{}\mc{D}_{-q,\de} \; , \quad
P\pa{\la} = \ex{-i\zeta_{q}} \pac{\zeta_{q}}^{2\nu\pa{\la}} \f{\ex{2i\pi\nu\pa{\la}}-1}{C^{\pa{R}}\!\pa{\la}}
\quad
\e{for} \; \la \in \Dp{}\mc{D}_{q,\de} \; ,
\label{definition fonction P time-like}
\enq
\beq
Q^{\pa{L}}\!\pa{\la}=C^{\pa{L}}\!\pa{\la} \, \ex{-i\zeta_{-q}} \pac{\zeta_{-q}}^{2\nu\pa{\la}} \pa{\ex{2i\pi\nu\pa{\la}}-1} \;  \qquad \e{and} \qquad
Q^{\pa{R}}\!\pa{\la}=  C^{\pa{R}}\!\pa{\la} \ex{i\zeta_{q}} \pac{\zeta_{q}}^{-2\nu\pa{\la}} \pa{\ex{2i\pi\nu\pa{\la}}-1} \; .
\label{definition fonction Q left and right time-like}
\enq
There $\zeta_{-q}=x\pa{u\pa{\la}-u\pa{-q}}$ and $\zeta_{q}=x\pa{u\pa{q}-u\pa{\la}}$ and we have set
\beqa
C^{\pa{L}}\!\pa{\la} &=& - \f{\kappa^{2}\!\pa{\la} \ex{-g\pa{\la}-ixu\pa{-q}} }{ \pac{x\pa{q-\la}}^{2\nu\pa{\la}} }
 \paf{\la+q}{u\pa{\la}-u\pa{-q}}^{2\nu\pa{\la}}  \pa{\ex{-2i\pi \nu\pa{\la}}-1} \; ,   \nonumber \\
C^{\pa{R}}\!\pa{\la} &=&  \f{ -  \kappa^{2}\!\pa{\la} }{ \ex{g\pa{\la}+ixu\pa{q}} }
\pa{ \f{u\pa{\la}-u\pa{q}}{\la-q} +i0^+ }^{2\nu\pa{\la}} \hspace{-4mm}\pac{x\pa{\la+q}}^{2\nu\pa{\la}} \pa{\ex{-2i\pi \nu\pa{\la}}-1} \; .
\label{definition fonction C left right timelike}
\eeqa

We now define a piecewise analytic matrix $\Ups$ according to Fig.~\ref{contour pour le RHP de Upsilon time-like}.
We will be more specific about the choice of the contours $\Ga^{\pa{L/R}}_{\ua/\da}$ around the points $\pm q$ and $\la_0$ when we will be constructing
the local parametrices. Here, we only precise that the jump contour for $\Ups$ remains in $U$ and that all jump curves 
are chosen so that, for a fixed $z \in \Ga^{\pa{L}}_{\ua}\cup \Ga^{\pa{R}}_{\ua} \setminus \paa{\pm q , \la_0} $ (resp. 
$z \in \Ga^{\pa{L}}_{\da}  \cup \Ga^{\pa{R}}_{\da} $), $\ex{i x u\pa{z} }$ (resp. $\ex{-ix u\pa{z}}$)
is exponentially small in $x$. 
The matrix $\Ups$ is discontinuous across the curve
$\Sg_{\Ups}=  \Ga^{\pa{L}}_{\ua}\cup\Ga^{\pa{L}}_{\da}  \cup \Ga^{\pa{R}}_{\ua}\cup\Ga^{\pa{R}}_{\da}$.
One readily checks that the matrix $\Ups$ is the unique solution of the below RHP (and hence there is a one-to-one correspondence between $\chi$ and
$\Ups$):
\begin{itemize}
\item $\Ups$ is analytic on $\mathbb{C}\setminus \Sg_{\Ups}$ and has continuous boundary values on $\Sg_{\Ups} \setminus \paa{\pm q}$ ;
\item $\Ups(\la) =  \e{O}\left(\ba{cc}
                                1 & 1 \\
                                1 & 1 \ea\right) \pa{\la-q}^{-\sg_3 \nu\pa{q}}
\pa{\la+q}^{\sg_3 \nu\pa{-q}}  \log\abs{\la^2-q^2}$ \; for \; $\la \tend \pm  q$ ;
\item $\Ups (\la)= I_2 + \la^{-1}\e{O}\left(\ba{cc}
                                1 & 1 \\
                                1 & 1 \ea\right)$ uniformly in $\la \tend \infty$ \; ;
\item $\Ups_{+}(\la)\, G_{\Ups}(\la)=\Ups_-(\la)   \quad\text{for}\; \;  \la \in \Sg_{\Ups} \setminus \paa{\pm q, \la_0} \; .$
\end{itemize}
With
\beq
G_{\Ups}\pa{\la}= M\pa{\la} \; \; \e{on} \quad  \Ga^{\pa{L}}_{\ua} \cup  \, \Ga^{\pa{R}}_{\da} \qquad , \qquad
G_{\Ups}\pa{\la}= N^{\pa{L}}\!\pa{\la} \; \; \e{on} \quad  \Ga^{\pa{L}}_{\da}
\quad \e{and} \quad  G_{\Ups}\pa{\la}= N^{\pa{R}}\!\pa{\la} \; \; \e{on} \;  \Ga^{\pa{R}}_{\ua} \;.
\enq
%
%
%


\subsubsection{The space-like regime}

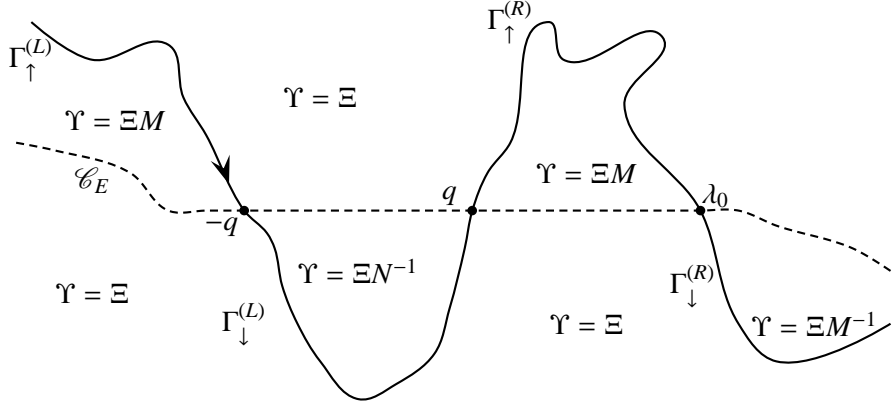
\begin{figure}[h]
\begin{center}

\begin{pspicture}(12,7)


\pscurve[linestyle=dashed, dash=3pt 2pt] (0,4.4)(0.5,4.3)(1.5,4)(2,3.5)(2.5,3.5)(3,3.5)(6,3.5)(9,3.5)(9.5,3.5)(10,3.3)(11,3)(11.5,2.7)

\rput(1,3.9){$\msc{C}_{E}$}


\psdots(3,3.5)(6,3.5)(9,3.5)
\rput(2.7,3.3){$-q$}
\rput(5.7,3.7){$q$}
\rput(9.2,3.7){$\la_0$}


\pscurve(0.2,6)(1,5.5)(2,5.7)(2.2,5)(2.5,4.5)(3,3.5)(3.3,3.2)(3.4,3)(3.5,2.5)
(3.7,2)(4,1.5)(4.5,1)(5,1.2)(5.5,1.6)(5.7,2.2)(5.9,3)(6,3.5)(6.2,4)(6.5,4.4)(7,6)(7.2,5.5)
(8.5,5.8)(8,4.9)(9,3.5)(9.5,2)(10,1.5)(11.5,2)


\psline[linewidth=3pt]{->}(2.75,4)(2.8,3.85)





\rput(0.2,5.5){$\Ga^{\pa{L}}_{\ua}$}
\rput(3,2){$\Ga^{\pa{L}}_{\da}$}

\rput(6.5,6){$\Ga^{\pa{R}}_{\ua}$}

\rput(8.9,2.5){$\Ga^{\pa{R}}_{\da}$}

\rput(1.3,4.7){$\Ups=\Xi M$}

\rput(4,5){$\Ups=\Xi$}
\rput(1,2.4){$\Ups=\Xi$}

\rput(4.5,2.7){$\Ups=\Xi N ^{-1}$}

\rput(7.5,4){$\Ups=\Xi M$}

\rput(10.5,2){$\Ups=\Xi  M^{-1}$}

\rput(7.5,2){$ \Ups=\Xi$}

\end{pspicture}

\caption{Contour $\Sg_{\Ups}=\Ga^{\pa{L}}_{\ua} \cup \Ga^{\pa{L}}_{\da} \cup \Ga^{\pa{R}}_{\ua} \cup \Ga^{\pa{R}}_{\da} $ appearing in the RHP for $\Ups$ (space-like regime).\label{contour pour le RHP de Upsilon space-like}}
\end{center}
\end{figure}
We introduce two matrices $M$ and $N$
\beq
M=\pa{ \ba{cc}
                            1  &   \a^{-2} e^{-2} \\
                             0 &  1
                            \ea }=I_2+P\pa{\la} \sg^+ \qquad \e{and} \qquad 
N=\pa{ \ba{cc}
                            1  &  0 \\
                             \a^2_- e^{2}\pa{\ex{-2i\pi\nu} -1}^2 &  1
                            \ea } = I_2+Q\pa{\la} \sg^{-}\;.
\label{definition des matrices M et N spacelike}
\enq
The matrix $M\!\pa{\la}$ has an analytic continuation to $U\setminus \intff{-q}{q}$ starting from
$\msc{C}_{E}\setminus \intff{-q}{q}$. The matrix $N\!\pa{\la}$ has an analytic continuation to $U\cap \mathbb{H}_-$.

%
%
%
%
%
%
%
%
%
%
This allows to write convenient local parameterizations around $\pm q$ for $P$ and $Q$:
\beq
P\pa{\la}=\ex{i\zeta_{-q}} \pac{\zeta_{-q}}^{-2\nu\pa{\la}} \f{\ex{2i\pi\nu\pa{\la}}-1}{C^{\pa{L}}\!\pa{\la}}
 = - \ex{i\zeta_{q}} \pac{\zeta_{q}}^{2\nu\pa{\la}} \f{\ex{-2i\pi\nu\pa{\la}}-1}{C^{\pa{R}}\!\pa{\la}}
\qquad \e{with} \quad \left\{ \ba{ccc} \zeta_{-q}&=&x\pa{u\pa{\la}-u\pa{-q}} \\
									\zeta_{q}&=&x\pa{u\pa{\la}-u\pa{q}} \ea \right. \;.
\label{definition fonction P spacelike}
\enq
Similarly,
\beq
Q\pa{\la}=C^{\pa{L}}\!\pa{\la}\, \ex{-i\zeta_{-q}} \pac{\zeta_{-q}}^{2\nu\pa{\la}} \pa{\ex{2i\pi\nu\pa{\la}}-1}
= - C^{\pa{R}}\!\pa{\la} \ex{-i\zeta_{q}} \pac{\zeta_{q}}^{-2\nu\pa{\la}} \pa{\ex{-2i\pi\nu\pa{\la}}-1} \; .
\label{definition Q spacelike}
\enq

Here, we bring to the reader's attention the difference of signs in
the definition of $\zeta_q$ in the time-like and space-like regimes. Also, the functions $C^{\pa{L/R}}\!\pa{\la}$ have been defined in 
\eqref{definition fonction C left right timelike}. The sole difference is that, in the space-like regime, the $+i0^+$ regularization plays no
role. 

The matrix $\Ups\pa{\la}$ defined in Fig.~\ref{contour pour le RHP de Upsilon space-like} in the unique solution to exactly the same RHP as formulated
for the time-like case but with the contours being defined in Fig.~\ref{contour pour le RHP de Upsilon space-like} and the jump matrix being now 
given by
%
%
%
%
%
%
%
%
%
%
%
%
%
\beq
G_{\Ups}\pa{\la}= M\pa{\la} \; \; \e{on} \quad  \Ga^{\pa{L}}_{\ua} \cup  \Ga^{\pa{R}}_{\ua}  \cup \Ga^{\pa{R}}_{\da} \qquad \e{and} \qquad
G_{\Ups}\pa{\la}= N\pa{\la} \; \; \e{on} \quad  \Ga^{\pa{L}}_{\da} \;.
\enq
%
%
%


\section{The local parametrices.}
\label{section parametrices}

We now build the parametrices 
around $\pm q$ and  $\la_0$. These will allows us to put the RHP for $\Ups$ (and hence the one for $\chi$) in correspondence 
with a RHP that has its jump matrices close to the identity,
uniformly on its whole jump contour (in the case of $\Ups\pa{\la}$ the jump matrices are close to the identity only uniformly away from the points $\la_0$
and $\pm q$). The role of the parametrices is to mimic the complicated local behavior of the solution $\chi$ near the stationary point $\la_0$ and the 
endpoints $\pm q$.  Once again, due to slight differences between the two regimes, we treat the space-like and the time-like regimes separately.

\subsection{The time-like regime}

We recall that for the time-like regime, the functions $P\pa{\la}$ and $Q^{\pa{L/R}}\!\pa{\la}$ appearing in the jump matrices are given respectively by
\eqref{definition fonction P time-like} and \eqref{definition fonction Q left and right time-like} with $C^{\pa{L/R}}\!\pa{\la}$ given by
\eqref{definition fonction C left right timelike}.
\subsubsection{The parametrix around $\la_0$}

It follows from the assumptions gathered in subsection \ref{soussection hypotheses}, that the function $u$
admits a local parameterization around $\la_0$, \textit{ie} there exists $\de>0$ such that $\ov{\mc{D}}_{\la_0,\de} \subset U$
and a holomorphic function $h$ on some open neighborhood of $\ov{\mc{D}}_{\la_0,\de}$ such that $u\pa{\la}-u\pa{\la_0}= - \om^2\pa{\la}$ with 
$\om\pa{\la}=\pa{\la-\la_0}h\pa{\la}$, and $h\pa{\ov{\mc{D}}_{\la_0,\de}\cap \mathbb{H}_{\pm}} \subset
\mathbb{H}_{\pm}$.

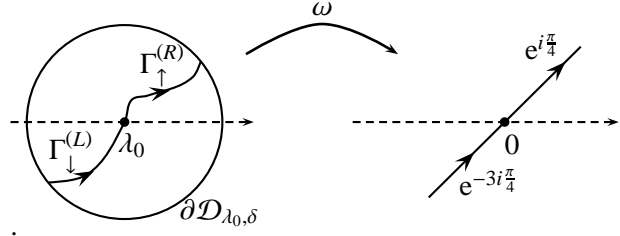
\begin{figure}[h]
\begin{center}

\begin{pspicture}(8,3)

\psline[linestyle=dashed, dash=3pt 2pt]{->}(0,1.5)(3.2,1.5)
\psline[linestyle=dashed, dash=3pt 2pt]{->}(4.5,1.5)(8,1.5)

\psdots(1.5,1.5)(6.5,1.5)
\rput(1.6,1.2){$\la_0$}
\rput(6.6,1.2){$0$}
\pscircle(1.5,1.5){1.3}


\pscurve(0.5,0.7)(1,0.8)(1.5,1.5)(1.6,1.8)(1.8,1.85)(2.4,2.1)(2.5,2.3)
\rput(0.8,1.1){$\Ga_{\da}^{\pa{L}}$}
\rput(2,2.25){$\Ga_{\ua}^{\pa{R}}$}

\psline(5.5,0.5)(7.5,2.5)

\rput(7,2.5){$\ex{i\f{\pi}{4}}$}
\rput(6.3,0.7){$\ex{-3i\f{\pi}{4}}$}

\rput(2.75,0.3){$\Dp{}\mc{D}_{\la_0,\de}$}


\psline[linewidth=2pt]{->}(6,1)(6.1,1.1)
\psline[linewidth=2pt]{->}(7.2,2.2)(7.3,2.3)

\psline[linewidth=2pt]{->}(1,0.8)(1.1,0.85)
\psline[linewidth=2pt]{->}(2,1.9)(2.1,1.92)


\pscurve[linewidth=1pt.]{->}(3,2.4)(4,2.8)(5,2.4)
\rput(4,3){$\om$}

\end{pspicture}

\caption{Contours appearing in the local RHP around $\la_0$ in the time-like case.\label{contour pour le RHP local en lambda 0 time-like}}
\end{center}
\end{figure}

The curves $\Ga^{\pa{L/R}}_{\da/\ua}$ in  $\ov{\mc{D}}_{\la_0,\de}$ are defined according to Fig.~\ref{contour pour le RHP local en lambda 0 time-like}.

The parametrix $\mc{P}_0$ around $\la_0$ reads
\beq
\mc{P}_0\!\pa{\la}= I_2 -  \ov{b}_{21}\!\pa{\la} \ex{-i\f{\pi}{4}} \f{ \om\pa{\la}\! \sqrt{\pi x} } {2i \pi} \Psi\pa{1, \f{3}{2} ; ix \om^2\!\pa{\la}} \; \sg^- \; .
\enq

There, $\Psi\pa{a,b;z}$ is Tricomi's confluent hypergeometric function whose definition is recalled in appendix \ref{Appendix Properties CHF and Barnes}.
The function $\ov{b}_{21}$ is defined piecewise:
\beqa
\ov{b}_{21}\!\pa{\la} &=&  \a^2\!\pa{\la} \ex{-ix u\pa{\la_0} -g\pa{\la}} \pa{\ex{-2i\pi \nu\pa{\la}} -1}^2  \qquad \;\quad\;\,  \e{for}  \; \la \in \mathbb{H}^-\cap
\mc{D}_{\la_0,\de}  \; , \\
\ov{b}_{21}\!\pa{\la} &=&  \a^2\!\pa{\la} \ex{4i\pi \nu\pa{\la}} \ex{-ix u\pa{\la_0}-g\pa{\la}} \pa{\ex{-2i\pi \nu\pa{\la}} -1}^2 \quad \e{for} \;  \la \in \mathbb{H}^+\cap \mc{D}_{\la_0,\de} \;.
\eeqa
It is holomorphic on $ \mc{D}_{\la_0,\de} $ due to the jump condition satisfied by $\a$ \eqref{ecriture condition saut alpha}.

\noindent The paramertix $\mc{P}_0$ solves the RHP:
\begin{itemize}
\item $\mc{P}_0$ is analytic in $\mc{D}_{\la_0,\de}\setminus \paa{\Ga_{\ua}^{\pa{R}} \cup\Ga_{\da}^{\pa{L}}}\cap\mc{D}_{\la_0,\de}$
with continuous boundary values on $\paa{\Ga_{\ua}^{\pa{R}} \cup\Ga_{\da}^{\pa{L}}}\cap\mc{D}_{\la_0,\de}$ ;
\item $\mc{P}_0 = I_2+ \f{1}{\sqrt{x}} \e{O}\pa{\sg^-}$\;  uniformly in $\la \in \Dp{}\mc{D}_{\la_0,\de}$ ;
\item $ \pac{\mc{P}_0}_+\!\pa{\la} \; \pa{I_2 + \ov{b}_{21}\!\pa{\la} \ex{ix\om^2\!\pa{\la}} \sg^- } \;  =  \; \pac{\mc{P}_0}_-\!\pa{\la} $ .
\end{itemize}

The first two points in the formulation of the RHP for $\mc{P}_0$ are obvious. The validity of the jump conditions can be checked with the help of
identity \eqref{relation de discontinuite pour CHF Erreur}.


\subsubsection{The parametrix at $-q$}
The parametrices for the local RHPs at $\pm q$ are well known. They have already appeared in a series of works
\cite{CheianovZvonarevZeroTempforFreeFermAndPureSine,
DeiftItsKrasovskyAsymptoticsofToeplitsHankelWithFHSymbols,KozKitMailSlaTerRHPapproachtoSuperSineKernel} 
and can be constructed  from the differential equation method \cite{ItsDifferentialMethodForParametrix}.
 Here, we recall their form.

The parametrix $\mc{P}_{-q}$ around $-q$  reads
\begin{equation}
\mc{P}_{-q}\pa{\la}=\Psi\pa{\la  } L\pa{\la} \pac{ x\pa{u\pa{\la}-u\pa{-q} } }^{\nu\pa{\la}\sg_3} \ex{-\f{i\pi\nu\pa{\la}}{2}}\,\; ,
\label{parametrice en -q time like}
\end{equation}
\begin{equation}
\Psi\pa{\la}=
              \begin{pmatrix}
                        \Psi\pa{\nu\pa{\la},1;-ix\pac{u\pa{\la}-u\pa{-q}}}
                                        &ib_{12}(\la)\, \Psi\pa{1-\nu\pa{\la},1;ix\pac{u\pa{\la}-u\pa{-q}}} \vspace{1mm}\\
                        -ib_{21}(\la)\, \Psi\pa{1+\nu\pa{\la},1;-ix\pac{u\pa{\la}-u\pa{-q} }}
                                        & \Psi\pa{-\nu\pa{\la},1;ix\pac{u\pa{\la}-u\pa{-q} }}
              \end{pmatrix}
%
,
\end{equation}

\beq
\ba{cc}     b_{12}\!\pa{\la}&=-i \f{ \sin \pac{\pi \nu\pa{\la}} }{ \pi C^{\pa{L}}\!\pa{\la} }   \Ga^{2}\pa{1-\nu\pa{\la}} \; \\
            b_{21}\!\pa{\la}&= -i\f{ \pi C^{\pa{L}}\!\pa{\la} }
                                {\sin\pac{\pi \nu\pa{\la}} \Ga^{2}\pa{-\nu\pa{\la}} }  \ea
\quad , \qquad \e{so} \; \e{that} \quad  b_{12}\!\pa{\la}b_{21}\!\pa{\la} = - \nu^{2}\!\pa{\la} \; .
\enq
%
%
%
%
%
%
%
%
%
%
%
$C^{\pa{L}}\!\pa{\la}$ is given by \eqref{definition fonction C left right timelike} and
\begin{equation}
L\pa{\la}= \left\{\ba{cc}  I_2                            & -\tf{\pi}{2}<\e{arg}\pac{u\pa{\la}-u\pa{-q}}< \tf{\pi}{2} \vspace{4mm},\\
                  \left( \ba{cc} 1  &0 \\
                                0& \ex{2i\pi \nu\pa{\la} }
                                    \ea \right)  & \tf{\pi}{2}<\e{arg}\pac{u\pa{\la}-u\pa{-q}}< \pi \vspace{4mm},\\
                  \left( \ba{cc} \ex{2i\pi \nu\pa{\la} }  &0 \\
                                     0& 1
                                    \ea \right)  & -\pi<\e{arg}\pac{u\pa{\la}-u\pa{-q}}< -\tf{\pi}{2}.
        \ea \right.
\label{matrice L constante -q time like}
\end{equation}

$\mc{P}_{-q}$ is an exact solution of the RHP:
\begin{itemize}
\item $\mc{P}_{-q}$ is analytic on $\mc{D}_{-q,\de^{\prime}}\setminus \paa{\Ga^{\pa{L}}_{\ua}\cup\Ga^{\pa{L}}_{\da}}$ 
with continuous boundary values on  $\paa{\Ga^{\pa{L}}_{\ua}\cup\Ga^{\pa{L}}_{\da}}\setminus\paa{-q}$ ;
%

\item $\mc{P}_{-q}(\la) = \e{O} \pa{\ba{cc}
                                1&1 \\
                                1&1 \ea}
          \pa{\la + q}^{ \sg_3 \nu\pa{-q} }  \log\abs{\la+q}  , \;
                            \la \longrightarrow -q  \; ;$
\item $\mc{P}_{-q}\pa{\la}= I_2+   \f{1}{x^{1-\rho_{\de}}} \e{O} \pa{\ba{cc}
                                1&1 \\
                                1&1 \ea} , \quad \text{uniformly in } \la \in \partial\mc{D}_{-q,\de^{\prime}} $\; ;
\item $\left\{ \ba{l c c l}
   \pac{\mc{P}_{-q}}_{+}\!\pa{\la} M\pa{\la}& =&\pac{\mc{P}_{-q}}_-\pa{\la} \quad
                &\text{for }\la \in \Ga^{\pa{L}}_{\ua} \cap \mc{D}_{-q,\de^{\prime}} \; ; \vspace{2mm}\\
   \pac{\mc{P}_{-q}}_{+}\!\pa{\la} N^{\pa{L}}\!\pa{\la} &=& \pac{\mc{P}_{-q}}_-\!\pa{\la} \quad
               &\text{for }\la \in \Ga^{\pa{L}}_{\da} \cap \mc{D}_{-q,\de^{\prime}} \; .
                \ea \right.$
\end{itemize}

\begin{figure}[h]
\begin{center}

\begin{pspicture}(8,3)

\psline[linestyle=dashed, dash=3pt 2pt]{->}(0,1.5)(3.2,1.5)
\psline[linestyle=dashed, dash=3pt 2pt]{->}(4.5,1.5)(8,1.5)

\psdots(1.5,1.5)(6.5,1.5)
\rput(1.3,1.2){$-q$}
\rput(6.7,1.7){$0$}
\pscircle(1.5,1.5){1.3}


\pscurve(0.5,2.3)(1,2.1)(1.3,1.85)(1.4,1.8)(1.5,1.5)(2.4,0.8)(2.5,0.7)
\rput(1.35,2.2){$\Ga_{\ua}^{\pa{L}}$}
\rput(1.85,0.75){$\Ga_{\da}^{\pa{L}}$}

\rput(2.9,0.3){$\Dp{}\mc{D}_{-q,\de^{\prime}}$}

\psline(6.5,0.5)(6.5,2.5)


\psline[linewidth=2pt]{->}(6.5,1)(6.5,0.9)

\psline[linewidth=2pt]{->}(1,2.1)(1.1,2)
\psline[linewidth=2pt]{->}(2.4,0.8)(2.45,0.75)


\pscurve[linewidth=1pt.]{->}(3,2.4)(4,2.8)(5,2.4)
\rput(4,3){$u-u\pa{-q}$}

\end{pspicture}

\caption{Contours for the local RHP around $-q$ in the time-like case.\label{contour pour le RHP local en -q time-like}}
\end{center}
\end{figure}
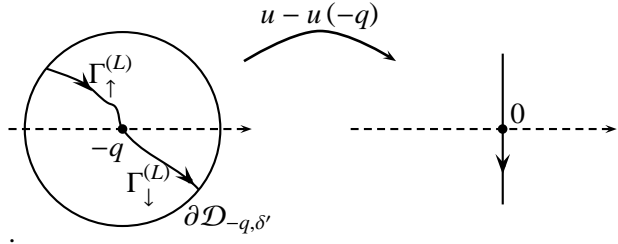

Here, we have set  
\beq
\rho_{\de^{\prime}}= 2 \sup \paa{ 
\abs{\Re\pa{\nu\pa{\la}} \; : \;  \la \in \Dp{} \mc{D}_{\pm q,\de^{\prime}} \cup \Dp{} \mc{D}_{- q,\de^{\prime}} } }  \; < 1 \; .
\label{definition rho delta}
\enq
The fact that $\rho_{\de^{\prime}}<1$ for sufficiently small $\de^{\prime}$ is a consequence of the assumptions
that $\abs{\Re\pa{\nu\pa{\pm q}}}<\tf{1}{2}$.
The canonically oriented contour $\Dp{} \mc{D}_{-q,\de^{\prime}}$ together with the definition of the contours $\Ga^{\pa{L}}_{\ua/\da}$
is depicted in Fig.~\ref{contour pour le RHP local en -q time-like}. $\de^{\prime}>0$ is chosen in such a way that
$\ov{\mc{D}}_{\pm q, \de^{\prime}} \subset \overset{\circ}{U}$,
$\ov{\mc{D}}_{\pm q, \de^{\prime}} \cap \ov{\mc{D}}_{\la_0,\de} = \emptyset$ and
$\ov{\mc{D}}_{ q, \de^{\prime}} \cap \ov{\mc{D}}_{- q, \de^{\prime}}= \emptyset$. Playing with the $\de$ entering in the definition of the
parametrix $\mc{P}_0$, one can tune it in such a way that $\de^{\prime}=\de$. We shall assume such a choice in the following.

\subsubsection{The parametrix at $q$}

The parametrix $\mc{P}_q$ around $q$ reads

\beq
\mc{P}_q\pa{\la}=\Psi\pa{\la  } L\pa{\la} \pac{ x\pa{u\pa{q} - u\pa{\la}} }^{-\nu\pa{\la}\sg_3} \ex{-\f{i\pi\nu\pa{\la}}{2}}\,\; . \;\;
\label{parametrice en q time like}
\enq
Here,
\beq
\Psi\pa{\la}=
            \pa{ \ba{cc}
                        \Psi\pa{-\nu\pa{\la},1;-ix\pac{u\pa{\la}-u\pa{q}}}
                                        & i \, \wt{b}_{12}\!\pa{\la}\, \Psi\pa{1+\nu\pa{\la},1;ix\pac{u\pa{\la}-u\pa{q}}} \vspace{1mm}\\
                        -i \, \wt{b}_{21}\!\pa{\la}\, \Psi\pa{1-\nu\pa{\la},1;-ix\pac{u\pa{\la}-u\pa{q} }}
                                        & \Psi\pa{\nu\pa{\la},1;ix\pac{u\pa{\la}-u\pa{q} }}
                        \ea}
,
\enq
\beq
\ba{ccc}         \wt{b}_{12}\pa{\la}&=&i  \f{ \pi  \pac{C^{\pa{R}}\! \pa{\la}}^{-1} }{    \Ga^{2}\pa{-\nu\pa{\la}} \sin \pac{\pi \nu\pa{\la}}  } \vspace{2mm}\\
            \wt{b}_{21}\pa{\la}&=& i \pi^{-1} \Ga^{2}\pa{1-\nu\pa{\la}} C^{\pa{R}}\!\pa{\la} \sin\pac{\pi \nu\pa{\la}}
    \ea \; , \quad  \wt{b}_{12}\!\pa{\la} \wt{b}_{21}\!\pa{\la} = - \nu^{2}\!\pa{\la} \; .
\enq
%
%
%
%
%
%
%
%
%
%
%
$C^{\pa{R}}\!\pa{\la}$ is given by \eqref{definition fonction C left right timelike} and
\begin{equation}
L\pa{\la}= \left\{\ba{cc}  I_2                            & -\tf{\pi}{2}<\e{arg}\pac{u\pa{q}-u\pa{\la}}< \tf{\pi}{2} \vspace{4mm} \; ,\\
                  \pa{ \ba{cc} \ex{2i\pi \nu\pa{\la}}  &0 \\
                                0& 1
                                    \ea }  & \tf{\pi}{2}<\e{arg}\pac{u\pa{q}-u\pa{\la}}< \pi \vspace{4mm} \; ,\\
                  \pa{ \ba{cc} 1  &0 \\
                                0 & \ex{2i\pi \nu\pa{\la} }
                                    \ea }  & -\pi<\e{arg}\pac{u\pa{q}-u\pa{\la}}< -\tf{\pi}{2} \; .
        \ea \right.
\label{matrice L en q time like}
\end{equation}

\begin{figure}[h]
\begin{center}

\begin{pspicture}(8,3)

\psline[linestyle=dashed, dash=3pt 2pt]{->}(0,1.5)(3.2,1.5)
\psline[linestyle=dashed, dash=3pt 2pt]{->}(4.5,1.5)(8,1.5)

\psdots(1.5,1.5)(6.5,1.5)
\rput(1.3,1.2){$q$}
\rput(6.7,1.7){$0$}
\pscircle(1.5,1.5){1.3}


\pscurve(0.5,2.3)(1,2.1)(1.3,1.85)(1.4,1.8)(1.5,1.5)(2.4,0.8)(2.5,0.7)
\rput(1.35,2.2){$\Ga_{\ua}^{\pa{R}}$}
\rput(1.85,0.75){$\Ga_{\da}^{\pa{R}}$}
\rput(2.8,0.3){$\Dp{}\mc{D}_{q,\de}$}

\psline(6.5,0.5)(6.5,2.5)


\psline[linewidth=2pt]{->}(6.5,1)(6.5,1.1)

\psline[linewidth=2pt]{->}(1,2.1)(1.1,2)
\psline[linewidth=2pt]{->}(2.4,0.8)(2.45,0.75)


\pscurve[linewidth=1pt.]{->}(3,2.4)(4,2.8)(5,2.4)
\rput(4,3){$u-u\pa{q}$}

\end{pspicture}

\caption{Contours for the local RHP around $q$ in the time-like case.\label{contour pour le RHP local en q time-like}}
\end{center}
\end{figure}
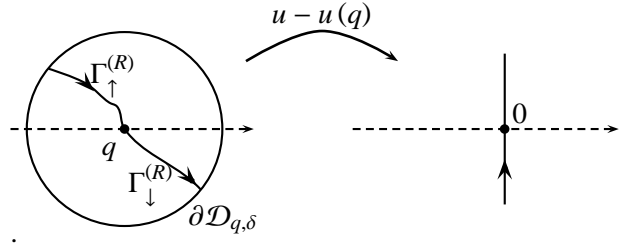

$\mc{P}_{q}$ is an exact solution of the RHP:
\begin{itemize}
\item $\mc{P}_{q}$ is analytic on $\mc{D}_{q,\de}\!\setminus \!\paa{\Ga^{\pa{R}}_{\ua}\!\cup\!\Ga^{\pa{R}}_{\da} } \cap \mc{D}_{q,\de}\;$ and has continuous boundary values on $\paa{\Ga^{\pa{R}}_{\ua}\!\cup\!\Ga^{\pa{R}}_{\da}} \cap \mc{D}_{q,\de}\! \setminus\!\paa{q}$ \; ;

\item $\mc{P}_{q}(\la) = \e{O} \pa{\ba{cc}
                                1&1 \\
                                1&1 \ea} \pa{\la-q}^{-\sg_3\nu\pa{q}} \log\abs{\la-q}  \;\; ,
                                \;  \la \longrightarrow   q  \; ;$
\item $\mc{P}_{q}(\la)= I_2+ \f{1}{x^{1-\rho_{\de}}} \e{O} \pa{\ba{cc}
                                1&1 \\
                                1&1 \ea} , \quad \text{uniformly in} \; \la \in \Dp{} \mc{D}_{q,\de} $ \; ;
\item $\left\{ \ba{lccl}
   \pac{\mc{P}_{q}}_{+}\!\pa{\la}N^{\pa{R}}\!\pa{\la}&=&\pac{\mc{P}_{q}}_-\!\pa{\la} \quad
               &\text{for }\la \in \Ga^{\pa{R}}_{\ua} \cap \mc{D}_{q,\de} \setminus\paa{q} \; , \vspace{3mm}\\
   \pac{\mc{P}_{q}}_{+}\!\pa{\la} M\!\pa{\la}&=&\pac{\mc{P}_{q}}_-\!\pa{\la} \quad
   &\text{for }\la \in \Ga^{\pa{R}}_{\da} \cap \mc{D}_{q,\de} \setminus\paa{q}.
                \ea \right.$
\end{itemize}
 The canonically
oriented contour $\Dp{} \mc{D}_{q,\de}$ together with the definition of the curves $\Ga^{\pa{R}}_{\ua/\da}$ in the vicinity of $q$
is depicted in
Fig.~\ref{contour pour le RHP local en q time-like}. Note the change of orientation of the jump curve due to $u^{\prime}\pa{q}<0$.
Also $\rho_{\de}$ is as given in \eqref{definition rho delta}.

\subsubsection{Asymptotically analysable RHP for $\Pi$}

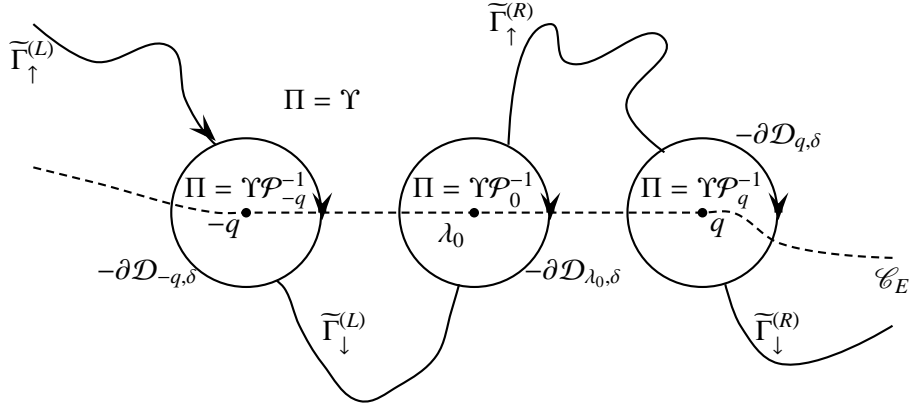
\begin{figure}[h]
\begin{center}

\begin{pspicture}(12,7)


\pscurve[linestyle=dashed, dash=3pt 2pt](0.2,4.1)(1,3.9)(2.6,3.5)(3,3.5)(6,3.5)(9,3.5)(9.4,3.5)(10,3.1)(11.5,2.9)


\psdots(3,3.5)(6,3.5)(9,3.5)
\rput(2.7,3.3){$-q$}
\pscircle(3,3.5){1}
\pscircle(6,3.5){1}
\pscircle(9,3.5){1}
\rput(5.7,3.2){$\la_0$}
\rput(9.2,3.3){$q$}

\rput(11.5,2.6){$\msc{C}_{E}$}


\pscurve(0.2,6)(1,5.5)(2,5.7)(2.2,5)(2.5,4.5)(2.6,4.4)

\pscurve(3.4,2.6)(3.5,2.5)(3.7,2)(4,1.5)(4.5,1)(5,1.2)(5.5,1.6)(5.7,2.2)(5.8,2.55)
\pscurve(6.45,4.4)(7,6)(7.2,5.5)(8.5,5.8)(8,4.9)(8.5,4.3)
\pscurve(9.3,2.55)(9.5,2)(10,1.5)(11.5,2)


\psline[linewidth=3pt]{->}(2.5,4.55)(2.6,4.45)

\psline[linewidth=3pt]{->}(4,3.5)(4,3.4)

\psline[linewidth=3pt]{->}(7,3.5)(7,3.4)

\psline[linewidth=3pt]{->}(10,3.5)(10,3.4)





\rput(0.2,5.5){$ \wt{\Ga}^{\pa{L}}_{\ua}$}
\rput(4.3,1.9){$ \wt{\Ga}^{\pa{L}}_{\da}$}

\rput(6.5,6){$ \wt{\Ga}^{\pa{R}}_{\ua}$}
\rput(10,1.9){$ \wt{\Ga}^{\pa{R}}_{\da}$}

\rput(1.7,2.7){$-\Dp{}\mc{D}_{-q,\de}$}

\rput(7.3,2.7){$-\Dp{}\mc{D}_{\la_0,\de}$}

\rput(10,4.5){$-\Dp{}\mc{D}_{q,\de}$}


\rput(4,5){$\Pi=\Ups$}

\rput(3,3.8){$\Pi= \Ups \mc{P}^{-1}_{-q}  $}

\rput(6,3.8){$\Pi= \Ups \mc{P}^{-1}_{0}  $}

\rput(9,3.8){$\Pi= \Ups \mc{P}^{-1}_{q}  $}

\end{pspicture}

\caption{Contour $\Sg_{\Pi}$ appearing in the RHP for $\Pi$, time-like regime.\label{contour pour le RHP de Pi time-like}}
\end{center}
\end{figure}

We now define a piecewise analytic matrix $\Pi$ in terms of $\Ups$ and the parametrices
according to Fig.~\ref{contour pour le RHP de Pi time-like}.  In particular one has $\Pi=\Ups$ everywhere outside of the disks.
The matrix $\Pi$ has its jump matrices uniformly close to the identity matrix in respect to the $x \tend +\infty$ limit. Hence, it can be computed
perturbatively in $x$ by the use \cite{DeiftZhouSteepestDescentForOscillatoryRHP} of Neumann series expansion for the solution of the singular integral equation equivalent to the RHP for $\Pi$. This matrix $\Pi$ is the unique solution to the RHP:
\begin{itemize}
\item $\Pi$ is analytic on $\mathbb{C}\setminus \Sg_{\Pi}$ and has continuous boundary values on $\Sg_{\Pi}$ ;
\item $\Pi (\la) =  I_2 +\la^{-1} \e{O} \pa{\ba{cc}
                                1&1 \\
                                1&1 \ea}\; $, uniformly in $\la \tend \infty$ ;
\item $\Pi_{+}(\la)\, G_{\Pi}(\la)=\Pi_-(\la)   \quad\text{for} \; \la \in \Sg_{\Pi} \; .$
\end{itemize}
The jump matrix $G_{\Pi}\pa{\la}$ for $\Pi\pa{\la}$ reads
\beq
G_{\Pi}\pa{\la}=G_{\Ups}\pa{\la}  \;\; \e{on} \;\; \wt{\Ga} = \wt{\Ga}_{\ua}^{\pa{L}}\cup\wt{\Ga}_{\da}^{\pa{L}}\cup\wt{\Ga}_{\da}^{\pa{R}}\cup\wt{\Ga}_{\ua}^{\pa{R}}
\qquad  \e{and} \qquad
G_{\Pi}\pa{\la} = \left\{ \ba{cc}   \mc{P}_{\pm q}^{-1}\pa{\la}  & \e{on} \;\; -\Dp{}\mc{D}_{\pm q ,\de}  \vspace{2mm} \\
							\mc{P}_{0}^{-1}\pa{\la}  & \e{on} \;\;  -\Dp{}\mc{D}_{\la_0 ,\de}
  \ea  \right. \; .
\label{definition matrice de saut pour RHP Pi}
\enq

\subsection{Asymptotic expansion for the algebraically small jump matrices}

Note that the jump matrices along $\wt{\Ga}$ are exponentially close to $I_2$ in $x$ and this in the
$L^1\pa{\Sg_{\Pi}}\cap L^{2}\pa{\Sg_{\Pi}} \cap L^{\infty}\pa{\Sg_{\Pi}}$ sense.
Only the jump matrices on the disks are algebraically in $x$ close to the identity matrix. The latter jump matrices
have the below asymptotic expansion into inverse powers of $x$, valid uniformly on the boundary of their respective domains of definition
($\Dp{}\mc{D}_{\pm q, \de}$ or $\Dp{}\mc{D}_{\la_0,\de}$):
%
%
%
%
%
%
%
%
%
%
%
%
%
%
%
%
%
%
%
%
%
%
%
%
%
%
%
%
%
%
%
%
%
%
%
%
\beqa
\mc{P}_{-q}^{-1}\!\pa{s} &\simeq & I_2 + \sul{n\geq 0}{} \f{ V^{\pa{-;n}}\!\pa{s} }{\pa{n+1}! \pac{x\pa{s+q}}^{n+1} }    \; , \nonumber\\
\mc{P}_{q}^{-1}\!\pa{s} &\simeq& I_2 + \sul{n\geq 0}{} \f{ V^{\pa{+;n}}\!\pa{s} }{\pa{n+1}! \pac{x\pa{s-q}}^{n+1} }  \; , \nonumber\\
\mc{P}_{0}^{-1}\!\pa{s} &\simeq& I_2 + \sg^- \sul{n \geq 0}{}  \f{ d^{\pa{n}} \! \pa{s} }{x^{n+\f{1}{2}} \pa{s-\la_0}^{2n+1} } \; .
\label{ecriture generique DA diverse parametrices time-like}
\eeqa
Where,
\beqa
V^{\pa{-;n}}\!\pa{s} &=& \pa{-i}^{n+1} \paf{ s+q }{  u\pa{s}-u\pa{-q} }^{n+1}
        \pa{  \ba{cc}
               \pa{-1}^{n+1} \pa{-\nu}^2_{n+1}     & i \pa{n+1} b_{12} \pa{-1}^{n+1} \pa{1-\nu}^2_{n} \\
                 -i\pa{n+1} b_{21}  \pa{1+\nu}^2_{n} &  \pa{\nu}^2_{n+1}   \ea }   \; , \\
V^{\pa{+;n}}\!\pa{s} &=& \pa{-i}^{n+1} \paf{ s-q }{  u\pa{s}-u\pa{q} }^{n+1}
        \pa{  \ba{cc}
               \pa{-1}^{n+1} \pa{\nu}^2_{n+1}     & i \pa{n+1} \wt{b}_{12} \pa{-1}^{n+1} \pa{1+\nu}^2_{n} \\
                 -i\pa{n+1} \wt{b}_{21}  \pa{1-\nu}^2_{n} &  \pa{-\nu}^2_{n+1}  \ea }   \; , \label{definition matrice V+n time like}\\
d^{\pa{n}}\!\pa{s} &=& - \f{ i^{n} \Ga\pa{\tf{1}{2}+n}}{2\pi}  \f{\ex{-i\f{\pi}{4}}}{ h^{2n+1}\!\pa{s}} \ov{b}_{21}\!\pa{s}  \; .
\eeqa
We remind that $\om\pa{\la}=\pa{\la-\la_0} h\pa{\la}$ and we have used the conditions $\ddet{}{\mc{P}_{\pm q}}=1=\ddet{}{\mc{P}_0}$
so as to invert the parametrices and then infer their asymptotic expansion from the one of CHF \eqref{asy-Psi}. Also, we have not made explicit that $b_{ij}$, $\wt{b}_{ij}$ and $\nu$ are functions of $s$.

\subsection{The space-like regime}

The discussion of the space-like regime resembles, up to minor subtelties, to the previous one.
Therefore, we make it as short as possible.

\subsubsection{The parametrix around $\la_0$}

The parametrix $\mc{P}_0$ on $\mc{D}_{\la_0,\de}$ for the local RHP around $\la_0$ reads
\beq
\mc{P}_0\pa{\la}= I_2 - \ov{b}_{12}\!\pa{\la} \ex{i\f{\pi}{4}} \f{ \om\pa{\la}\!\! \sqrt{\pi x} } {2i \pi} \Psi\pa{1, \f{3}{2} ; -ix \om^2\!\pa{\la}}
\sg^+ \;
\qquad \e{with} \quad
\ov{b}_{12}\!\pa{\la}=   \a^{-2}\!\pa{\la} \ex{ix u\pa{\la_0} +g\pa{\la}}  \; .
\enq
$\ov{b}_{12}$ is holomorphic on $ \mc{D}_{\la_0,\de} $. The parametrix $\mc{P}_0$ is a solution to the RHP
\begin{itemize}
\item $\mc{P}_0$ is analytic in $\mc{D}_{\la_0,\de}\setminus \paa{\Ga_{\ua}^{\pa{R}} \cup\Ga_{\da}^{\pa{R}}}\cap\mc{D}_{\la_0,\de}$
and has continuous boundary values on  $\paa{\Ga_{\ua}^{\pa{R}} \cup\Ga_{\da}^{\pa{R}}}\cap\mc{D}_{\la_0,\de}$;
%
%
%
%
\item $\mc{P}_0 = I_2\; +\; \f{1}{ \sqrt{x}} \e{O}\pa{\sg^+}$ \;\; uniformly in  $\la \in \Dp{}D_{\la_0,\de}$ \; ;
\item $ \pac{\mc{P}_0}_+\!\pa{\la} \; \pa{I_2+\ov{b}_{12}\!\pa{\la} \ex{-ix\om^2\!\pa{\la}} \sg^+ }  \; =  \; \pac{ \mc{P}_0 }_-\!\pa{\la}$ \; .
\end{itemize}

\begin{figure}[h]
\begin{center}

\begin{pspicture}(8,3)

\psline[linestyle=dashed, dash=3pt 2pt]{->}(0,1.5)(3.2,1.5)
\psline[linestyle=dashed, dash=3pt 2pt]{->}(4.5,1.5)(8,1.5)

\psdots(1.5,1.5)(6.5,1.5)
\rput(1.3,1.2){$\la_0$}
\rput(6.7,1.7){$0$}
\pscircle(1.5,1.5){1.3}


\pscurve(0.5,2.3)(1,2.1)(1.3,1.85)(1.4,1.8)(1.5,1.5)(2.4,0.8)(2.5,0.7)
\rput(1.35,2.2){$\Ga_{\ua}^{\pa{R}}$}
\rput(1.85,0.75){$\Ga_{\da}^{\pa{R}}$}

\psline(5.5,2.5)(7.5,0.5)

\rput(6.3,2.3){ $\ex{3i\f{\pi}{4}}$ }
\rput(6.7,0.7){ $\ex{-i\f{\pi}{4}}$ }
\rput(2.85,0.35){$\Dp{}\mc{D}_{\la_0,\de}$}


\psline[linewidth=2pt]{->}(6,2)(6.1,1.9)
\psline[linewidth=2pt]{->}(7,1)(7.1,0.9)

\psline[linewidth=2pt]{->}(1,2.1)(1.1,2)
\psline[linewidth=2pt]{->}(2.4,0.8)(2.45,0.75)


\pscurve[linewidth=1pt.]{->}(3,2.4)(4,2.8)(5,2.4)
\rput(4,3){$\om$}

\end{pspicture}

\caption{Contours for the local RHP around $\la_0$ in the space-like case.\label{contour pour le RHP local en lambda 0 space-like}}
\end{center}
\end{figure}
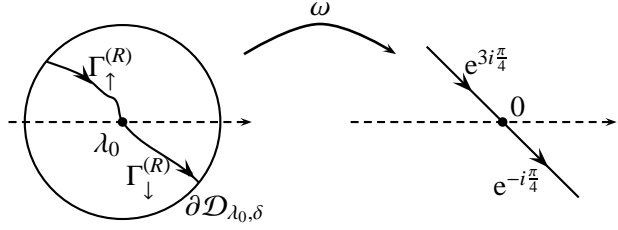

The jump curve for the parametrix $\mc{P}_0$ is depicted on Fig.~\ref{contour pour le RHP local en lambda 0 space-like}.
%
%
%


\subsubsection{The parametrix around $-q$}
This parametrix $\mc{P}_{-q}$ is exactly the same as in the time-like regime. Hence, we do not present it here.

\subsubsection{The parametrix around $q$}

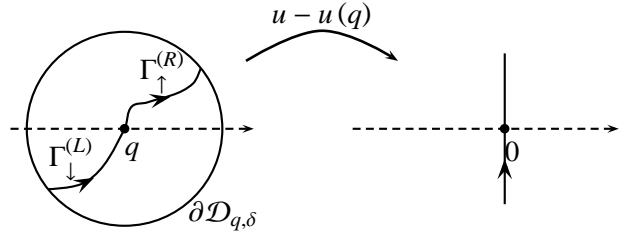
\begin{figure}[h]
\begin{center}

\begin{pspicture}(8,3)

\psline[linestyle=dashed, dash=3pt 2pt]{->}(0,1.5)(3.2,1.5)
\psline[linestyle=dashed, dash=3pt 2pt]{->}(4.5,1.5)(8,1.5)

\psdots(1.5,1.5)(6.5,1.5)
\rput(1.6,1.2){$q$}
\rput(6.6,1.2){$0$}
\pscircle(1.5,1.5){1.3}


\pscurve(0.5,0.7)(1,0.8)(1.5,1.5)(1.6,1.8)(1.8,1.85)(2.4,2.1)(2.5,2.3)
\rput(0.8,1.1){$\Ga_{\da}^{\pa{L}}$}
\rput(2,2.25){$\Ga_{\ua}^{\pa{R}}$}
\rput(2.8,0.35){$\Dp{}\mc{D}_{q,\de}$}

\psline(6.5,0.5)(6.5,2.5)


\psline[linewidth=2pt]{->}(6.5,1)(6.5,1.1)

\psline[linewidth=2pt]{->}(1,0.8)(1.1,0.85)
\psline[linewidth=2pt]{->}(2,1.9)(2.1,1.92)


\pscurve[linewidth=1pt.]{->}(3,2.4)(4,2.8)(5,2.4)
\rput(4,3){$u-u\pa{q}$}

\end{pspicture}

\caption{Contours for the parametrix around  $q$ in the space-like regime.\label{contour pour le RHP local en q space-like}}
\end{center}
\end{figure}

The parametrix  $\mc{P}_q$ around $q$ reads
\beq
\mc{P}_q\pa{\la}=\Psi\pa{\la  } L\pa{\la} \pac{ x\pa{ u\pa{\la}-u\pa{q}} }^{-\nu\pa{\la}\sg_3} \ex{\f{i\pi\nu\pa{\la}}{2}}\,\; . \;\;
\label{parametrice en q space-like}
\enq
Here,
\beq
\Psi\pa{\la}=
            \pa{ \ba{cc}
                        \Psi\pa{-\nu\pa{\la},1;-ix\pac{u\pa{\la}-u\pa{q}}}
                                        & i \, \wt{b}_{12}\pa{\la}\, \Psi\pa{1+\nu\pa{\la},1;ix\pac{u\pa{\la}-u\pa{q}}} \vspace{1mm}\\
                        -i \, \wt{b}_{21}\pa{\la}\, \Psi\pa{1-\nu\pa{\la},1;-ix\pac{u\pa{\la}-u\pa{q}}}
                                        & \Psi\pa{\nu\pa{\la},1;ix\pac{u\pa{\la}-u\pa{q}}}
                        \ea}
,
\enq
\beq
\ba{cc}         \wt{b}_{12}\!\pa{\la}&= i  \f{ \Ga^{2}\pa{1+\nu\pa{\la}}}{ \pi C^{\pa{R}}\!\pa{\la}} \sin\pac{\pi \nu\pa{\la}}  \; \\
            \wt{b}_{21}\!\pa{\la}&=  \f{i \pi  C^{\pa{R}}\!\pa{\la} }{    \Ga^{2}\pa{\nu\pa{\la}} \sin \pac{\pi \nu\pa{\la}}  }
    \ea \; , \quad  \wt{b}_{12}\!\pa{\la} \wt{b}_{21}\!\pa{\la}  = - \nu^{2}\!\pa{\la}  \; .
\label{definition des b tilde space like}
\enq
$C^{\pa{R}}$ is given by \eqref{definition fonction C left right timelike} and
\begin{equation}
L\pa{\la}= \left\{\ba{cc}  I_2                            & -\tf{\pi}{2}<\e{arg}\pac{ u\pa{\la}-u\pa{q} }< \tf{\pi}{2} \vspace{4mm} \; ,\\
                  \pa{ \ba{cc} 1  &0 \\
                                0& \ex{-2i\pi\nu\pa{\la} }
                                    \ea }  & \tf{\pi}{2}<\e{arg}\pac{ u\pa{\la}-u\pa{q} }< \pi \vspace{4mm} \; ,\\
                  \pa{ \ba{cc} \ex{-2i\pi \nu\pa{\la}}  &0 \\
                                0 & 1
                                    \ea }  & -\pi<\e{arg}\pac{ u\pa{\la}-u\pa{q} }< -\tf{\pi}{2} \; .
        \ea \right.
\label{matrice L constante}
\end{equation}
$\mc{P}_{q}$ is an exact solution of the RHP:
\begin{itemize}
\item $\mc{P}_{q}$ is analytic on $\mc{D}_{q,\de}\setminus \paa{\Ga^{\pa{R}}_{\ua}\cup\Ga^{\pa{R}}_{\da}}  \cap \mc{D}_{q,\de}\;$
with continuous boundary values on $\paa{\Ga^{\pa{R}}_{\ua}\cup\Ga^{\pa{R}}_{\da}}  \cap \mc{D}_{q,\de}\setminus \paa{q}$ ;
\item $\mc{P}_{q}(\la) = \e{O} \pa{\ba{cc}
                                1&1 \\
                                1&1 \ea} \pa{\la-q}^{-\sg_3 \nu\pa{q}} \log\abs{\la-q} \;  ,
                                \;\;  \la \longrightarrow   q  \; ;$
\item $\mc{P}_{q}(\la)= I_2+ \f{1}{x^{1-\rho_{\de}}} \e{O} \pa{\ba{cc}
                                1&1 \\
                                1&1 \ea}, \quad \text{uniformly in } \la \in \Dp{} \mc{D}_{q,\de} $ \; ;
\item $\left\{ \ba{lccl}
   \pac{\mc{P}_{q}}_{+}\!\pa{\la} M\pa{\la} &=& \pac{\mc{P}_{q}}_-\!\pa{\la} \quad
               &\text{for }\la \in \Ga^{\pa{R}}_{\ua} \cap \mc{D}_{q,\de} \; , \vspace{3mm}\\
   \pac{\mc{P}_{q}}_{+}\!\pa{\la} N\pa{\la} &=& \pac{\mc{P}_{q}}_-\!\pa{\la} \quad
   &\text{for }\la \in \Ga^{\pa{R}}_{\ua} \cap \mc{D}_{q,\de} \; .
                \ea \right.$
\end{itemize}
 The canonically
oriented contour $\Dp{} \mc{D}_{q,\de}$ as well as the definition of the jump curves $\Ga^{\pa{L/R}}_{\da/\ua}$ is depicted in
Fig.~\ref{contour pour le RHP local en q space-like}. Finally, $\rho_{\de}$ has been defined in \eqref{definition rho delta}.

\subsubsection{The RHP for $\Pi$}

\begin{figure}[h]
\begin{center}

\begin{pspicture}(12,7)


\pscurve[linestyle=dashed, dash=3pt 2pt](0.2,4.1)(1,3.9)(2.6,3.5)(3,3.5)(6,3.5)(9,3.5)(9.4,3.5)(10,3.1)(11.5,2.9)


\psdots(3,3.5)(6,3.5)(9,3.5)
\rput(2.7,3.3){$-q$}
\pscircle(3,3.5){1}
\pscircle(6,3.5){1}
\pscircle(9,3.5){1}
\rput(5.7,3.2){$q$}
\rput(9.2,3.3){$\la_0$}
\rput(11,2.7){$\msc{C}_{E}$}


\pscurve(0.2,6)(1,5.5)(2,5.7)(2.2,5)(2.5,4.5)(2.6,4.4)

\pscurve(3.4,2.6)(3.5,2.5)(3.7,2)(4,1.5)(4.5,1)(5,1.2)(5.5,1.6)(5.7,2.2)(5.8,2.55)
\pscurve(6.45,4.4)(7,6)(7.2,5.5)(8.5,5.8)(8,4.9)(8.5,4.3)
\pscurve(9.3,2.55)(9.5,2)(10,1.5)(11.5,2)


\psline[linewidth=3pt]{->}(2.5,4.55)(2.6,4.45)

\psline[linewidth=3pt]{->}(4,3.5)(4,3.4)

\psline[linewidth=3pt]{->}(7,3.5)(7,3.4)

\psline[linewidth=3pt]{->}(10,3.5)(10,3.4)





\rput(0.2,5.5){$ \wt{\Ga}^{\pa{L}}_{\ua}$}
\rput(4.3,1.9){$ \wt{\Ga}^{\pa{L}}_{\da}$}

\rput(6.5,6){$ \wt{\Ga}^{\pa{R}}_{\ua}$}
\rput(10,1.9){$ \wt{\Ga}^{\pa{R}}_{\da}$}

\rput(1.7,2.7){$-\Dp{}\mc{D}_{-q,\de}$}

\rput(7.3,2.7){$-\Dp{}\mc{D}_{q,\de}$}

\rput(10,4.5){$-\Dp{}\mc{D}_{\la_0,\de}$}


\rput(4,5){$\Pi=\Ups$}

\rput(3,3.8){$\Pi= \Ups \mc{P}^{-1}_{-q}  $}

\rput(6,3.8){$\Pi= \Ups \mc{P}^{-1}_{q}  $}

\rput(9,3.8){$\Pi= \Ups \mc{P}^{-1}_{0}  $}

\end{pspicture}

\caption{Contour $\Sg_{\Pi}$ appearing in the RHP for $\Pi$, space-like regime.\label{contour pour le RHP de Pi space-like}}
\end{center}
\end{figure}
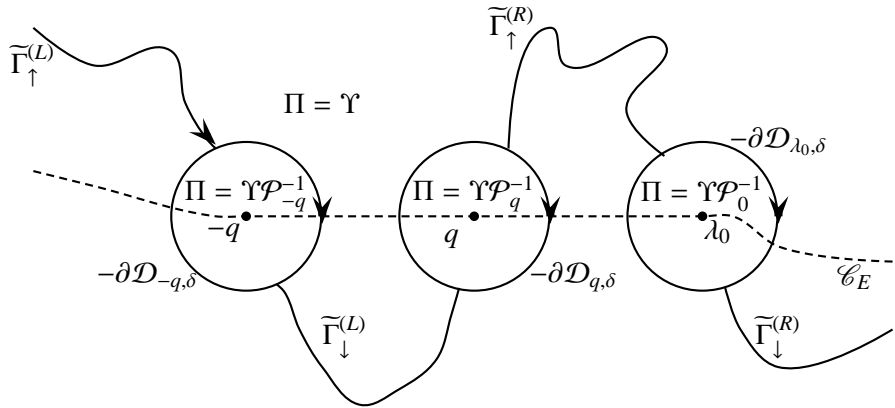

The matrix $\Pi$ is defined according to Fig.~\ref{contour pour le RHP de Pi space-like} and is the unique solution to the RHP
formulated in exactly the same way as for the time-like regime. The difference consists in the precise form of the contours
due to the fact that in the space-like regime $\la_0>q$.
%
%
%
%
%
%
%
%
%
%
%
%
%
%
%
%
%
%
%
%
%
%
%

%
%
%
%

\subsection{The asymptotic expansion for the parametrices}
The jump matrices $\mc{P}_{\pm q }^{-1}$ have the same asymptotic expansion as in the time-like regime
\eqref{ecriture generique DA diverse parametrices time-like} with the sole exception that the coefficients $\wt{b}_{12}$, $\wt{b}_{21}$
entering in the definition of $V^{\pa{+;n}}$ \eqref{definition matrice V+n time like} are now given by 
\eqref{definition des b tilde space like}. The matrix $\mc{P}_{0}^{-1}$ has the below asymptotic expansion
%
%
%
%
%
%
%
%
%
%
%
%
%
%
%
%
%
%
\beq
\mc{P}_{0}^{-1}\pa{s}  \simeq  I_2 + \sg^+ \sul{n \geq 0}{}  \f{ d^{\pa{n}}\!\pa{s} }{x^{n+\f{1}{2}} \pa{s-\la_0}^{2n+1} }  \qquad \e{with} \qquad 
d^{\pa{n}}\!\pa{s} = \ov{b}_{12}\!\pa{s} \pa{-i}^n \f{\Ga\pa{\tf{1}{2}+n}}{2\pi}  \f{ \ex{i\f{\pi}{4}} }{ h^{2n+1}\!\pa{s}} \; .
\enq
%
%
%


\section{Asymptotic expansion of the Fredholm determinant}
\label{section DA determinant}

Starting from now on, we will treat both regimes (space and time-like) simultaneously.

\subsection{The asymptotic expansion for $\Pi$}

In this subsection we present two ways of writing down the asymptotic expansion for the matrix $\Pi$. The first, given in proposition \ref{proposition DA
Pi}, traces back all the different fractional powers of $x$ and oscillating terms that appear in the asymptotic expansion of $\Pi$.
It also provides one with a sharp and quite optimal control of the remainders. 
The second one, given in proposition \ref{proposition DA Pi heuristique}, is considerably less explicit 
and, by far, does not provide optimal estimates for the remainders. However, it is easier to implement form the 
computational point of view, especially when one is interested in calculating only a couple of terms in the asymptotics. 
One can then build on the first asymptotic expansion so as to on the one hand argue for a sharper form of the estimates for the remainders
and on the other hand identify which among the computed terms are relevent and which are not. 
We start this section by presenting the Neumann series expansion for $\Pi$.

\begin{defin}
\label{Definition Matrices Delta Nabla Sur contour encastre}
Let $\Sg_{\Pi}$ be the jump contour for the matrix $\Pi$. We define 
the contour $\Sg_{\Pi}^{ \pa{N} }$ as being the inslotted version of the N-fold Carthesian product $\Sg_{\Pi} \times \dots \times \Sg_{\Pi}$.
Namely it is obtained from $\Sg_{\Pi} \times \dots \times \Sg_{\Pi}$ by putting the contour for $z_{k+1}$
slightly shifted to the right from the contour for $z_{k}$. We have depicted the inslotted contour for $N=2$ on
Fig.~\ref{contour integration encastree for finale a deux N}.

\begin{figure}[h]
\begin{center}

\begin{pspicture}(12,7)

\psline[linestyle=dashed, dash=3pt 2pt]{->}(0.2,3.5)(11.5,3.5)


\psdots(3,3.5)(6,3.5)(9,3.5)


\pscircle(3,3.5){1}
\pscircle(6,3.5){1}
\pscircle(9,3.5){1}


\pscircle[linestyle=dashed, dash=3pt 2pt](3,3.5){1.3}
\pscircle[linestyle=dashed, dash=3pt 2pt](6,3.5){1.3}
\pscircle[linestyle=dashed, dash=3pt 2pt](9,3.5){1.3}


\pscurve(0.2,6)(1,5.5)(2,5.7)(2.2,5)(2.5,4.5)(2.6,4.4)

\pscurve(3.4,2.6)(3.5,2.5)(3.7,2)(4,1.5)(4.5,1)(5,1.2)(5.5,1.6)(5.7,2.2)(5.8,2.55)
\pscurve(6.45,4.4)(7,6)(7.2,5.5)(8.5,5.8)(8,4.9)(8.5,4.3)
\pscurve(9.3,2.55)(9.5,2)(10,1.5)(11.5,2)


\pscurve[linestyle=dashed](0.2,6.4)(1,5.9)(2,6)(2.5,5.5)(2.4,4.6)

\pscurve[linestyle=dashed](3.55,2.35)(3.7,2.3)(4,1.9)(4.5,1.3)(5,1.5)(5.3,1.8)(5.7,2.25)

\pscurve[linestyle=dashed](6.45,4.7)(6.8,6.4)(7.2,6)(8.5,6.1)(8.6,5.2)(8.25,4.55)

\pscurve[linestyle=dashed](9.4,2.25)(10,1.8)(11.5,2.3)


\psline[linewidth=3pt]{->}(1.15,5.5)(1.2,5.5)

\psline[linewidth=3pt]{->}(4,3.5)(4,3.4)

\psline[linewidth=3pt]{->}(1.7,3.5)(1.7,3.6)

\psline[linewidth=3pt]{->}(7,3.5)(7,3.4)

\psline[linewidth=3pt]{->}(10,3.5)(10,3.4)





\rput(2,6.4){$\wt{\Ga}^{\pa{L}}_{\ua}\pac{z_1}$}
\rput(1.2,5){$\wt{\Ga}^{\pa{L}}_{\ua}\pac{z_2}$}


\rput(2.3,2){$-\Dp{}\mc{D}\pac{z_1}$}
\rput(3,3.8){$-\Dp{}\mc{D}\pac{z_2}$}

\end{pspicture}

\caption{The inslotted contour for $N=2$. The integration over $z_1$ runs through the dotted contour
whereas the one over $z_2$ runs through the full one. $\Dp{}\mc{D}\pac{z_i}$ refers to the three disks over which the variable $z_i$
is integrated.\label{contour integration encastree for finale a deux N}}
\end{center}
\end{figure}
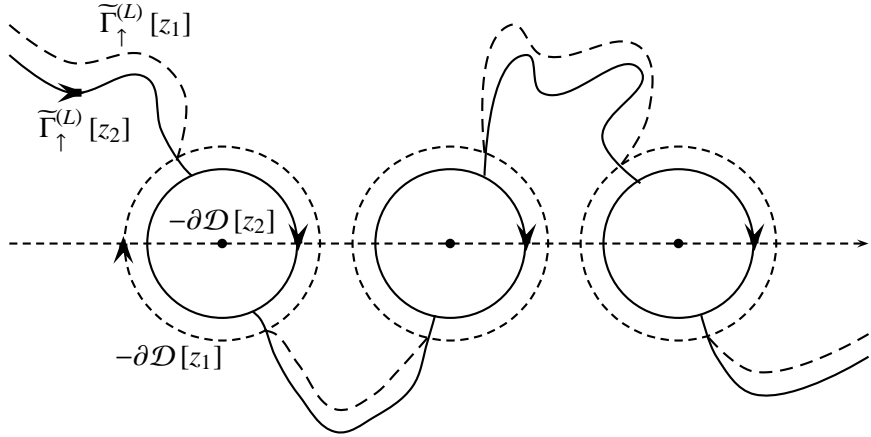

Let $\e{pr}_k$ stands for the projection on the $k^{\e{th}}$ factor of an N-fold Carthesian product, \textit{ie}  given $\bs{z}=\pa{z_1,\dots,z_N}$
one has $\e{pr}_k\pa{\bs{z}}=z_k$. The contour $\Sg_{\Pi}^{\pa{N}}$ thus defines N curves $\Sg_{\Pi}\pac{z_k}\equiv \e{pr}_k\pa{\Sg_{\Pi}^{\pa{N}}}$,
 $k=1,\dots,N$. Each of these can be interpreted as giving rise to the jump contour for the RHP problem associaed with the matrix $\Pi$. 
In the following whenever $\De$, resp. $\nabla$, is inegrated along $\Sg_{\Pi}\pac{z_k}$, it should be understood as originating from the jump matrix 
$I_2+\De$, resp. $I_2+\nabla$, appearing in the RHP for $\Pi$, resp. $\Pi^{-1}$, when the latter is formulated on the jump contour $\Sg_{\Pi}\pac{z_k}$.

\end{defin}

\begin{lemme}
Let $I+\De$ be the jump matrix for $\Pi$ and $\nabla=\e{Comat}\pa{\De}^t $ be the transpose of the adjugate matrix to $\De$.
Then, for $x$-large  enough, the matrices $\Pi$ and $\Pi^{-1}$ admit the below uniformly convergent Neumann series
\beqa
\Pi\!\pa{\la} &=& I_2 + \sul{N \geq 1}{} \Int{ \Sg_{\Pi}^{\pa{N}} }{}  \f{ \dd^N z}{ \pa{2i\pi}^N }
\f{ \De\pa{z_N}\dots \De\pa{z_1} }{\pa{\la-z_1} \pl{s=1}{N-1}\pa{z_s-z_{s+1}}} \label{ecriture serie Neumann Pi}\\
\Pi^{-1}\!\pa{\la} &=& I_2 + \sul{N \geq 1}{} \Int{ \Sg_{\Pi}^{\pa{N}} }{}  \f{ \dd^N z}{ \pa{2i\pi}^N }
\f{ \nabla\pa{z_1}\dots \nabla\pa{z_N} }{\pa{\la-z_1} \pl{s=1}{N-1}\pa{z_s-z_{s+1}}} \label{ecriture serie Neumann Pi-1} \;.
\eeqa
The convergence holds in $L^{\infty}\pa{O}$ sense for $\la\in O$, with $O$ any subset of $\Cx$  such that $\e{d}\!\pa{ O,\Sg_{\Pi} } >0$. 
Also, it holds for $\la_{\pm} \in \Sg_{\Pi}$ in the $L^{2}\!\pa{\Sg_{\Pi}}$ sense. Finally, the matrices $\De$ and $\nabla$ that are 
integrated along the inslotted contour $\Sg_{\Pi}^{\pa{N}}$ should be understood
according to definition \ref{Definition Matrices Delta Nabla Sur contour encastre}. 

\end{lemme}

\Proof

We define two linear operators on $2\times2$ $L^{2}\!\pa{\Sg_{\Pi}}$-valued matrices
\beq
\mc{C}_{\Sg_{\Pi}}^{\De}\!\pac{M}\pa{\la} = \Int{\Sg_{\Pi}}{} \!\! \f{\dd s}{2i\pi \pa{\la_+ - s} }  M\pa{s} \De\pa{s}
\qquad \e{and} \qquad
 ^{\bs{t}}\mc{C}_{\Sg_{\Pi}}^{\nabla}\!\pac{M}\pa{\la} = \Int{\Sg_{\Pi}}{} \!\!  \f{\dd s}{2i\pi \pa{\la_+ - s} } \nabla\pa{s} M\pa{s} \; .
\label{definition operateur Cauchy + cas matrice L2}
\enq
Using that for sufficiently regular, not necessarily bounded, contours $\Sg_{\Pi}$, the $\pm$ limits of the Cauchy transform
with support on $\Sg_{\Pi}$ are continuous operators on
$L^{2}\!\pa{\Sg_{\Pi}}$ with norm $c\pa{\Sg_{\Pi}}$ \cite{KhvedelidzePreciseStudyPropertiesCauchyTransform},
it is easy to see that that the two above operators are also continuous\symbolfootnote[2]{By intechanging the roles of $\De$ and $M$, it is easy to see that 
$\mc{C}^{\De}_{\Sg_{\Pi}}$ is continuous on $\msc{M}_2\pa{L^{\infty}\!\pa{\Sg_{\Pi}}}$ since $\De \in L^{2}\!\pa{\Sg_{\Pi}}$.} 
on the space $\msc{M}_2\pa{L^{2}\!\pa{\Sg_{\Pi}}}$ of $2\times2$ matrices with $L^{2}\!\pa{\Sg_{\Pi}}$ entries:
\beqa
\norm{\mc{C}_{\Sg_{\Pi}}^{\De}\pac{M} }_{L^{2}\pa{\Sg_{\Pi}}} &\leq&  2 c\pa{\Sg_{\Pi}}  \norm{\De}_{L^{\infty}\pa{\Sg_{\Pi}}}  \norm{M}_{L^{2}\pa{\Sg_{\Pi}}}  \label{ecriture continuite operateur Cauchy matriciel} \\
\norm{ ^{\bs{t}} \mc{C}_{\Sg_{\Pi}}^{\nabla}\pac{M} }_{L^{2}\pa{\Sg_{\Pi}}} &\leq&  2 c\pa{\Sg_{\Pi}}  \norm{\De}_{L^{\infty}\pa{\Sg_{\Pi}}}  \norm{M}_{L^{2}\pa{\Sg_{\Pi}}}
\label{ecriture continuite operateur Cauchy matriciel transpose}
\eeqa

There we made use of the fact that $\nabla$ is the transpose of the adjugate matrix to $\De$
so that $\norm{\De}_{L^{\infty}\pa{\Sg_{\Pi}}} =\norm{\nabla}_{L^{\infty}\pa{\Sg_{\Pi}}}$ and
$\norm{\De}_{L^{2}\pa{\Sg_{\Pi}}} =\norm{\nabla}_{L^{2}\pa{\Sg_{\Pi}}}$.

It is a standard fact \cite{ClanceyGohbergFactorizationMatrixFunctionSingIntEqnRHPreference} that there is a one-to-one correspondence between the 
solution to the RHP for $\Pi$ (or $\Pi^{-1}$) and the unique solution to the singular integral equations
\beq
\Pi_+ - \mc{C}_{\Sg_{\Pi}}^{\De}\pac{\Pi_+}  = I_2 \qquad \qquad  \e{and} \qquad  \qquad
\Pi_+^{-1} - ^{\bs{t}}\! \mc{C}_{\Sg_{\Pi}}^{\nabla} \pac{\Pi^{-1}_+}  = I_2   \;.
\label{equation Singuliere Pi et Pi-1}
\enq
Indeed, provided that $\Pi_+$ is known, the matrix $\Pi$ (or $\Pi^{-1}$) admits the below integral representation for $\la$ away from $\Sg_{\Pi}$
\beq
\Pi\pa{\la} = I_2  + \Int{\Sg_{\Pi}}{}  \f{ \dd s }{ 2i\pi \pa{\la-s } } \Pi_{+}\pa{s}\De\pa{s}
\qquad \e{and} \qquad 
\Pi^{-1}\!\pa{\la} = I_2  + \Int{\Sg_{\Pi}}{}  \f{ \dd s }{ 2i\pi \pa{\la-s } } \nabla\pa{s} \Pi_{+}^{-1}\pa{s} \;.
\enq
The estimates for the jump matrices on the boundary of the discs $\Dp{}\mc{D}_{\pm q , \de} $ and  $\Dp{} \mc{D}_{\la_0,\de}$ and the specific choice
for the shape of the contour $\Sg_{\Pi}$ at infinity lead to 
$\norm{\De}_{L^{2}\pa{\Sg_{\Pi}}} + \norm{\De}_{L^{\infty}\pa{\Sg_{\Pi}}}= \e{O}\pa{x^{-w} }$
with $  w= \min\pa{\tf{1}{2}, 1-\rho_{\de}}>1 $ and $\rho_{\de}$ defined in \eqref{definition rho delta}. 
This implies that for $x$-large enough the operators $I-\mc{C}_{\Sg_{\Pi}}^{\De}$ and
$I- ^{\bs{t}} \mc{C}_{\Sg_{\Pi}}^{\nabla}$ are invertible and that their inverse can be computed by a Neumann series expansion
 converging in $L^{2}\pa{\Sg_{\Pi}}$:
\beqa
\Pi_+\!\pa{\la} &=& 
I_2+\sul{N\geq 1}{}  \paa{\mc{C}_{\Sg_{\Pi}}^{\De} }^{N} \!\!\pac{I_2}\pa{\la} =
I_2 + \sul{N\geq 1}{}
\Int{\Sg_{\Pi}}{} \f{\dd^N z}{\pa{2i\pi}^N}  \f{ \De\pa{z_N} \dots \De\pa{z_1}  }
{\pa{\la_{+}-z_1}  \pl{s=1}{N-1}\pa{ \pac{z_{s}}_{+}-z_{s+1}}  } \; .
\label{serie de Neumann Pi+} \\
\Pi^{-1}_+\!\pa{\la} &=& 
I_2+\sul{N\geq 1}{}  \,  \paa{^{\bs{t}} \mc{C}_{\Sg_{\Pi}}^{\nabla}}^{N} \!\! \pac{I_2} \pa{\la} =
I_2 + \sul{N\geq 1}{}
\Int{\Sg_{\Pi}}{} \f{\dd^N z}{\pa{2i\pi}^N}  \f{ \nabla\pa{z_1} \dots \nabla\pa{z_N}  }{\pa{\la_{+}-z_1}
  \pl{s=1}{N-1}\pa{ \pac{z_{s}}_{+}-z_{s+1}}   } \; .
\label{serie de Neumann Pi+ inverse}
\eeqa
Where $\paa{\mc{C}_{\Sg_{\Pi}}^{\De}}^{N}\!\!= \mc{C}_{\Sg_{\Pi}}^{\De} \circ \dots \circ \mc{C}_{\Sg_{\Pi}}^{\De}$ stands for the composition of
$N$ operators $\mc{C}_{\Sg_{\Pi}}^{\De}$. In \eqref{serie de Neumann Pi+}-\eqref{serie de Neumann Pi+ inverse}
the integration runs across the Carthesian product of $N$ copies of $\Sg_{\Pi}$:
$\Sg_{\Pi}\times \dots \times \Sg_{\Pi}$. 

The fact that $\Pi^{\pm 1}\pa{\la}$ admits a uniformly convergent Neumann series for $\la$ belonging to any open set O at finite distance from $\Sg_{\Pi}$
follows from the $L^{2}\pa{\Sg_{\Pi}}$ convergence of the series \eqref{serie de Neumann Pi+}-\eqref{serie de Neumann Pi+ inverse}, 
the fact that $\De \in \msc{M}_2\pa{L^{2}\pa{\Sg_{\Pi}}\cap L^{1}\pa{\Sg_{\Pi}} }$, and that $\e{d}\pa{O,\Sg_{\Pi}}>0$.

Finally, it is easy to check that one gets
the expression for $\Pi\!\pa{\la}$ (resp. $\Pi^{-1}\!\pa{\la}$) on $\Cx\setminus \Sg_{\Pi}$ by replacing the $+$ type regularization $\la_+$ of $\la$ in \eqref{serie de Neumann Pi+} (resp. \eqref{serie de Neumann Pi+ inverse}) by $\la \in\Cx\setminus \Sg_{\Pi}$. 
 
The $N^{\e{th}}$ summand of the Neumann series for $\Pi_{+}^{\pm 1}$ can be expressed in terms of regularized 
by deforming the original contour $\Sg_{\Pi}\times \dots \times \Sg_{\Pi}$ to the inslotted one $\Sg_{\Pi}^{\pa{N}}$. The latter manipulation is 
possible due to the properties of the locally analytic matrices $\De\pa{z}$ and $\nabla\pa{z}$. It allows one to get rid of the $+$ regularization in the 
integrals.

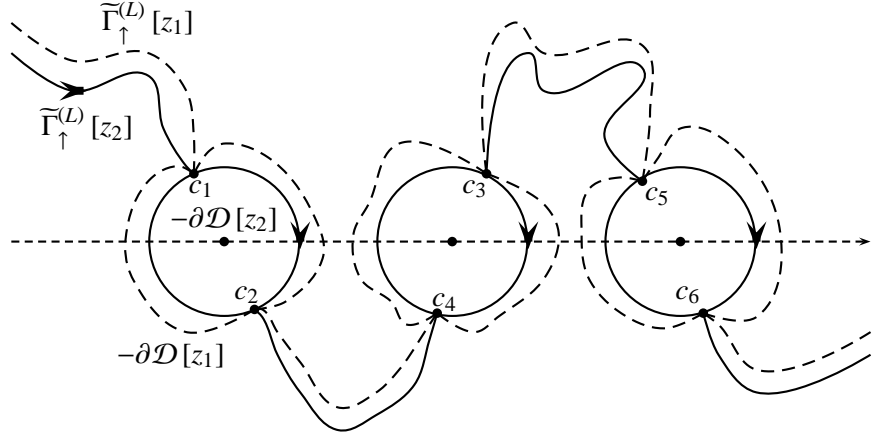
\begin{figure}[h]
\begin{center}

\begin{pspicture}(12,7)

\psline[linestyle=dashed, dash=3pt 2pt]{->}(0.2,3.5)(11.5,3.5)


\psdots(3,3.5)(6,3.5)(9,3.5)

\pscircle(3,3.5){1}
\pscircle(6,3.5){1}
\pscircle(9,3.5){1}


\pscurve(0.2,6)(1,5.5)(2,5.7)(2.2,5)(2.5,4.5)(2.6,4.4)

\pscurve(3.4,2.6)(3.5,2.5)(3.7,2)(4,1.5)(4.5,1)(5,1.2)(5.5,1.6)(5.7,2.2)(5.8,2.55)
\pscurve(6.45,4.4)(7,6)(7.2,5.5)(8.5,5.8)(8,4.9)(8.5,4.3)
\pscurve(9.3,2.55)(9.5,2)(10,1.5)(11.5,2)


\pscurve[linestyle=dashed](0.2,6.4)(1,5.9)(2,6)(2.5,5.5)(2.6,4.9)(2.6,4.4)

\pscurve[linestyle=dashed](3.4,2.6)(3.5,2.6)(3.7,2.3)(4,1.9)(4.5,1.3)(5,1.5)(5.3,1.8)(5.7,2.4)(5.8,2.55)

\pscurve[linestyle=dashed](6.45,4.4)(6.8,6.4)(7.2,6)(8.5,6.1)(8.6,5.2)(8.5,4.3)

\pscurve[linestyle=dashed](9.3,2.55)(9.5,2.4)(10,1.8)(11.5,2.3)


\pscurve[linestyle=dashed](2.6,4.4)(2.8,4.7)(3,4.8)(3.6,4.6)(4.2,3.8)(4.3,3.3)(3.8,2.7)(3.4,2.6)
\pscurve[linestyle=dashed](2.6,4.4)(2.4,4.5)(2,4.3)(1.7,3.5)(1.8,3)(2.5,2.3)(3.4,2.6)

\pscurve[linestyle=dashed](5.8,2.55)(6.3,2.3)(6.6,2.5)(7.2,3)(7.4,3.6)(7.2,4)(6.45,4.4)
\pscurve[linestyle=dashed](5.8,2.55)(5.4,2.4)(5.2,2.7)(4.8,3)(4.7,3.6)(4.9,4)(5.3,4.8)(6.45,4.4)

\pscurve[linestyle=dashed](8.5,4.3)(8.7,4.6)(9,5)(9.5,4.9)(10.3,3.7)(10,2.5)(9.3,2.55)

\pscurve[linestyle=dashed](8.5,4.3)(8.2,4.35)(7.8,4.2)(7.7,3.6)(7.8,3)(8.3,2.5)(8.8,2.35)(9.3,2.55)


\psline[linewidth=3pt]{->}(1.15,5.5)(1.2,5.5)

\psline[linewidth=3pt]{->}(4,3.5)(4,3.4)

\psline[linewidth=3pt]{->}(7,3.5)(7,3.4)

\psline[linewidth=3pt]{->}(10,3.5)(10,3.4)





\psdots(2.6,4.4)(3.4,2.6)(5.8,2.55)(6.45,4.4)(8.5,4.3)(9.3,2.55)

\rput(2.7,4.2){$c_1$}
\rput(3.3,2.8){$c_2$}

\rput(5.9,2.7){$c_4$}
\rput(6.3,4.2){$c_3$}

\rput(9.1,2.8){$c_6$}
\rput(8.7,4.1){$c_5$}


\rput(2,6.4){$\wt{\Ga}^{\pa{L}}_{\ua}\pac{z_1}$}
\rput(1.2,5){$\wt{\Ga}^{\pa{L}}_{\ua}\pac{z_2}$}


\rput(2.3,2){$-\Dp{}\mc{D}\pac{z_1}$}
\rput(3,3.8){$-\Dp{}\mc{D}\pac{z_2}$}

\end{pspicture}

\caption{Construction of the inslotted contour.\label{contour deforme pour integration encastree}}
\end{center}
\end{figure}

 The construction of the inslotted contour $\Sg_{\Pi}^{\pa{N}}$
is depicted in Fig.~\ref{contour deforme pour integration encastree} and \ref{deformation contour vers cercles
encastres}. Initially, the integral is performed with the use of the
$+$ boundary value of $z_1$ on the integration contour for $z_2$. Hence, away from the points of triple intersections $c_i$,
 we can deform the integration contour for $z_1$
to the $+$ side of the integration contour for $z_2$. 
One ends up with a contour as depicted in Fig.~\ref{contour deforme pour integration encastree}. There, the dotted lines
correspond to the integration contour for $z_1$ whereas the full lines give the integration contour for $z_2$.
One then proceeds inductively in this way up to $z_N$. As $\la$ is assumed to lie uniformly away from the original contour $\Sg_{\Pi}$,
there is no problem to deform the integration contour for $z_1$ in the vicinity of $\Sg_{\Pi}$ as the pole at $z_1=\la$ is lying
"far" away. 

\begin{figure}[h]
\begin{center}

\begin{pspicture}(8,4)

\pscircle(1.5,2){1}

\pscircle[linestyle=dashed](6.5,2){1.4}

\pscircle(6.5,2){0.8}


\pscurve[linestyle=dashed](1.5,3)(1.6,3.2)(2,3.3)(2.7,2.8)(3,2)(2.8,1.5)(1.7,0.8)(1.5,1)

\pscurve[linestyle=dashed](1.5,3)(1.3,3.2)(1,3.3)(0.5,2.8)(0.3,2)(0.4,1.5)(1.3,0.8)(1.5,1)

\pscurve(1.5,3)(1.4,3.3)(1.2,3.5)(0.5,3.7)
\pscurve[linestyle=dashed](1.5,3)(1.5,3.4)(1.2,3.8)(0.5,4)

\pscurve[linestyle=dashed](1.5,1)(1.6,0.7)(1.8,0.5)(2.5,0.3)
\pscurve(1.5,1)(1.5,0.6)(1.8,0.2)(2.5,0)

\pscurve(6.5,1.2)(6.5,0.6)(6.8,0.3)(8,0)
\pscurve(6.5,2.8)(6.5,3.4)(6.2,3.6)(5.5,3.8)

\pscurve[linestyle=dashed](6.5,0.6)(6.8,0.4)(8,0.5)
\pscurve[linestyle=dashed](6.5,3.4)(6.4,3.8)(6.2,4)(5.5,4)



\pscurve{->}(4,3.2)(3.8,3)(4.4,3)
\pscurve{->}(4,2.2)(3.8,2)(4.4,2)
\pscurve{->}(4,1.2)(3.8,1)(4.4,1)


\rput(8.1,3){$\Dp{}\mc{D}\pac{z_1}$}
\rput(6.5,1.8){$\Dp{}\mc{D}\pac{z_2}$}


\end{pspicture}
\caption{Deformation of the circles.\label{deformation contour vers cercles encastres}}
\end{center}
\end{figure}
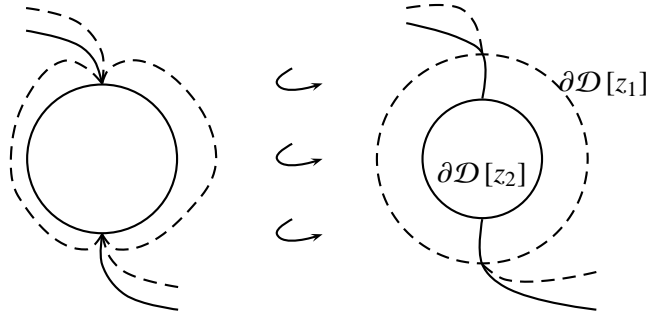

It remains to threat the integration on the intersection points of the disks $\Dp{}\mc{D}\big[z_j \big]$ with the curves
$\wt{\Ga}^{\pa{L/R}}_{\ua/\da} \big[z_j \big]$. We first reduce the most interior disc (the one over which $z_N$ is integrated and then the procedure is
repeated by induction) to smaller a one. The jump matrices $\De$ have different analytic continuations from the right and left of  the points $c_i$ 
(this corresponds to the discontinuity lines of $\mc{P}_0$ and $\mc{P}_{\pm}$). Taking this difference into account produces the small extensions of the 
contours $\Ga^{\pa{L/R}}_{\ua/\da}\pac{z_N}$ as depicted on the right part of
Fig.~\ref{deformation contour vers cercles encastres} together with smaller discs $\Dp{}\mc{D}\pac{z_N}$. It is in this way that the matrix $\De$
integrated over $\Sg_{\Pi}\pac{z_N}$ is identified with the one stemming from the jump matrix for $\Pi$ when the latter is defined as in 
\eqref{definition matrice de saut pour RHP Pi} but with jumps on $\Sg_{\Pi}\pac{z_N}$ (what corresponds to slight deformations of the curves
$\wt{\Ga}^{\pa{L/R}}_{\ua/\da}$).  \qed


\begin{prop}
\label{proposition DA Pi}
The matrix $\Pi$ admits the series expansion
\beq
\Pi\pa{\la}= I_2 +\sul{N \geq 1 }{ \infty } \f{ \Pi_{N}\!\pa{\la} }{ x^{N} } \; ,
\label{ecriture Pi comme serie asymptotique detaille}
\enq
that is valid uniformly away from $\Sg_{\Pi}$ and also on the boundary $\Sg_{\Pi}$ in the sense of $L^{2}\pa{\Sg_{\Pi}}$ boundary values.
The coefficients $\Pi_{N}$ of this expansion take the form
\bem
\Pi_N \! \pa{\la} = A_N\! \pa{\la} \; + \;   \sul{m=-\pac{\f{N}{2}} }{ \pac{\tf{N}{2}} }  
\f{ \ex{i x m \pac{u\pa{q}-u\pa{-q}}}}{ x^{2m\pac{\nu\pa{q}+\nu\pa{-q}}} }
\Pi_{N}^{\pa{m}}\!\pa{\la}  \\
+\sul{b=1}{\pac{\tf{N}{2}}} \;  \sul{p=0}{ b } \; \sul{m=b-\pac{\f{N}{2}} }{ \pac{\f{N}{2}}-b } \;
\f{ \ex{i x m \pac{u\pa{q}-u\pa{-q}}}}{ x^{2m\pac{\nu\pa{q}+\nu\pa{-q}}} } \cdot
 x^{ \f{b}{2}} \f{ \ex{ix \bs{\eta}\pac{ b u\pa{\la_0}- p u\pa{q} +\pa{p-b}u\pa{-q} }}  }
{ x^{  2 \bs{\eta} \pa{b-p}\nu\pa{-q}   - 2p \bs{\eta} \nu\pa{q} } }
  \Pi_{N}^{\pa{m,\, b,\, p}}\!\pa{\la} \; ,
\label{ecriture form detaillee AE}
\end{multline}
 and one should set  $\bs{\eta}=1$ in the space-like regime and $\bs{\eta}=-1$ in the time-like.

\vspace{2mm}
The matrix $A_N\pa{\la}$ contains only exponentially small corrections, \textit{ie} $\pac{A_N}_{ij}\pa{\la}=\e{O}\pa{x^{-\infty}}$
with a  $\e{O}$ that is uniform for $\la$-uniformly away from $\Sg_{\Pi}$.

\vspace{2mm}
The matrices  $\Pi_{N}^{\pa{m}}\!\pa{\la}$ and $\Pi_{N}^{\pa{m,b, p}}\!\pa{\la}$ admit the asymptotic expansion
\beqa
\Pi_{N}^{\pa{m}}\!\pa{\la} &=& \sul{r \geq 0}{} \Pi_{N;r}^{\pa{m}}\!\pa{\la} \qquad \e{with} \qquad
 \Pi_{N;r}^{\pa{m}}\!\pa{\la} = \e{O}\pa{ \f{ \pa{\log x}^{N +r - \de_{m, 0}- 2m } }{ x^r } M } \;\; , \\
\Pi_{N}^{\pa{m,\, b,\, p}}\!\pa{\la} &=& \sul{r \geq 0}{} \Pi_{N;r}^{\pa{m,\, b,\, p}}\!\pa{\la} \qquad \e{with} \qquad
 \Pi_{N;r}^{\pa{m,\, b,\, p}}\!\pa{\la} = \e{O}\pa{ \f{ \pa{\log x}^{N +r - 2\pa{m+b} } }{ x^r } M}\; ,
\eeqa
The estimates hold for $\la$ uniformly away from $\Sg_{\Pi}$.

The matrix $M$ appearing in the various $\e{O}$ estimates takes the form:
\beq
M=\pa{\ba{cc}
  1 & m_{+} \f{\ex{i x u\pa{q}}}{x^{2\nu\pa{q}} } + m_{-} x^{2\nu\pa{-q}}  \ex{ix u\pa{-q}}  + m_{0} \sqrt{x} \ex{ix u\pa{\la_0}}   \\
  \wt{m}_{+} \f{x^{2\nu\pa{q}}}{\ex{i x u\pa{q}}} + \wt{m}_{-}  \f{\ex{-ix u\pa{-q}}}{x^{2\nu\pa{-q}}}  + \wt{m}_{0} 
 \f{ \sqrt{x} }{ \ex{ix u\pa{\la_0}} } & 1
\ea }
\label{ecriture forme generale matrice M pour estimation Pi exacte}
\enq
There $m_{\pm}$, $m_0$, $\wt{m}_{\pm}$ and $\wt{m}_0$ are $x$-independent coefficents. Moreover, necessarily, $m_0=0$ in the time-like regime
and $\wt{m}_0=0$ in the space-like one.

\end{prop}

We postpone the proof of this asymptotic expansion to appendix \ref{Appendix Proofs asymptotic expansion for Pi and determinant}
as it is rather cumbersome and long. However, at this point, we would like to make several comments in respect to the form of the
asymptotic expansion.

The above asymptotic  expansion is in a form very similar to the one of the functionals $\mc{H}_n\pac{\nu,\ex{g},u}$
given in Theorem \ref{theorem representation serie de Natte}. 
In fact, the large-$x$ behavior of the matrix $\Pi_N$ contains various fractional
powers of $x$, each appearing with its own oscillating pre-factor.
Once that one has fixed a given phase factor and fractional power of $x$, then
the corresponding matrix coefficients $\Pi_{N}^{\pa{m}}$ or $\Pi_N^{\pa{m,\, b,\, p}}$ admit an asymptotic expansion
in the more-or-less standard sense. That is to say, each of their entries admits an asymptotic expansion
into a series whose $n^{\e{th}}$ term can be written as $\tf{P_{N+n}\pa{\log x}}{x^n}$ with $P_{N+n}$ being a polynomial of degree at most $N+n$.
One of the consequences of such a structure is that an oscillating term present in $x^{-n}\Pi_n\pa{\la}$ may be in fact dominant in respect to, say, a
non-oscillating term present in $x^{-n^{\prime}}\Pi_{n^{\prime}}\pa{\la}$ where $n^{\prime}<n$.

We would also like to point out that the asymptotic expansions of $\Pi_N$ and $\Pi_{N+1}$ share many oscillating terms at equal frequencies
(\textit{eg} $\ex{ix\pac{u\pa{q}-u\pa{-q}}}$ is present in $\Pi_{N}$ and $\Pi_{N+1}$ for any $N \geq 2$).
However, those issued from $\Pi_{N+1}$ have an additional dumping pre-factor $\log x \cdot x^{-1}$ in respect to the same ones issued from $\Pi_N$.
Finally, there may also appear additional oscillatory terms $\ex{ \pm i x u\pa{z}}$, $z=\pm q$ or $\la_0$ (and their associated
fractional powers of $x$) in the off-diagonal parts of $\Pi_{N}^{\pa{m}}$ and
$\Pi_N^{\pa{m,\, b,\, p}}$, \textit{cf} \eqref{ecriture forme generale matrice M pour estimation Pi exacte}.



\vspace{3mm}

There is also another way of writing down the asymptotic expansion of $\Pi\pa{\la}$. Although it is more compact, it is also less explicit
and provides one with weaker estimates for the remainders.

\begin{prop}
\label{proposition DA Pi heuristique}
The matrix $\Pi$ admits the asymptotic expansion
\beq
\Pi\pa{\la}= I_2 +\sul{n \geq 0 }{ N } \f{ \Pi^{\pa{n}}\!\pa{\la} }{ x^{\f{1+n}{2}} }  +
\e{O}\pa{ \ba{cc} 1 & 1 \\ 1 & 1 \ea  }    x^{-\pa{N+1}w}  \quad \e{with} \quad
w=\min\pa{\f{1}{2}, 1-\rho_{\de}}\; ,
\label{ecriture Pi comme serie asymptotique heuristique}
\enq
that is valid uniformly away from $\Sg_{\Pi}$.

For $\la$ belonging to any connected component of $\infty$ in $\Cx\setminus \Sg_{\Pi}$, the first few terms appearing in this expansion read
\beqa
\Pi^{\pa{0}}\pa{\la} &=& - \f{ d^{\pa{0}}\!\pa{\la_0} }{ \la-\la_0 } \sg   \;\;  , \qquad  \quad
\Pi^{\pa{1}}\pa{\la} = - \sul{\eps=\pm}{} \f{ V^{\pa{\eps; 0}}\!\pa{\eps q} }{ \la-\eps q }   \; , \\
\Pi^{\pa{2}}\pa{\la} &=&  \sul{\eps=\pm}{} \f{ d^{\pa{0}}\!\pa{\la_0} }{\la_0-\eps q}
    \paa{ \f{ V^{\pa{\eps; 0}} \!\pa{\eps q} \sg }{\la-\la_0}  -\f{ \sg V^{\pa{\eps; 0}}\!\pa{\eps q}  }{\la-\eps q} }
- \f{ \sg }{ 2 } \f{\Dp{}^2 }{\Dp{}s^2}  \paa{  \f{d^{\pa{1}} \!\pa{s}}{\la-s} }_{s=\la_0} \hspace{-4mm} .
\eeqa
The expression for $\Pi^{\pa{3}}$ is a bit more involved.
\bem
\Pi^{\pa{3}}\pa{\la} =  \f{ \pac{ d^{\pa{0}}\!\pa{\la_0}  }^2 }{ \la-\la_0}
\sul{\eps=\pm}{} \f{ \sg V^{\pa{\eps; 0}}\!\pa{\eps q} \sg  }{ \pa{\la_0-\eps q}^{2} }
+ \sul{\eps=\pm}{} \eps \f{  V^{\pa{-\eps; 0}}\!\pa{-\eps q} V^{\pa{\eps; 0}}\!\pa{\eps q}  }{ 2q\pa{ \la-\eps q } } \\
- \f{1}{2}\sul{\eps=\pm}{} \f{ \Dp{} }{ \Dp{}s } \paa{ \f{ V^{\pa{\eps; 1}}\pa{s}  - 2 V^{\pa{\eps; 0}}\!\pa{\eps q}V^{\pa{\eps; 0}}\pa{s} }{\la-s} }_{s=\eps q} \hspace{-4mm}.
\end{multline}
There $\sg=\sg^+$ in the space-like regime and $\sg=\sg^{-}$ in the time-like regime.
%
%
%
%
%
%
%
%
%
%
\end{prop}

This form of the asymptotic expansion is the closest, in spirit, to the one appearing in the literature, \textit{cf} \textit{eg}
\cite{DeiftKriechMcLaughVenakZhouOrthogonalPlyExponWeights}.
However, it does not represent a "well-ordered" asymptotic expansion in the sense that each matrix $\Pi^{\pa{n}}$ depends on the various
fractional powers of $x$ and oscillating corrections. Some terms present in the entries of $\Pi^{\pa{p}}$ are dominant in respect to
the ones present $\Pi^{\pa{\ell}}$, $\ell < p$. Moreover, the expansion \eqref{ecriture Pi comme serie asymptotique heuristique}
does not provide one with a precise identification of these terms. 
This form is however very convenient from the computational point of view, and having explicit expressions for the matrices 
$\Pi^{\pa{\ell}}$ easily
allows one to identify the various matrices entering in the "well-ordered" asymptotic expansion
\eqref{ecriture Pi comme serie asymptotique detaille}-\eqref{ecriture form detaillee AE}.

\Proof

 The unique solution $\Pi_+$ to the singular integral equation \eqref{equation Singuliere Pi et Pi-1} equivalent to 
the uniquely solvable RHP for $\Pi$ provides an integral representation for $\Pi$  on $\Cx\setminus\Sg_{\Pi}$. Namely,
\beq
\Pi\pa{\la} = I_2 + \Int{\Sg_{\Pi}}{} \f{ \dd s }{2i\pi \pa{\la-s}} \Pi_+\pa{s} \De\pa{s}  \qquad \e{for} \quad  \la \in \Cx\setminus \Sg_{\Pi} \;.
\label{equation pour la representation integrable Pi off axis}
\enq

The only places where the jump matrix for $\Pi$ is  not exponentially close to the identity are the three boundaries of the discs
$-\Dp{}\mc{D}_{\pm q , \de}$ and $-\Dp{}\mc{D}_{\la_0 , \de}$. There one has
\beq
\Pi_{+} \mc{P}_{\pm q}^{-1} = \Pi_- \quad \e{on} \;\;  - \Dp{}\mc{D}_{\pm q,\de} \qquad \e{and} \qquad
 \Pi_{+} \mc{P}_{0}^{-1} = \Pi_- \quad \e{on} \;\;  - \Dp{}\mc{D}_{\la_0,\de} \; .
\enq
Note that the minus sign refers  to the clockwise orientation of the boundary of the discs in Fig.~\ref{contour pour le RHP de Pi space-like}
and \ref{contour pour le RHP de Pi time-like}.

By using the estimate
\beq
\mc{N}\pa{\De} = \norm{\De}_{L^{1}\pa{ \Sg_{\Pi} }} +  \norm{\De}_{L^{\infty}\pa{ \Sg_{\Pi} }}  = \e{O}\pa{ x^{-w}  }
\quad \e{with} \quad w=\min\pa{\f{1}{2}, 1-\rho_{\de}} 
\enq
and equation \eqref{equation Singuliere Pi et Pi-1},  one shows by standard methods (see \textit{eg} 
\cite{DeiftKriechMcLaughVenakZhouOrthogonalPlyExponWeights}) the existence of the asymptotic
expansion \eqref{ecriture Pi comme serie asymptotique heuristique} for $\Pi\pa{\la}$.
This expansion
 is valid uniformly away from the jump curve $\Sg_{\Pi}$.

We would like to stress that for computing the coefficients of the asymptotic expansion, we can drop the integration contours 
other then the boundaries of the disks $\Dp{}\mc{D}_{\pm q/ \la_0; \de}$. Indeed as $\pa{\Pi_{+}-I_2 }\in L^{2}\pa{\Sg_{\Pi}}$,
\textit{cf} \eqref{ecriture serie Neumann Pi}   , and
$\norm{\De}_{ L^{2}\cap L^1 \pa{\wt{\Ga}} }= \e{O}\pa{ x^{-\infty} }$, it is clear that the integration along $\wt{\Ga}$
in \eqref{equation Singuliere Pi et Pi-1} can only produce exponentially small corrections. Thence,
it cannot contribute to the asymptotic expansion \eqref{ecriture Pi comme serie asymptotique heuristique}.
As a consequence, the matrix coefficients $\Pi^{\pa{n}}$ in \eqref{ecriture Pi comme serie asymptotique heuristique} can be computed
by plugging\symbolfootnote[2]{It is possible to insert the asymptotic expansion, which a priori is valid only uniformly away from $\Sg_{\Pi}$ in
\eqref{ecriture Pi comme serie asymptotique heuristique} in as much as one slightly deforms the contour $\Sg_{\Pi}$ in the $+$ direction what is 
allowed in virtue of the analytic properties of $\Pi_+$ and the local analyticity of $\De$.} the asymptotic series into the integral equation \eqref{equation pour la representation integrable Pi off axis},  dropping
there all the exponentially small corrections (stemming from the integration along $\wt{\Ga}$)
and replacing the jump matrices $\mc{P}_{0}^{-1}$, $\mc{P}_{\pm q}^{-1}$ by their asymptotic expansions which are valid uniformly
on the boundaries of the three discs. This leads to the formal (in the sense that valid order by order in $x$) equation,
\beq
\Pi_+\!\pa{\la} \simeq I_2 - \f{1}{2i\pi} \Int{\Dp{}\mc{D}_{\la_0,\de}}{} \f{\dd s  }{\la_+ - s } \sul{n \geq 0}{}
            \f{ \Pi_+\!\pa{s}  \,  d^{\, \pa{n}}\!\pa{s}\sg }{ \pa{s-\la_0}^{2n+1} x^{n+\f{1}{2}} }
- \f{1}{2i\pi} \sul{\eps=\pm}{} \Int{\Dp{}\mc{D}_{\eps q,\de}}{} \f{\dd s  }{\la_+ - s }
\sul{n \geq 0}{} \f{ \Pi_+\!\pa{s} \, V^{\pa{\eps;n}}\!\pa{s} }{ \pa{n+1}! \pa{s-\eps q}^{n+1} x^{n+1} } \; .
\enq
It now remains to equate the coefficients of equal inverse powers in $x$.
This yields sets of recurrence relations between the various terms appearing in the asymptotic expansion for $\Pi$.
A straightforward residue computation leads to the result for $\Pi^{\pa{n}}, \; n=0, \dots, 3$, for $\la$ belonging to any connected component of 
$\infty$ in $\Cx\setminus \Sg_{\Pi}$. \qed

\subsection{Proof of the leading asymptotics of the determinant}
\label{sous-section DA determinant preuve}
We now prove theorem \ref{theorem DA determinant}. We divde the proof into three part. First, we obtain a modified version of the 
integral representation \eqref{formule derviee x log det} for $\Dp{x} \log \ddet{}{I+V}$ that will be more suited for our further
computations. Then, we use this integral representation so as to compute the first few $x$-dependent terms in
the asymptotics. Finally, we fix the constant, $x$-independent part of the asymtptotics.

\subsubsection*{ $\bullet$ Modification of the integral representation}
The first few terms of the asymptotic expansion of $\ddet{}{I+V}$ can be obtained by using the identity \eqref{formule derviee x log det} between the $x$-derivative of $\log \ddet{}{I+V}$
and the RHP data $\chi$, together with the asymptotic expansion for $\Pi$. As a starting remark, we observe that one can always choose
the contour $\Ga\pa{\msc{C}_{E}}$ appearing in \eqref{formule derviee x log det} in such a way
that it only passes in the region where
\beq
\chi\pa{\la}=\Pi\pa{\la} \a^{-\sg_3}\!\pa{\la} \pa{I_2+C\pac{e^{-2}}\pa{\la} \sg^+} \; .
\enq
Then, plugging this exact expression for $\chi$ into the trace appearing in \eqref{formule derviee x log det}, one gets that
\beq
\e{tr}\paa{\Dp{\la}\chi\pa{\la}\pac{ \sg_3 +2 C\pac{e^{-2}}\pa{\la} \sg^+ }  \chi^{-1}\!\pa{\la}}
= \e{tr}\pac{\Dp{\la}\Pi\pa{\la}\sg_3\Pi^{-1}\pa{\la}}- 2 \Dp{\la}\pa{ \log \a}\pa{\la} \;.
\label{equation pour l'identite des traces}
\enq
It remarkable, but also important from the computational point of view, that the matrix allowing one to simplify the complicated functions $E\pa{\la}$
\eqref{definition fonction E+} appearing in the formulation of the initial RHP, does not play a direct a role in the computation of the
asymptotics of the determinant. In particular, one does not have
to deal with integrations on $\Ga\pa{\msc{C}_{E}}$ of Cauchy transforms $C\pac{e^{-2}}\pa{\la}$.
Inserting \eqref{equation pour l'identite des traces} into \eqref{formule derviee x log det}, one obtains that the contribution of
$- 2 \Dp{\la}\pa{ \log \a}\pa{\la}$
can be separated from the rest, so that
\beq
\Dp{x}\log \ddet{}{I+V}\pac{\nu,u,g}= a_{-1}
-i \f{ \Dp{} }{ \Dp{}\eta }
\paa{ \;   \Int{  \Ga\pa{ \msc{C}_{E} }  }{} \hspace{-2mm} \f{\dd \la}{4\pi} \,  \ex{i\eta u\pa{\la}} \,   \e{tr}\pac{\Dp{\la} \Pi\pa{\la} \sg_3  \Pi^{-1}\pa{\la} }   }_{\eta=0^+}   \hspace{-3mm} .
\label{ecriture decomposee leading and subleading derivee x log det}
\enq
where we have set
\beq
a_{-1} = \Int{-q}{q} \f{\dd \la}{2\pi} \,  u^{\prime}\pa{\la} \log \paf{\a_-\pa{\la}}{\a_+\pa{\la}} 
= i \Int{-q}{q}  u^{\prime}\!\pa{\la} \nu\pa{\la}  \dd \la \;.
\enq

\begin{figure}[h]
\begin{center}

\begin{pspicture}(12,7)



\rput(1.9,5.1){$P_1$}
\rput(10.4,1.9){$P_2$}
\psdots(2.25,4.9)(10.6,1.6)


\psdots(3,3.5)(6,3.5)(9,3.5)
\pscircle[linestyle=dashed, dash=3pt 2pt](3,3.5){1}
\pscircle[linestyle=dashed, dash=3pt 2pt](6,3.5){1}
\pscircle[linestyle=dashed, dash=3pt 2pt](9,3.5){1}


\pscurve[linestyle=dashed, dash=3pt 2pt](0.2,6)(1,5.5)(2,5.7)(2.2,5)(2.5,4.5)(2.6,4.4)

\pscurve[linestyle=dashed, dash=3pt 2pt](3.4,2.6)(3.5,2.5)(3.7,2)(4,1.5)(4.5,1)(5,1.2)(5.5,1.6)(5.7,2.2)(5.8,2.55)
\pscurve[linestyle=dashed, dash=3pt 2pt](6.45,4.4)(7,6)(7.2,5.5)(8.5,5.8)(8,4.9)(8.5,4.3)
\pscurve[linestyle=dashed, dash=3pt 2pt](9.3,2.55)(9.5,2)(10,1.5)(11.5,2)

\pscurve(2,4.8)(2.8,5)(4,5.5)(6,6)(8,6.5)(9,6)(10,5)(11,3)(10.5,1.5)(9,1.2)(6,0.5)(4,1)(3,1.3)(2.5,2)(1.5,3)(2,4.8)

\pscurve(2.25,4.85)(2.4,5.8)(1.8,6)(0.2,6.5)

\pscurve(10.6,1.6)(11,1.3)(11.5,1.6)


\psline[linewidth=3pt]{->}(2.5,4.55)(2.6,4.45)

\psline[linewidth=3pt]{->}(6,6)(6.1,6.05 )

\psline[linewidth=3pt]{->}(1.8,6)(1.9,5.95)

\psline[linewidth=3pt]{->}(11.3,1.4)(11.5,1.6)

\psline[linewidth=3pt]{->}(4,3.5)(4,3.4)

\psline[linewidth=3pt]{->}(7,3.5)(7,3.4)

\psline[linewidth=3pt]{->}(10,3.5)(10,3.4)


\rput(6,5){$\Sg_{\Pi}$}

\rput(2.3,6.5){$\ga^{\pa{L}}$}

\rput(9,6.5){$\ga^{\pa{0}}$}

\rput(11.1,1){$\ga^{\pa{R}}$}

\rput(1,5.1){$\wt{\Ga}_{\ua}^{\pa{L}}$}

\rput(11.8,2.2){$\wt{\Ga}^{\pa{R}}_{\da}$}

\end{pspicture}

\caption{ Contour $\ga=\ga^{\pa{L}} \cup \ga^{\pa{0}} \cup \ga^{\pa{R}}$. The contour $\Sg_{\Pi}$ is depicted in dotted lines. \label{contour ga pour l'integration serie Natte}}
\end{center}
\end{figure}
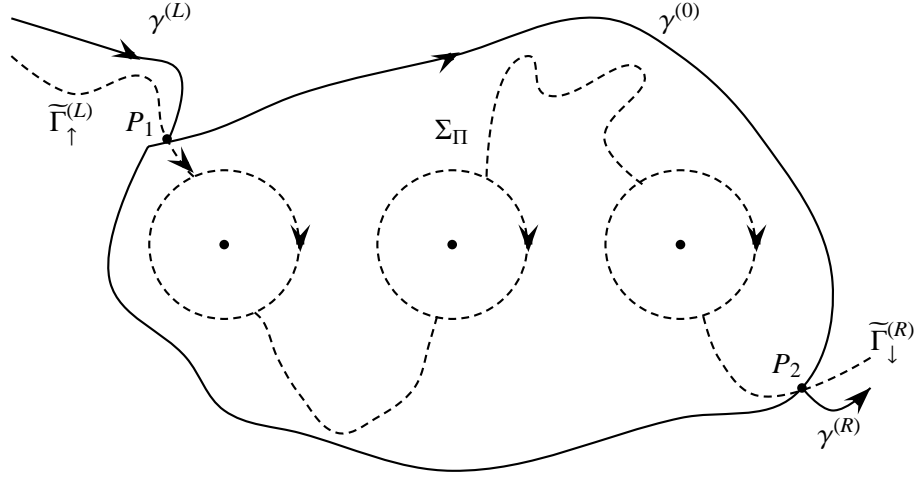

Note that one cannot exchange the $\eta$-derivation and the $\la$-integration symbols in \eqref{ecriture decomposee leading and subleading
derivee x log det} yet. To be able to do so, we
deform the most exterior parts of the contour $\Ga\pa{ \msc{C}_{E} }$ in \eqref{ecriture decomposee leading and subleading derivee x log det} to
$\wt{\Ga}^{\pa{L}}_{\ua}$ and $\wt{\Ga}^{\pa{R}}_{\da}$, \textit{cf} Fig.~\ref{contour ga pour l'integration serie Natte}.
 We denote $\ga^{\pa{0}}$ the resulting interior loop.
The integrand along $\ga^{\pa{0}}$ remains unchanged. However, when integrating along
the contours $\wt{\Ga}^{\pa{L/R}}_{\ua/\da}$, one should replace $\e{tr}\pac{\Dp{\la}  \Pi\pa{\la} \sg_3  \Pi^{-1}\pa{\la} } $
by the difference between the two boundary values:
\beq
\e{tr}\pac{ \Dp{\la} \Pi_-\pa{\la} \sg_3  \Pi_-^{-1}\pa{\la} } -\e{tr}\pac{ \Dp{\la} \Pi_+\pa{\la} \sg_3  \Pi_+^{-1}\pa{\la} }     \;.
\label{ecriture saut Pi sur contour externe}
\enq
We remind that not only the $\pm$ boundary values themselves, but also theses of $\Pi$'s derivatives do exist on 
$\wt{\Ga}^{\pa{L/R}}_{\ua/\da}$. This is a consequence of the
fact the jump matrix $I_2+\De$ for $\Pi$ on $\wt{\Ga}^{\pa{L/R}}_{\ua/\da}$ admits an analytic continuation to a neighborhood of these curves. 
This fact allows one for a local deformation of the jump contour $\Sg_{\Pi}$,  meaning that
$\Pi_+$ (resp. $\Pi_-$) admits an analytic continuation to some neighborhood of $\Sg_{\Pi}$
located on its $-$ (resp. $+$) side.

The difference in \eqref{ecriture saut Pi sur contour externe} can be estimated with the help of the jump condition for $\Pi$ along
$\wt{\Ga}^{\pa{L/R}}_{\ua/\da}$:
$\Pi_+ M = \Pi_-$ where $M$ is given by \eqref{definition matrice M time like} :
\bem
\e{tr}\pac{ \Dp{\la} \Pi_-\pa{\la} \sg_3  \Pi_-^{-1}\pa{\la} }   =
\e{tr}\pac{  \paa{ \pac{\Dp{\la} \Pi^{}_+\pa{\la} }M^{}\pa{\la} + \Pi_+\pa{\la} \Dp{\la}M\pa{\la}} \sg_3  M^{-1}\pa{\la}\Pi_+^{-1}\pa{\la} }   \\
= \e{tr}\pac{ \Pi_+^{-1} \pa{\Dp{\la} \Pi_+\pa{\la} }M^2\pa{\la} \sg_3} +
\e{tr}\pac{  \Dp{\la}M\pa{\la}  \sg_3  M^{-1}\pa{\la}   }  \;.
\end{multline}
Using that $M=I_2 + P \sg^+$, with $P$ being defined in \eqref{definition fonction P time-like} , we obtain the jump formula
\beq
\e{tr}\pac{ \Dp{\la} \Pi_-\pa{\la} \sg_3  \Pi_-^{-1}\pa{\la} } -\e{tr}\pac{ \Dp{\la} \Pi_+\pa{\la} \sg_3  \Pi_+^{-1}\pa{\la} }    =
2 \a^{-2}\pa{\la} e^{-2}\!\pa{\la}  \e{tr}\pac{ \Dp{\la} \Pi_+\pa{\la} \sg^+  \Pi_+^{-1}\pa{\la} }  \;.
\enq
Using, once again, the jump condition on $\wt{\Ga}^{\pa{L/R}}_{\ua/\da}$, we see that  $\e{tr}\pac{ \Dp{\la} \Pi\pa{\la} \sg^+  \Pi^{-1}\pa{\la} }$ has no
discontinuity across those parts of $\wt{\Ga}^{\pa{L/R}}_{\ua/\da}$ that we focus on. It can thus be extended to a holomorphic function in some neighborhood of this curve. Thence,
we can deform the contours of integration $\wt{\Ga}^{\pa{L/R}}_{\ua/\da}$ to $\ga^{\pa{L/R}}$ as depicted on Fig.~\ref{contour ga pour l'integration serie Natte}.  Once
that this has been done, there is no problem anymore to exchange
the $\eta$-derivation with the $\la$-integration.
Indeed,  $\e{tr}\pac{ \Dp{\la} \Pi\pa{\la} \sg^+  \Pi\pa{\la} }$ is bounded when $\Re\pa{\la} \tend \pm \infty$ along $\ga^{\pa{L/R}}$, and
the function $G$ given by
\beq
G\pa{\la}= u\pa{\la} \paa{   {\bf 1}_{\ga^{\pa{0}}}\!\pa{\la} + 2 \a^{-2}\pa{\la} e^{-2}\pa{\la}   {\bf 1}_{\ga^{\pa{L}}\cup \ga^{\pa{R}} }\!\pa{\la}  }
\label{definition fonction G}
\enq
is integrable. Here, ${\bf 1}_A$ stands for the characteristic function of the set $A$. Once that the $\eta$-derivative is computed, we get
the below integral representation
\beq
\Dp{x} \log \ddet{}{I+V}\pac{\nu,u,g} = a_{-1} +  \Oint{ \ga }{} \f{\dd \la}{4\pi} G\pa{\la}
 \tr\pac{ \Dp{\la} \Pi\pa{\la} \sg\pa{\la} \Pi^{-1}\pa{\la} } \; .
\label{representation derivee log det pour analysis asymptotique}
\enq
The final contour $\ga$ is depicted in Fig.~\ref{contour ga pour l'integration serie Natte}
and the matrix-valued function reads 
$\sg\pa{\la}= \sg_3  {\bf 1}_{\ga^{\pa{0}}}\!\pa{\la} +  \sg^+  {\bf 1}_{\ga^{\pa{L}}\cup\ga^{\pa{R}} }\!\pa{\la}$.

\subsubsection*{ $\bullet$ Extracting the first few $x$-dependent terms}

$\e{tr}\pac{ \Dp{\la} \Pi\pa{\la} \sg^{+} \Pi^{-1}\! \pa{\la} }$ is bounded on $\ga^{\pa{L}}\cup \ga^{\pa{R}}$ and
$\norm{G}_{L^{1}\pa{\ga^{\pa{R}}\cup \ga^{\pa{L}} }}= \e{O}\pa{x^{-\infty}}$. Hence, we can drop the part of integration over 
$\ga^{\pa{L}}\cup \ga^{\pa{R}}$ when computing the asymptotic expansion of $\log \ddet{}{I+V}$. It thus remains to treat 
the integration along $\ga^{\pa{0}}$.

As follows from proposition \ref{proposition DA Pi}, $\Pi$ has a uniform  asymptotic expansion on $\ga^{\pa{0}}$
given by \eqref{ecriture Pi comme serie asymptotique detaille}.
In order to obtain the leading asymptotic expansion
for the $x$-derivative of the determinant, it is readily seen that it is enough to plug in the more compact expansion
\eqref{ecriture Pi comme serie asymptotique heuristique} to the desired order and then drop all the terms that are irrelevant.
This is simpler from the point of view of computations and justified \textit{a posteriori} by the form of the 
well ordered asymptotic expansion \eqref{ecriture Pi comme serie asymptotique detaille}. Therefore we get
\bem
\e{tr}\pac{\Dp{\la}\Pi\pa{\la}\sg_3\Pi^{-1}\pa{\la}} = \f{1}{x} \e{tr}\paa{\pac{\Pi^{\pa{1}}}^{\prime}\!\pa{\la}\sg_3 }
 + \f{1}{x^{\f{3}{2}}} \e{tr}\paa{\pac{\Pi^{\pa{2}}}^{\prime}\!\pa{\la} - \Pi^{\pa{0}}\!\pa{\la}\pac{\Pi^{\pa{1}}}^{\prime}\!\pa{\la}
 - \Pi^{\pa{1}}\!\pa{\la}\pac{\Pi^{\pa{0}}}^{\prime}\!\pa{\la}}  \sg_3 \\
 + \f{1}{x^2} \e{tr}\paa{\pac{\Pi^{\pa{3}}}^{\prime}\!\pa{\la} - \Pi^{\pa{1}}\!\pa{\la}\pac{\Pi^{\pa{1}}}^{\prime}\!\pa{\la}}  \sg_3
 +  \e{o}\pa{ \f{\ex{\pm ix \pac{u\pa{q}-u\pa{-q}} } }{ x^{\pm 2\pac{\nu\pa{q}+2\nu\pa{-q}}+2} } ,
\f{\ex{i \bs{\eta} x\pac{u\pa{\la_0}-u\pa{\pm q}}  } }{x^{\f{3}{2} \mp 2\nu\pa{\pm q} }  }    , \f{\pa{\log x} }{x^{2}} } \;,
\label{developpement asymptotique trace Pi}
\end{multline}
uniformly on $\ga^{\pa{0}}$. There the $\e{o}$ refers to sub-leading terms that have been ignored. It distinguishes between the various oscillating and
non-oscillating corrections that have been ignored. Also one should set $\bs{\eta}=1$ in the space-like regime and $\bs{\eta}=-1$ in the time-like regime.
Note that due to the compactness of $\ga^{\pa{0}}$, the order of the $\e{o}$-remainder
is preserved by the integration along $\ga^{\pa{0}}$.

Note that, in \eqref{developpement asymptotique trace Pi}, we have been able to simplify certain products by exploiting that, regardless of the time or 
space like regimes $\pac{\Pi^{\pa{0}}}^2=0$ and that traces of matrices proportional to $\sg$ (with $\sg=\sg^{\pm}$ depending on the space
ot time-like regime) vanish ( \textit{eg} $\Pi^{\pa{0}} \Pi^{\pa{1}} 
\pac{\Pi^{\pa{0}}}^{\prime} \propto \sg$).

We now insert the explicit form of the first few matrix coefficients appearing in the expansion of $\Pi$  and then
integrate the expansion \eqref{developpement asymptotique trace Pi} along $\ga^{\pa{0}}$ with the appropriate weight. At the end of the day,
by using the precise estimates provided by the expansion \eqref{ecriture Pi comme serie asymptotique detaille}, we get
\beq
\Dp{x} \log \ddet{}{I+V} =a_{-1} + \f{ a_{0} }{ x }  + \f{ a_{1} }{ x^{\f{3}{2}} } \pa{ 1\!+\e{O}\paf{\log x}{x} } +
\f{ a_{2}^{\e{osc}}  }{ x^2 } \pa{ 1\! +\e{O}\paf{\log x}{x} }
+
\f{ a_{2}^{\e{no}}  }{ x^2 }  \pa{ 1\! +\e{O}\paf{\log x}{x} }
+ \e{O}\paf{a_1}{x^{w+\f{3}{2}}} + \e{O}\paf{a_2^{\e{osc}}}{x^{w+2}} .
\nonumber
\enq
Above the last $\e{O}$ corresponds to higher order oscillating correction with bigger phases than those involved in the definition of $a_1$
and $a_2^{\e{osc}}$. The term responsible for the logarithmic contribution to the determinant 
coincides with the one appearing in the time-independent case (the so-called generalized sine kernel)
considered in \cite{KozKitMailSlaTerRHPapproachtoSuperSineKernel}:
\beq
a_0 = -\pa{\nu^2\!\pa{q} +  \nu^{2}\!\pa{-q}} \; .
\enq
The first $\la_0$ dependent term is an oscillating correction
\beq
a_{1} =  \f{i}{2} \sul{\eps = \pm }{}  d^{\pa{0}}\!\pa{\la_0} \f{ u\pa{\la_0}-u\pa{\eps q} }{\pa{\la_0-\eps q}^2} \e{tr}\paa{V^{\pa{\eps;0}}\!\pa{\eps q} \pac{\sg_3,\sg}}  \; .
\enq
$a_2^{\e{osc}}$ contains the oscillating term coming from the boundaries $\pm q$:
\beq
a_{2}^{\e{osc}}  =  -  \f{u\pa{q}-u\pa{-q}}{2i\pa{2q}^2}  \e{tr}\paa{ \pac{V^{\pa{+;0}}\!\pa{q},V^{\pa{-;0}}\!\pa{-q}  } \sg_3}  \; .
\enq
Finally, $a_2^{\e{no}}$ corresponds to the first non-oscillating corrections issued from the endpoints $\pm q$:
\beq
a_2^{\e{no}}=\f{i}{4} \sul{\eps = \pm }{}  u^{\prime\prime}\!\pa{\eps q} \e{tr}\paa{V^{\pa{\eps;1}}\!\pa{\eps q}- \pac{V^{\pa{\eps;0}}\!\pa{\eps q}}^2 } \sg_3
+  u^{\prime}\!\pa{\eps q} \e{tr} \paa{  \pac{V^{\pa{\eps;1}}}^{\prime}\!\pa{\eps q}
- 2V^{\pa{\eps;0}}\!\pa{\eps q} \pac{V^{\pa{\eps;0}}}^{\prime}\!\pa{\eps q} } \sg_3
\label{formule intermediare pour correction a2}
\enq
Here, we precise that $u^{\prime}= \Dp{\la} u$, $u^{\prime \prime}= \Dp{\la}^2 u$ and $\pac{V^{\pa{\eps,a}}}^{\prime}= \Dp{\la} V^{\pa{\eps,a}} $. 

It now remains to insert the explicit expressions for $V^{\pa{\pm,k}}$ as well as $d^{\pa{n}}$ so as to obtain the expressions for the coefficients
$a_k$, $k=1,2$.

We get that, independently of the time-like or space-like regime,
\beq
\f{a_{2}^{\e{osc}}}{x^2} =  i\pac{u\pa{q}-u\pa{-q}} \cdot  \f{\nu\pa{q}\nu\pa{-q} }{u^{\prime}\!\pa{q} u^{\prime}\!\pa{-q} \pa{2q x}^2 }
\pa{ \f{\mc{S}_-}{\mc{S}_+} - \f{\mc{S}_+}{\mc{S}_-} }
\label{formule pour a2 Osc}\;.
\enq
%
%
%
%
%
%
%
%
%
%
%
%
%
As for $a_1$, we have
\beq
\f{a_1}{x^{\f{3}{2}}}=
 \f{1}{2\sqrt{\pi} h\pa{\la_0} x^{\f{3}{2}}}
 \paa{ i \nu\pa{-q}\cdot \f{  \pac{u\pa{-q}-u\pa{\la_0} } }{u^{\prime}\!\pa{-q} \pa{\la_0+q}^2} \f{\mc{S}_0}{\mc{S}_-}
 - i \nu\pa{q} \cdot \f{ \pac{u\pa{q}-u\pa{\la_0}}  }{u^{\prime}\pa{q} \pa{\la_0-q}^2} \f{\mc{S}_0}{\mc{S}_+}     }
\label{formule pour a1 time like}
\enq
in the time-like regime, and
\beq
\f{ a_1 }{ x^{\f{3}{2}} }=
 \f{ 1 }{ 2\sqrt{\pi} h\pa{\la_0} x^{\f{3}{2}} }
\paa{ i \nu\pa{-q} \cdot \f{ \pac{u\pa{\la_0}-u\pa{-q}} }{u^{\prime}\!\pa{-q} \pa{\la_0+q}^2} \f{\mc{S}_-}{\mc{S}_0}
-i \nu\pa{q} \cdot \f{ \pac{u\pa{\la_0}-u\pa{q}} }{u^{\prime}\!\pa{q} \pa{\la_0-q}^2} \f{\mc{S}_+}{\mc{S}_0}   }
\label{formule pour a1 space like}
\enq
in the space-like one.
We remind that $\mc{S}_{\pm}$ and $\mc{S}_0$ have been defined in \eqref{definition fonction S plus et moins}
and \eqref{definition fonction S zero et kappa}

%
%
%
%
%
%
%
%

\subsubsection*{$\bullet$ The constant term}

The $x$-derivative cannot fix the constant in $x$ part of the leading asymptotics.
We use the $\la_0$-derivative identity so as to fix the $\la_0$ dependent part of this constant.
Then, in the space-like regime one obtains the $\la_0$-independent part of the constant term by sending $\la_0 \tend \infty$ (the asymptotic
expansion is uniform in $\la_0$ lying uniformly away to the right from $q$).
In such a limit, the determinant can be related, up to $\e{O}\pa{x^{-\infty}}$ corrections, to the generalized sine kernel determiant studied in
\cite{KozKitMailSlaTerRHPapproachtoSuperSineKernel}. In this way, we are able to fix the constant in this regime.
In the time-like regime, in order to fully fix the constant, one has also to compute the $q$-derivative of the determinant asymptotically.

We already know from the above analysis 
that $\log \ddet{}{I+V}=x a_{-1} + \log x a_0 + C\pac{\nu,u, g} +\e{o}\pa{1} $. Using \eqref{formule derviee lambda 0 log det}, we get
\beq
\Dp{\la_0} C\pac{\nu, u , g} = \Oint{ \ga^{\pa{0}} }{} \f{\dd z}{4\pi} \pa{\Dp{\la_0}u}\pa{z}
\sul{\eps=\pm}{}  \f{  \tr\pac{ V^{\pa{\eps;0}}\pa{\eps q} \sg_3 } }{ \pa{z-\eps q }^2 }  = 
%
%
%
%
%
%
- \Dp{\la_0} \paa{ \nu^2\pa{q} \log \abs{u^{\prime}\pa{q}} + \nu^2\pa{-q} \log u^{\prime}\pa{-q}} \; .
\enq
 The absolute value has been chosen so as to treat the space-like and time-like
regimes simultaneously.

In the space-like regime, the asymptotics are uniform in $\la_0$, as long as $\la_0$ remains uniformly away from $q$.
Hence, one can set $\la_0=\infty$ in the asymptotics so as to fix the constant term.
When $\la_0=+\infty$, the function $u$ has no saddle-point, a straightforward computation shows that 
$V\pa{\la,\mu}=V_{GSK}\pa{\la,\mu} + \e{O}\pa{x^{-\infty}}$, with
\beq
V_{GSK}\pa{\la,\mu}= - \paa{1-\ex{2i\pi \nu\pa{\la}}}^{\f{1}{2}} \paa{1-\ex{2i\pi \nu\pa{\mu}}}^{\f{1}{2}}
\cdot \f{ \wt{e}^{-1}\!\pa{\la} \wt{e}\pa{\mu}  -  \wt{e}^{-1}\!\pa{\mu} \wt{e}\pa{\la}  }{ 2i \pi \pa{\la-\mu} }   \qquad \e{with} \quad
\wt{e}\pa{\la}=e\pa{\la} \pa{\ex{-2i\pi \nu\pa{\la}}-1}^{\f{1}{2}} .
\nonumber
\enq
Moreover, the big O symbol is uniform on $\intff{-q}{q}$. This means that
\beq
\ddet{\intff{-q}{q}}{I+V} = \ddet{ \intff{-q}{q} }{ I+V_{GSK} } \cdot \pa{1+\e{O}\pa{x^{-\infty}}} \; .
\enq
 This last identity stems from the fact that the resolvent of a generalized sine kernel is
polynomially bounded in $x$, and this uniformly on $\intff{-q}{q}$, \textit{cf} \cite{KozKitMailSlaTerRHPapproachtoSuperSineKernel}.  Using the $x\tend +\infty $ asymptotic behavior of $\ddet{ \intff{-q}{q} }{ I+V_{GSK} } $ obtained in \cite{KozKitMailSlaTerRHPapproachtoSuperSineKernel}, we get that
\bem
C\pac{\nu, u, g}=-\nu^2\pa{q}\log\pac{2 q \pa{u^{\prime}\!\pa{q} +i0^+} } -\nu^2\pa{-q}\log\pac{2 q u^{\prime}\pa{-q}} 
+ \log G\pa{1,\nu\pa{q}}G\pa{1,\nu\pa{-q}} + C_1\pac{\nu} \\
+\Int{-q}{q} \dd \la g^{\prime}\pa{\la} \nu\pa{\la}\;
-\Int{-q}{q} \dd \la \nu\pa{\la} \log^{\prime}\pa{ \ex{-2i\pi \nu\pa{\la}} - 1 } \; .
\label{formule pour le terme constant de log det}
\end{multline}
The functional $C_1$ has been defined in \eqref{definition fonctionnelle C1} and we agree upon the shorthand notation $G\pa{1,z}=G\pa{1+z}G\pa{1-z}$
for the product of two Barnes functions. Note that the $i0^+$ regularization only matters in the time-like regime where
$u^{\prime}\pa{q}<0$. Of course, for the moment we have only proven the value of the constant term in the space-like regime.
To see that the constant term is indeed given by $C\pac{\nu, u, g}$ \eqref{formule pour le terme constant de log det}
in the time-like regime as well, we apply the so-called $q$ derivative method \cite{KozKitMailSlaTerRHPapproachtoSuperSineKernel}. Namely, starting from the identity
\beq
\Dp{q}\log \ddet{}{I+V} = R\pa{q,q} + R\pa{-q,-q} \; ,
\enq
one replaces the resolvent $R$ by its leading in $x$ part corresponding to sending $\Pi=I_2$ in the reconstruction formula
for $\ket{F^{R}\pa{\la}}$ in terms of $\chi$. The leading resolvent around $\pm q$
is then expressed in terms of CHF with the use of identities \eqref{Appendix Special Functions Phi s'ecrit comme Psi}.
Then, following word for word the steps described in \cite{KozKitMailSlaTerRHPapproachtoSuperSineKernel}
one obtains that, in the time-like regime, $\Dp{q}C\pac{\nu,u,g}$ is indeed given by the
partial $q$-derivative of \eqref{formule pour le terme constant de log det}.
This fixes the $\la_0$ and $q$ dependent part of the constant term in this regime.
As the remaining $\la_0$ and $q$ independent part has to be the same in both regimes, the constant term is fully fixed.

The form of the asymptotic expansion given in theorem \ref{theorem DA determinant} follows once upon
applying the identity
\beq
\ex{- \Int{-q}{q} \log^{\prime}\pa{\ex{-2i\pi \nu\pa{\la}}-1} \nu\pa{\la} \dd \la }  G\pa{1,\nu\pa{q}}G\pa{1,\nu\pa{-q}} =
\ex{i\f{\pi}{2} \pa{\nu^2\!\pa{q}-\nu^2\!\pa{-q}}} \pa{2\pi}^{\nu\pa{-q} - \nu\pa{q}} G^2\pa{1+\nu\pa{q}} G^2\pa{1-\nu\pa{-q}} \;.
\enq
This identity is a direct consequence of \eqref{Appendix special functions Int rep Barnes}. \qed


\section{Natte Series for the determinant}
\label{section Series de Natte}

In this section, we derive a new series representation, that we call the Natte series, for $\ddet{}{I+V}$.
Just as a Fredholm series is well adapted for computing the determinant of the operator $I+V$ perturbatively
when the kernel $V$ is small, the Natte series is built in such a way that it is immediately fit for an asymptotic analysis of the determinant.
The form and existence of the series is closely related to the fact that the asymptotic behavior of this determinant
can be obtained by an application of the Deift-Zhou steepest-descent method.

Let $I+\De$ be the jump matrix for $\Pi$. Then, according to sections \ref{section transfos RHP} and \ref{section parametrices}, $\De$
has an asymptotic expansion that is valid uniformly on the contour $\Sg_{\Pi}$:
\beq
\De\pa{z} \simeq  \sul{n \geq 0}{} \f{\De^{\pa{n}}\!\pa{z;x}}{x^{n +1}}  \qquad \e{where} \quad
\left\{ \ba{cccc}  \De^{\pa{n}}\!\pa{z;x}  &=& \sqrt{x} \f{d^{\pa{n}}\!\pa{z}  }{ \pa{z-\la_0}^{2n+1} }  \sg & \qquad \e{for} \;\; z \in \Dp{} \mc{D}_{\la_0,\de}
\vspace{2mm} \\
				\De^{\pa{n}}\!\pa{z;x}  &=& \f{ V^{ \pa{\pm; n}  }\!\pa{z}  }{ \pa{n+1}! \pa{z \mp q }^{n+1} } 	& \qquad \e{for} \;\;  z \in \Dp{} \mc{D}_{\pm q ,\de}  \ea  \right.
\label{formule DA matrice Delta}
\enq
and everywhere else $\De^{\pa{n}}\pa{z;x}=0$. In other words $\De\pa{z;x}$ is a $\e{O}\pa{x^{-\infty}}$ everywhere else on the contour.
Moreover, one can convince oneself that  this  $\e{O}\pa{x^{-\infty}}$ holds in the $L^{1}\cap L^{\infty}\!\pa{\msc{C}}$ sense,
 for any curve $\msc{C}$ that is lying sufficiently close to
$\msc{C}_{E}$. Finally, we remind that $\sg=\sg^+$ in the space-like regime and  $\sg=\sg^-$ in the time-like regime.

\subsection{The leading Natte series}

We start the derivation of the Natte series by providing a convenient integral representation for $\log\ddet{}{I+V}$.

\begin{lemme}

Let $V$ be the kernel defined in \eqref{definition noyau GSK FF} and $\Pi\pa{\la} \equiv \Pi\pa{\la;x}$ be the unique solution to the associated 
RHP. Then, the logarithm of the Fredholm determinant admits the below representation
\beq
\log \ddet{}{I+V}\pac{\nu,u,g}=  \log \ddet{}{I+V}^{\pa{0}}\pac{\nu,u,g} \; + \;  \log \ddet{}{I+V}^{\pa{\e{sub}}}\pac{\nu,u,g}
\enq
where
\beq
\log \ddet{}{I+V}^{\pa{0}} \pac{\nu,u,g} = ix \Int{-q}{q} u^{\prime}\!\pa{\la} \nu\pa{\la} \dd \la
\; - \; \pa{\nu^2\!\pa{q}+\nu^2\!\pa{-q}} \log x + C\pac{\nu,u,g}
\enq
and $C\pac{\nu,u,g}$ has been defined in \eqref{formule pour le terme constant de log det}. Also
\beq
\log \ddet{}{I+V}^{\pa{\e{sub}}}\pac{\nu,u,g} = \Int{+\infty}{x} \dd x^{\prime} \Oint{ \ga }{} \f{\dd \la}{4\pi} G\pa{\la}
\paa{ \tr\pac{ \Dp{\la} \Pi\pa{\la;x^{\prime}} \sg\pa{\la} \Pi^{-1}\!\pa{\la;x^{\prime}} }
  +  \f{1}{x^{\prime}} \Int{\Dp{} \mc{D} }{} \f{\dd z}{2i\pi} \f{\tr\pac{\De^{\pa{1}}\pa{z;x^{\prime}}\sg_3  }}{\pa{\la-z}^2}  } \; .
\label{formule explicite exacte pour logdet sub}
\enq
The contour $\ga$ is as defined in Fig. \ref{contour ga pour l'integration serie Natte} and
$\Dp{}\mc{D}=-\Dp{}\mc{D}_{q,\de} \cup -\Dp{}\mc{D}_{-q,\de} \cup -\Dp{}\mc{D}_{\la_0,\de}$. The function $G$ has been defined in \eqref{definition
fonction G}, $\De^{\pa{1}}$ in \eqref{formule DA matrice Delta}
and we remind that $\sg\pa{\la}= \sg_3  {\bf 1}_{\ga^{\pa{0}}}\!\pa{\la} +  \sg^+  {\bf 1}_{\ga^{\pa{R}}\cup\ga^{\pa{L}} }\!\pa{\la}$.

The convergence of this integral representation is part of the conclusion of the lemma.

\end{lemme}

\Proof

The formula for the $x$-derivative of the determinant \eqref{formule derviee x log det} is the starting point of the proof.
By re-ordering the terms we get, exactly as in the proof of theorem \ref{theorem DA determinant},
\beq
\Dp{x}\log \ddet{}{I+V}\pac{\nu,u,g}= \Dp{x} \log \ddet{}{I+V}^{\pa{0}}\!\pac{\nu,u,g} + \mc{R} \;,
\enq
in which
\beq
\mc{R} = -i  \f{\Dp{}}{\Dp{} \eta }  \pac{\;   \Int{\Ga\pa{\msc{C}_{E}}}{} \hspace{-3mm} \f{\dd \la}{4\pi} \ex{i\eta u\pa{\la}}   \paa{
\e{tr}\pac{\Dp{\la} \Pi\pa{\la;x} \sg_3  \Pi^{-1}\!\pa{\la;x} }
 + \Int{\Dp{} \mc{D} }{} \f{\dd z}{2i\pi}   \f{ \e{tr}\pac{\De^{\pa{1}}\pa{ z;x} \sg_3  } } { x \pa{\la-z}^2}  }   }_{\eta=0^+}
  \hspace{-4mm}.
\label{equation Reste x exact}
\enq
Here the matrix $\De^{\pa{1}}$ appears in \eqref{equation Reste x exact} as its contribution has already been taken into account in 
$\Dp{x}\log\ddet{}{I+V}^{\pa{0}}$. It had thus to be substracted.

Now, performing exactly the same steps as in the proof of theorem \ref{theorem DA determinant}, we recast the integral in such a way 
that the $\eta$-derivative can be moved inside of the integration symbol.
Note that the operation of squeezing the contour $\Ga\pa{\msc{C}_E}$ in \eqref{equation Reste x exact} to $\ga$ does not affect the term coming from $\De^{\pa{1}}$ as it is holomorphic outside
of $\Dp{}\mc{D}$. Once that the $\eta$-derivative has been computed, the result follows by an $x$-integration. This integration is licit as,
due to the presence of $\De^{\pa{1}}$, the $\eta$-differentiated integrand behaves as $\e{O}\pa{ \tf{\log x}{x^2} }$ , for $x \tend +\infty$,
in what concerns the non-oscillating contributions and as
$\ex{i x v}x^{-w}$ for the oscillating terms. Here, $\nu$ and $w$ are constants such that  $v\in \R$ and $\Re\pa{w}>0$.
The oscillating contributions are thus also integrable, at least in the Riemann-sense.
\qed

We are now in position to derive the logarithmic Natte series representation for $\log\ddet{}{I+V}$.

\begin{theorem}
There exists a sequence of functionals $\mc{F}_N\pac{\nu,u,\ex{g}}\pa{x}$, such that
\beq
\log \ddet{}{I+V}\pac{\nu,u,g}=  \log \ddet{}{I+V}^{\pa{0}}\pac{\nu,u,g} + \sul{N\geq 1}{} \mc{F}_N\pac{\nu,u,\ex{g}}\pa{x} \; .
\label{ecriture serie Natte pour le log}
\enq
There exists a positive  N-independent constant  $m\pa{x}$ such that  $\abs{ \mc{F}_N\pac{\nu,u,\ex{g}}\pa{x}  } \leq \pac{m\pa{x}}^{N}$. 
$m\pa{x}$ is such that $m\pa{x}=\e{O}\pa{x^{-w}}$ where, for $\de>0$  but small enough
\beq
w=\f{3}{4} \min\pa{ \tf{1}{2} , 1 - \wt{w}-2 \max_{\eps=\pm}  \abs{\Re\pac{\nu\pa{\eps q}}}} \quad \e{and} \quad 
\wt{w}= 2 \max_{\eps=\pm} \paa{  \sup_{\Dp{}\mc{D}_{\eps q \, , \de }} \abs{ \Re\pac{\nu-\nu\pa{\eps q}} }  }
\label{definition constante w dans estimation de FN}
\enq
The functionals $\mc{F}_N\pac{\nu,u,\ex{g}}\pa{x}$ admit the integral representation
\beq
\mc{F}_N\pac{\nu,u,\ex{g}}\pa{x} = \sul{r = 1 }{N} \; \sul{ \substack{ \Sg \eps_k=0 \\ \eps_k \in \paa{\pm 1, 0}} }{}  \sul{\tau =\da,\ua}{}
\Int{+\infty}{x} \hspace{-1mm} \dd x^{\prime}  \Oint{ \ga^{\tau}  }{}  \f{ \dd \la}{4\pi} \hspace{-3mm}
\Int{ \paa{\Sg_{\Pi}^{\tau}}^{\pa{r,N}} }{} \hspace{-3mm} \f{ \dd^N z}{\pa{2i\pi}^N} \;
\cdot H_{N,r}\bigg( \la, \{ z_j \}_{j=1}^{N}, x^{\prime}; \{\eps_j\}_{j=1}^N  \bigg) \pac{\nu,u} \cdot  \pl{p=1}{N} \ex{  \eps_p g\pa{z_p} } 
\label{ecriture FN grosse somme qui pue}
\enq
in terms of the auxiliary functionals
\beq
H_{N,r}\bigg( \la, \{ z_j \}_{j=1}^{N}, x; \{\eps_j\}_{j=1}^N  \bigg) \pac{\nu,u}  =
\f{ - G \pa{\la}  D_{N,r} \bigg( \la, \{ z_j \}_{j=1}^{N}, x; \{\eps_j\}_{j=1}^N  \bigg) \pac{\nu,u}  }
{ \pa{\la-z_1}^2 \pa{\la-z_{r+1} } \pl{p=1}{r-1} \pa{z_{p}-z_{p+1}}  \pl{p=r+1}{N} \pa{z_{p}-z_{p+1} }  } \; .
\label{definition des fonctionnelles H_N}
\enq
The first summation in \eqref{ecriture FN grosse somme qui pue} runs through all possible choices of the variables $\eps_k \in \paa{\pm 1 , 0}$ subject 
to the constraint $\sum_{}^{} \eps_k = 0$. Then, one sums over integrals running over the upper/lower part $\ga^{\ua/\da}$  of the contour $\ga$  and 
also over the associated inslotted contour $\paa{\Sg_N^{\ua/\da}}^{\pa{r,N}}$.

For $N \geq 2$, the functionals $D_N$ are defined as the functionals of $\nu$ and $u$ that appear in the expansion of
\beq
\tr\pac{\De\pa{z_r}\dots \De\pa{z_1} \sg\pa{\la} \nabla\pa{z_{r+1}} \dots \nabla\pa{z_N}  } =
\sul{ \substack{ \Sg \eps_k=0 \\ \eps_k \in \paa{\pm 1, 0}} }{}
D_{N,r} \bigg( \la, \{ z_j \}_{j=1}^{N}, x; \{\eps_j\}_{j=1}^N  \bigg) \pac{\nu,u} \; \ex{ \sul{p=1}{N} \eps_p g\pa{z_p} } \; ,
\label{definition des fonctionnelles DN}
\enq
into different powers of $\ex{g\pa{z_{\ell}}}$.  For $N=1$ one has
\beq
 D_1\pa{\la, z_1, x} = \tr\pac{ \pa{\De\pa{z_1}- x^{-1}\De^{\pa{1}}\!\pa{z_1} }\sg_3}  \bs{1}_{\ga^{\pa{0}}}\!\pa{\la}  \; .
\enq
Above, $\nabla$ is the adjugate matrix to $\De$: $\nabla = \e{Comat}\pac{\De}^{t}$,  so that
$I_2+\nabla$ corresponds to the jump matrix for $\Pi_+^{-1}$.
Finally, $\ga^{\ua/\da}$ denotes that part of the curve $\ga$ which lies above/below of $\Sg_{\Pi}$.
Let $P_1$, $P_2$ stand for the two intersection points between $\ga$ and $\Sg_{\Pi}$, \textit{cf} Fig.~\ref{contour ga pour l'integration serie Natte}. 
Then, $\Sg_{\Pi}^{\ua/\da}$ is the contour equal eveywhere to $\Sg_{\Pi}$ except in a small vicinity of the points $P_{k}$, where it avoids
these points by below/above. Then $\paa{ \Sg_{\Pi} ^{\ua / \da} } ^{\pa{r,N}}$ is realized as the Carthesian product of two inslotted 
contours of length $r$ and $N-r$: 
$\paa{ \Sg_{\Pi}^{\ua/\da} }^{\pa{r,N}} = \paa{\Sg_{\Pi}^{\ua/\da}}^{ \pa{r} } \times \paa{\Sg_{\Pi}^{\ua/\da}}^{ \pa{N-r} }$.

\end{theorem}

Starting from the definition \ref{Definition Matrices Delta Nabla Sur contour encastre} of the matrices $\De$ and $\nabla$
on inslotted contours $\Sg_{\Pi}^{\pa{N}}$, one defines the matrices $\De$ and $\nabla$ on $\Sg^{\ua/\da}_{\Pi}\pac{z_k}$
as the analytic continuations of $\De$ from $\Sg_{\Pi}\pac{z_k}$.

\Proof

The functional $\mc{F}_N\pac{\nu,u,g}$ will be constructed by merging the integral representation \eqref{formule explicite exacte pour logdet sub}
with the Neumann series for $\Pi$ \eqref{ecriture serie Neumann Pi}
and $\Pi^{-1}$ \eqref{ecriture serie Neumann Pi-1}. 
These series converge uniformly in $\la$ (and every finite order $\la$-derivative) on every open set $O$ such that $\e{d}\!\pa{O,\Sg_{\Pi}}>0$.
 However, in \eqref{formule explicite exacte pour logdet sub}, one integrates 
$\tr\pac{\Dp{\la} \Pi\pa{\la}  \sg\pa{\la} \Pi^{-1}\!\pa{\la} }$
with a weigth along $\ga$, where the contour $\ga$ is depicted in Fig.~\ref{contour ga pour l'integration serie Natte}.
The latter contour intersects $\Sg_{\Pi}$.  Hence, it contains points that are not uniformly away from $\Sg_{\Pi}$.
However, we have already argued that $\tr\pac{\Dp{\la} \Pi_{\pm}\pa{\la}  \sg\pa{\la} \Pi_{\pm}^{-1}\!\pa{\la} }$
can be analytically continued to a small neighborhood of $\Sg_{\Pi}$ located to the right/left of $\Sg_{\Pi}$.
Such an analytic continuation can be also performed on the level of the Neumann series for $\Pi^{\pm 1}$.

In order to have a Neumann series representation for $\Pi$ or $\Pi^{-1}$ that is uniformly convergent in 
$\la \in \ga^{\ua/\da}\setminus\paa{P_1,P_2}$, we use the local analyticity of the jump matrices for $\Pi$
so as to deform the original jump contour $\Sg_{\Pi}$ appearing in the RHP for $\Pi$ and $\Pi^{-1}$
\eqref{ecriture serie Neumann Pi}-\eqref{ecriture serie Neumann Pi-1} into the contour $\Sg_{\Pi}^{\ua/\da}$: 

\beq
\Pi\pa{\la}= I_2 + \sul{N\geq 1}{} \Int{\Sg^{\ua/\da}_{\Pi}}{} \hspace{-1mm}\f{\dd z}{2i\pi\pa{\la-z}}
\paa{\mc{C}_{\Sg_{\Pi}^{\ua/\da}}^{\De}}^{N-1}\hspace{-5mm}\pac{I_2}\pa{z} \De\pa{z}
\quad \e{and} \quad
\Pi^{-1}\pa{\la}= I_2 + \sul{N\geq 1}{} \Int{\Sg^{\ua/\da}_{\Pi}}{} \hspace{-1mm} \f{\dd y}{2i\pi\pa{\la-y}}
 \nabla\pa{y}  \paa{ ^{\bs{t}}\mc{C}_{\Sg_{\Pi}^{\ua/\da}}^{\nabla}}^{N-1}\hspace{-5mm}\pac{I_2}\pa{y}  \;.
\nonumber
\enq
According to these formulae $\Pi$, $\Pi^{-1}$ are holomorphic on some small vicinity of $\ga^{\ua/\da}$. However, we do stress
that these analytic continuation from above and below  $P_k$ differ at $P_k$.  Hence, for $\la \in \ga^{\ua/\da}\setminus\paa{P_1,P_2}$, we get 
\beq
\tr\pac{\Dp{\la} \Pi\pa{\la}  \sg\pa{\la} \Pi^{-1}\!\pa{\la} } =  \sul{N \geq 1}{}  f_N\pa{\la,x} 
\label{ecriture decomposition trace Pi prime sg Pi}
\enq
with
\bem
f_N\pa{\la,x}= \sul{r=1}{N-1} \Int{\Sg^{\ua/\da}_{\Pi}}{} \hspace{-1mm}\f{ - \dd z \dd y }{\pa{2i\pi}^2 \pa{\la-z}^2  \pa{\la-y} }
\cdot \tr\pac{ \paa{\mc{C}_{\Sg_{\Pi}^{\ua/\da}}^{\De}}^{r-1}\hspace{-4mm}\pac{I_2}\pa{z} \De\pa{z,x} \sg\pa{\la}
\nabla\pa{y,x} \paa{ ^{\bs{t}}\mc{C}_{\Sg_{\Pi}^{\ua/\da}}^{\nabla}}^{N-r-1}\hspace{-4mm}\pac{I_2}\pa{y}   }  \\
 +  \Int{\Sg^{\ua/\da}_{\Pi}}{} \hspace{-1mm}\f{ - \dd z }{ 2i\pi \pa{\la-z}^2   }
\cdot \tr\pac{ \paa{\mc{C}_{\Sg_{\Pi}^{\ua/\da}}^{\De}}^{N-1}\hspace{-4mm}\pac{I_2}\pa{z} \De\pa{z,x} \sg\pa{\la} } \;. 
\label{definition fN}
\end{multline}
Above, we have insisted on the dependence on $x$ of the matrices $\De$ and $\nabla$. 
The representation \eqref{ecriture decomposition trace Pi prime sg Pi} allows us to define the functional $\mc{F}_N\pac{\nu,u,\ex{g}}$:
\beq
\mc{F}_1\pac{\nu,u, \ex{g}}=   \Int{+\infty}{ x} \dd x^{\prime} \Oint{ \ga }{}  \paa{ f_1\pa{\la,x^{\prime}}+  \f{1}{x^{\prime}} \Int{\Dp{} \mc{D} }{} \f{\dd z}{2i\pi} \f{\tr\pac{\De^{\pa{1}}\pa{z;x^{\prime}}\sg_3  }}{\pa{\la-z}^2}  }  G\pa{\la} \f{\dd \la}{4\pi}
\label{ecriture rep int F1}
\enq
and, for $N\geq 2$,
\beq
\mc{F}_N\pac{\nu,u,\ex{g}}=   \Int{+\infty}{ x} \dd x^{\prime} \Oint{ \ga }{}   f_N\pa{\la,x^{\prime}} G\pa{\la,x^{\prime}} \f{\dd \la}{4\pi} \;.
\label{ecriture rep int FN}
\enq
We remind that $G$ is given by \eqref{definition fonction G} and above, we have explicitly insisted on its $x$-dependence. 

In the following we justify that \eqref{ecriture rep int F1}-\eqref{ecriture rep int FN} are well defined and that 
one can exchange the integrals over $\ga$ and $\intfo{x}{+\infty}$ in \eqref{formule explicite exacte pour logdet sub} 
with the summation \eqref{ecriture decomposition trace Pi prime sg Pi}. 
Then, we provide explicit bounds for $\mc{F}_N\pac{\nu,u,\ex{g}}$ and finally outline the steps leading to the derivation of the representation 
\eqref{ecriture FN grosse somme qui pue} for $\mc{F}_N$. 

\subsubsection*{Exchange of symbols}

Building on the identities:
\beq
\abs{\e{tr}\pa{AB}} \leq  \e{max}_{j,k} \pa{\abs{B_{jk}}} \sul{j,k}{} \abs{ A_{jk} } \quad \e{and} \quad 
\sul{j,k}{} \norm{ \pa{AB}_{jk}  }_{L^{1}\pa{\msc{C}}} \leq 
\sul{j,k,\ell}{} \norm{A_{j\ell}}_{L^{2}\pa{\msc{C}}} \norm{B_{\ell k}}_{L^{2} \pa{\msc{C}} } \leq 
4 \norm{A}_{L^{2}\pa{\msc{C}}} \norm{B}_{L^{2}\pa{\msc{C}}} \;, 
\nonumber
\enq
 and after some algebra one obtains that 
\bem
\norm{f_N}_{L^{\infty}\pa{\ga}} \leq 4 \max_{ \substack{ \tau \in \paa{\ua,\da} \\ k=2,3 } } \paa{ \e{d}^{-k}\!\pa{\ga^{\tau},\Sg_{\Pi}^{\tau} }}
\sul{\tau = \ua, \da}{} \sul{r=1}{N-1} \left\{ 
\norm{  \paa{ \mc{C}_{\Sg_{\Pi}^{\tau}}^{\De}}^{r-1}\hspace{-4mm}\pac{I_2} \f{\De}{2\pi}  }_{L^{2}\pa{\Sg_{\Pi}^{\tau}}} 
 \norm{ \f{\nabla}{2\pi} \paa{\, ^{\bs{t}} \mc{C}_{\Sg_{\Pi}^{\tau} }^{\nabla}}^{N-1-r} \hspace{-4mm}\pac{I_2}   }_{L^{2}\pa{\Sg_{\Pi}^{\tau}}}
  \right. \\ 
\left.  + \; \; \; \norm{  \paa{ \mc{C}_{\Sg_{\Pi}^{\tau}}^{\De}}^{N-1}\hspace{-4mm}\pac{I_2}   }_{L^{2}\pa{\Sg_{\Pi}^{\tau}}} 
\norm{ \f{\De}{2\pi} }_{L^{2}\pa{\Sg_{\Pi}^{\tau}}}   \right\} \;.
\nonumber
\end{multline}
In the intermediate calculation we have used $\norm{\sg\pa{\la}}_{L^{\infty}\pa{\ga}}=1$.  
By using the estimates \eqref{ecriture continuite operateur Cauchy matriciel}-\eqref{ecriture continuite operateur Cauchy matriciel transpose}, 
one gets that for $\tau = \ua$ or $\da$
\beq
\norm{ \paa{\mc{C}_{\Sg^{\tau}_{\Pi}}^{\De}}^{r}\hspace{-2mm}\pac{I_2}   }_{L^{2}\pa{\Sg_{\Pi}^{\tau}}} \!\!\leq
\paa{2 c\pa{\Sg_{\Pi}^{\tau}} \norm{\De}_{L^{\infty}\pa{\Sg^{\tau}_{\Pi}}} }^{r-1}  
\norm{\mc{C}^{I_2}_{\Sg^{\tau}_{\Pi}}\pac{\De}}_{L^{2}\pa{\Sg_{\Pi}^{\tau}}}     \;\; \leq 
\paa{2 c\pa{\Sg_{\Pi}^{\tau}} }^{r} \norm{\De}_{L^{\infty}\pa{\Sg^{\tau}_{\Pi}}} ^{r-1}  \norm{\De}_{L^{2}\pa{\Sg^{\tau}_{\Pi}}}  
\nonumber
\enq
and 
\beq
\norm{ \paa{ ^{\bs{t}} \mc{C}_{\Sg^{\tau}_{\Pi}}^{\nabla}}^{r}\hspace{-2mm}\pac{I_2}   }_{L^{2}\pa{\Sg_{\Pi}^{\tau}}} \!\!\leq
\paa{2 c\pa{\Sg_{\Pi}^{\tau}} }^{r} \norm{\De}_{L^{\infty}\pa{\Sg^{\tau}_{\Pi}}} ^{r-1}  \norm{\De}_{L^{2}\pa{\Sg^{\tau}_{\Pi}}} 
 \; . \nonumber
\enq
Also, one has 
\beq
\norm{  \paa{ \mc{C}_{\Sg_{\Pi}^{\tau}}^{\De}}^{r}\hspace{-2mm} \pac{I_2} \De  }_{L^{2}\pa{\Sg_{\Pi}^{\tau}}}  \leq 2 
\norm{\De}_{L^{\infty}\pa{\Sg_{\Pi}^{\tau}}} 
\norm{  \paa{ \mc{C}_{\Sg_{\Pi}^{\tau}}^{\De}}^{r}\hspace{-2mm}\pac{I_2} }_{L^{2}\pa{\Sg_{\Pi}^{\tau}}} \; .
\enq
The estimates and asymptotic expansions of $\De$ ensure that there exists an $x$-independent constant $C_2$ such that 
\beq
\underset{\tau \in \paa{\ua,\da}}{\max} \paa{ \norm{\De}_{L^{\infty}\pa{\Sg_{\Pi}^{\tau} }} 
+ \norm{\De}_{L^{2}\pa{\Sg_{\Pi}^{\tau}  }}   } \leq  \f{C_2}{c\pa{\Sg_{\Pi}}} x^{-w}
\qquad \e{where} \qquad
c\pa{\Sg_{\Pi}} = \underset{\tau \in \paa{\ua,\da}}{\max}c\pa{\Sg_{\Pi}^{\tau} } \;.
\enq
Hence,
\beq
\norm{f_N}_{L^{\infty}\pa{\ga}} \leq  \f{2 N}{\pi c\!\pa{\Sg_{\Pi}}  } \max_{ \substack{ \tau \in \paa{\ua,\da} \\ k=2,3 } } \paa{ \e{d}^{-k}\!\pa{\ga^{\tau},\Sg_{\Pi}^{\tau} }}
\paf{ 2 C_2 }{ x^w }^{N}  \pa{ \pi^{-1} + c\pa{\Sg_{\Pi}} } \quad \e{with} \quad 
w=\f{3}{4}\min\pa{\tf{1}{2}, 1-\wt{w}-2 \e{max}_{\pm}\abs{\Re\nu\pa{\pm q}}} \;.
\label{ecriture majoration fonctionfN}
\enq

It follows that for $x$ large enough and for $N\geq N_0$ (with $N_0 w > 1 $) $\pa{\la,x^{\prime}} \mapsto 
\sum_{p=N_0}^{N} G\pa{\la,x^{\prime}} f_p\pa{\la,x^{\prime}}$ is bounded on $\ga\times \intoo{x}{+\infty}$
by an integrable function. The terms corresponding to $p=1,\dots,N_0-1$ are also integrable as will be shown below. Hence, by the 
dominated convergence theorem one can exchange the summation and the integration symbols leading to \eqref{ecriture serie Natte pour le log}. 
It now remains to provide sharper estimates for each summand. 

\subsubsection*{Sharper estimates for $\mc{F}_N\pac{\nu,u,\ex{g}}$}

It follows from proposition \ref{proposition fine bounds for PiN} applied to the jump contour $\Sg_{\Pi}^{\ua/\da}$, that 
for $\la \in \ga^{\ua/\da}$ one has the representation 
\beq
\Pi_N\pa{\la} = A_N\pa{\la} \; + \; \sul{b=0}{ \pac{\tf{N}{2}} } \sul{ p=0 }{ b }  \sul{m= b - \pac{\tf{N}{2}} }{ \pac{\tf{N}{2}}-b }
\paf{ \mf{e}\pa{q;x} }{ \mf{e}\pa{-q;x} }^{m-\bs{\eta} p}  \paf{ \mf{e}\pa{\la_0;x} }{ \mf{e}\pa{-q;x} }^{\bs{\eta} b}
\times \sul{\eps \in \paa{\pm 1, 0}}{} \pac{\mf{e}\pa{v_{\eps};x}}^{\f{\sg_3}{2}} 
\Pi_{N;\eps}^{\pa{m,b,p}}\pa{\la} \pac{\mf{e}\pa{v_{\eps};x}}^{-\f{\sg_3}{2}} 
\nonumber
\enq
where $\mf{e}\pa{q,x}=\ex{ix u\pa{q}} x^{-2\nu\pa{q}}$,  $\mf{e}\pa{-q,x}=\ex{ix u\pa{-q}} x^{2\nu\pa{-q}}$ and 
$\mf{e}\pa{\la_0,x}=\ex{ix u\pa{\la_0}} x^{-\f{\bs{\eta}}{2}}$. Also, we remind that $\bs{\eta}=1$ in the space-like regmie and $\bs{\eta}=-1$
in the time-like and we agree upon $v_0=\la_0$ and $v_{\pm q} = \pm q$. The matrix $A_N\!\pa{\la}$ contains exponentially small corrections in $x$ and 
the remaining part represents the algebraically small ones. 

This representation ensures that for $\la \in \ga$
\beq
f_N\pa{\la,x} = \f{a_N\pa{\la,x}}{x^N}  \; + \;   \f{1}{x^N}\sul{b=0}{ \pac{\tf{N}{2}} + 1 } \sul{p=0}{b} 
\sul{m=b-\pac{\tf{N}{2}}-1 }{ \pac{\tf{N}{2}}+1-b }
\paf{ \mf{e}\pa{q,x} }{ \mf{e}\pa{-q,x}  }^{m-\bs{\eta} p }  \paf{ \mf{e}\pa{\la_0,x} }{ \mf{e}\pa{-q,x}  }^{\bs{\eta} b }  
c^{\pa{m,b,p}}_{N}\pa{\la,x} \;.
\enq

The functions $c^{\pa{m,b,p}}_{N}\pa{\la,x}$ and $a_N\pa{\la,x}$ can be expressed as traces involving appropriate combinations of the 
matrices $A_N$ and $\Pi_{N;\eps}^{\pa{m,b,p}}$. We have included all the exponentially small corrections stemming from the $A_j$'s, $j=1,\dots,N$
into $a_N\pa{\la,x}$. 

It follows from the properties of $A_j\!\pa{\la}$ and $\Pi_{j;\eps}^{\pa{m,b,p}}\pa{\la} $, $j=1,\dots,N$ that these functions 
are smooth in $\la \in \ga$ and $x$. Moreover, by using the estimates for the $L^{\infty}$ norms of the aforementioned matrices 
$A_N$ and $\Pi_{N;\eps}^{\pa{m,b,p}}$ \eqref{ecriture estimation norme AN et PIN et tout le bla bla}, 
after some algebra one shows that, for $x$-large enough, given any $k \in \mathbb{N}$ there exists an $N$-independent constant $C>0$ such that 
\beq
\abs{a_N\pa{\la,x}} \leq \f{C^N}{x^k} \quad \e{and} \quad 
\abs{c^{\pa{m,b,p}}_{N}\pa{\la,x} } \leq C^N x^{N \wt{w}} \;, \qquad \e{uniformly} \; \e{in} \; \la \in \ga^{\ua/\da} \;.
\enq
These estimates remain unchanged when considering first order partial derivatives in respect to $x$ of these functions. 
Hence for all integers $m,b,p$ of interest the function 
\beq
\pa{\la,y} \mapsto \phi_{m,b,p}\pa{\la,y}=
y^{-N} \paf{ \mf{e}\pa{q,y} }{ \mf{e}\pa{-q,y}  }^{m-\bs{\eta} p }  \paf{ \mf{e}\pa{\la_0,y} }{ \mf{e}\pa{-q,y}  }^{\bs{\eta} b }   
 c^{\pa{m,b,p}}_{N}\pa{\la,y} G\pa{\la;y}
\enq
is Riemann--integrable on $\ga \times \intoo{x}{+\infty}$. Suppose that $m,b$ or $p$ is non-zero. Then, for $N\geq 2$ 
an integration by parts leads to the estimate
\beq
\abs{  \Int{x}{+\infty} \dd y \phi_{m,b,p}\pa{\la,y}   } \leq \f{ \big[ \wt{C} \, \big]^N }{x^{Nw}} \abs{G\pa{\la;x}} \;,
\enq
with $w$ being defined as in \eqref{ecriture majoration fonctionfN}.
When $m=b=p=0$ one simply deals with a non-oscillating integral. In that case, 
\beq
\abs{  \Int{x}{+\infty} \dd y \phi_{m,b,p}\pa{\la,y}   } \leq \f{ \big[ \wt{C} \, \big]^N }{x^{N\pa{1-\wt{w}} -1}} \abs{G\pa{\la;x}} \;.
\enq
There are two cases of interest to consider. If $w=\tf{3}{8}$, then since $\wt{w}=\e{O}\!\pa{\de}$, taking $\de$ sufficiently small 
we get that $N w  \leq N\pa{1-\wt{w}} - 1$. It remains to treat the case when, for all $\de>0$ small enough $w \leq \tf{3}{8}$.
In other words,  $1-\wt{w}-2\max_{\pm} \abs{\Re \nu\!\pa{\pm q}}<\tf{1}{2}$. Therefore
\beq
\f{1}{4}- \f{\wt{w}}{2} \leq \max_{\pm} \abs{\Re \nu\!\pa{\pm q}} \;.
\enq
Thus, taking $\de$  small enough,  so that $\wt{w}\leq \tf{1}{10}$ one gets $\max_{\pm} \abs{\Re \nu\!\pa{\pm q}} \geq 1/5$.  Hence, for $N\geq 2$
\beq
\abs{  \Int{x}{+\infty} \dd y \phi_{m,b,p}\pa{\la,y}   } \leq \f{ \big[ \wt{C} \, \big]^N }{x^{Nw}} \abs{G\pa{\la;x}} \;.
\enq
Thus, once upon the integration over $\la\in\ga^{\ua}\cup\ga^{\da}=\ga$, we get that there exists a constant $m\pa{x}=\e{O}\pa{x^{-w }}$ such that 
$\abs{\mc{F}_N\pac{\nu,u,g}\!\pa{x} } \leq \pac{m\pa{x}}^N$ for $N\geq 2$. The fact that $\abs{\mc{F}_1\pac{\nu,u,g}\pa{x}}\leq m\pa{x}$
from a direct calculation based on the representation \eqref{ecriture rep int F1} 
and the first few terms of the asymptotic expansion of the matrix $\De$.

\subsubsection*{Justification of \eqref{ecriture FN grosse somme qui pue}} 
 
 We conclude this proof by explaining how one can obtain a slightly 
more convenient representation for each individual functional $\mc{F}_N\pac{\nu,u,g}\pa{x}$.
Starting from the Neumann series representations \eqref{serie de Neumann Pi+} and \eqref{serie de Neumann Pi+ inverse}, it follows that
for $\la \in \ga^{\ua/\da}$
\beq
\tr\pac{ \Dp{\la} \Pi\pa{\la} \sg\pa{\la} \Pi^{-1}\pa{\la} } = - \sul{N\geq 1}{} \sul{r = 1 }{N}
\Int{\paa{\Sg_{\Pi}^{\ua/\da}}^{\pa{r,N}} }{} \f{ \dd^N z}{\pa{2i\pi}^N}
\f{ \tr\pac{\De\pa{z_r}\dots \De\pa{z_1} \sg\pa{\la} \nabla\pa{z_{r+1}} \dots \nabla\pa{z_N}  }  }
{ \pa{\la-z_1}^2 \pa{\la-z_{r+1} } \pl{p=1}{r-1} \pa{z_{p}-z_{p+1}}  \pl{p=r+1}{N-1} \pa{z_{p}-z_{p+1} }  } \;.
\label{equation donnat trace derive Pi Pi inverse}
\enq
Above, $\paa{\Sg_{\Pi}^{\ua/\da}}^{\pa{r,N}}$ is the Cartesian product of two inslotted contours
$\paa{\Sg_{\Pi}^{\ua/\da}}^{\pa{r}} \times \paa{\Sg_{\Pi}^{\ua/\da}}^{\pa{N-r}}$\!\!.
To obtain \eqref{equation donnat trace derive Pi Pi inverse} it is enough to multiply out the two series
for $\Dp{\la}\Pi\pa{\la}$,  $\Pi^{-1}\pa{\la}$ and then take the trace. One can convice oneself that the matrices $\De$ and $\nabla$ are such that
\beq
\De\pa{z}= \ex{\f{g\pa{z}\sg_3}{2} } \De_{\mid g=0}\pa{z} \ex{-\f{g\pa{z}\sg_3}{2} }  \quad \qquad \e{and} \qquad  \quad
\nabla\pa{z} =  \ex{\f{g\pa{z}\sg_3}{2} } \nabla_{\mid g=0} \pa{z} \ex{-\f{g\pa{z}\sg_3}{2} } \; .
\enq
This means that there exists functionals $D_{N,r}^{} \Big(  \la, \{z_j\}_{j=1}^N, x; \{\eps_j\}_{j=1}^N \Big)\pac{\nu,u}$ such that
\beq
\tr\pac{\De\pa{z_r}\dots \De\pa{z_1} \sg\pa{\la} \nabla\pa{z_{r+1}} \dots \nabla\pa{z_N}  } =
\sul{ \substack{ \Sg \eps_k=0 \\ \eps_k \in \paa{\pm 1, 0}} }{}
D_{N,r} \Big( \la, \{ z_j \}_{j=1}^{N}, x;\{ \eps_j \}_{j=1}^N  \Big)\pac{\nu, u}  \, \exp\bigg\{ \sul{p=1}{N} \eps_p g\pa{z_p} \bigg\}  \;.
\label{Formule pour DN}
\enq
Above, the sum is taken over all possible choices of $N$ integers $\eps_k \in \paa{\pm 1, 0}$, such that $\sum_{k=1}^{N} \eps_k=0$.
We replace the trace in \eqref{equation donnat trace derive Pi Pi inverse}  by \eqref{Formule pour DN}
and then insert the result into the integral representation for $\log \ddet{}{I+V}^{\e{sub}}$. 
One can exchange the $\ga$ and $x^{\prime}$-integrals with the summation over $N$ in virtue of the previous discussion. 
One can also pull-out the finite sum over $\eps_k \in \paa{\pm 1, 0}$ out of the integrals. 
Indeed, given any choice of $\paa{\eps_k}$, the function $\pa{\la,x^{\prime}} \mapsto G\pa{\la, x^{\prime}} D_{N,r}$ is 
Riemann--integrable along $\ga\times \intoo{x}{+\infty}$ .
This stems from the fact that $D_{N,r}$ is bounded in $\la$, and as follows from the previous discussion, 
is at least Riemann-integrable in $x$ as an oscillatory integral. 
Moreover, it is clear that by harping on the steps that allow one to prove the expansion for $\Pi_N$ given in proposition \ref{proposition DA Pi}, 
one can just as well prove a similar type of expansion for $\mc{F}_N\pac{\nu,u,\ex{g}}$. 
\qed

\begin{lemme}
\label{lemme form plus explicite serie Natte et fonctionnelle}
Under the assumptions of section \ref{soussection hypotheses},  the Fredholm determinant of $I+V$, with $V$ being given
by \eqref{definition noyau GSK FF}, admits the below, absolutely convergent for $x$-large
enough, Natte series representation:
\beq
\ddet{}{I+V}\pac{ \nu,u,g }= \ddet{}{I+V}^{\pa{0}}\pac{\nu,u,g}
\Bigg\{  1 + \sul{n \geq 1}{}  \sul{\bs{k} \in \mc{K}_n }{}  \sul{ \paa{\eps_{\bs{t}}} \in \mc{E}\pa{\bs{k}} }{}
  H_n   \pa{  x, \paa{k}, \paa{\eps_{\bs{t}}}_{\bs{t}\in \J{k}} }  
 \pac{ \nu,u , \Pi_{ \bs{t}\in \J{k} }^{} \ex{\eps_{\bs{t}} g\pa{z_{\bs{t}} } }  }   \Bigg\} 
\; .
\label{ecriture serie Natte pour determinant}
\enq

The $n^{\e{th}}$ term of this series is a $\e{O}\pa{\pac{m\pa{x}}^{-n}}$, with $m\pa{x}=\e{O}\pa{x^{-w}}$
and $w$ being given as in \eqref{definition constante w dans estimation de FN}.  
The functional $H_n$ appearing above is a linear functional in respect to the funtion $\prod_{ \bs{t}\in \J{k} }^{} \ex{\eps_{\bs{t}} g\pa{z_{\bs{t}} } }$ 
of the $n$-variables $z_{\bs{t}}$. It produces a weighted integration of this function
over curves lying in some small neighborhood of the real axis:
\bem
 H_n   \pa{  x, \paa{k}, \paa{\eps_{\bs{t}}}_{\bs{t}\in \J{k}} }   \Big[ \nu,u , \pl{ \bs{t}\in \J{k} }{} \ex{\eps_{\bs{t}} g\pa{z_{\bs{t}} } }  \Big]   =
  \sul{  \substack{ r_{\bs{d}}=1   \\ \bs{d} \in \D{k} }    }{  \bs{d}_{1}  }  \hspace{1mm}
\sul{\tau_{\bs{d}}=\ua/\da }{}
\hspace{2mm} \pl{ \bs{d}\in \D{k} }{}  \; \Int{+\infty}{ x } \dd x_{\bs{d}}  \Oint{ \ga^{\tau_{\bs{d}}} }{}  \f{ \dd \la_{\bs{d}} }{ 4\pi }
\Int{  \paa{\Sg_{\Pi}^{\tau_{\bs{d}}}} ^{  \pa{  r_{\bs{d}} , \bs{d}_1  } }  }{ } \hspace{-2mm}\f{ \dd^{\bs{d}_1} z_{\bs{d},j} }{ \pa{2i\pi}^{ \bs{d}_1 } } \\
\times \pl{j=1}{n}\f{1}{k_{j}!} \cdot \pl{\bs{d}\in \D{k}}{}  H_{\bs{d}_1,r_{\bs{d}}}
\Big( \la_{\bs{d}} , \{ z_{\bs{d},j} \}_{j=1}^{\bs{d}_1}, x_{\bs{d}} ; \{ \eps_{\bs{d},j} \}_{j=1}^{\bs{d}_1} \Big) \pac{ \nu ,u}
\cdot \pl{ \bs{t}\in \J{k} }{} \ex{\eps_{\bs{t}} g\pa{z_{\bs{t}} } } \;.
\label{definition fonctionnelle complexe HN pour determinant}
\end{multline}
In \eqref{ecriture serie Natte pour determinant}, the sum is carried out over all the possible choices of $n$-uples of integers $\bs{k}=\pa{k_1,\dots,k_n}$
belonging to  
\beq
\mc{K}_n = \Bigg\{ \bs{k}=\pa{k_1,\dots,k_n} \in \mathbb{N}^n \; : \; \sul{s=1}{n} s k_s =n \Bigg\}
\enq
Each such $n$-uple of integers defines a set of triplets
\beq
\J{k}= \Big\{ \pa{s,p,j} \, , \, s\in \intn{1}{n} \, , \, p \in \intn{1}{k_s} \, , \, j \in \intn{1}{s} \Big\}
\enq
and a set of doublets
\beq
\D{k}=\Big\{ \pa{s,p} \, , \, s\in \intn{1}{n} \, , \, p \in \intn{1}{k_s} \Big\} \; .
\enq
A triplet $\pa{s,p,j}$ belonging to $\J{k}$ is denoted by $\bs{t}$ and a doublet $\pa{s,p}$ belonging to $\D{k}$ is denoted by $\bs{d}=\pa{s,p}$.
The notation $\bs{d}_1$ stands for the first coordinate of $\bs{d}$, \textit{ie} if $\bs{d}=\pa{s,p}$, then $s=\bs{d}_1$.
Once that a choice of $\bs{k}$ is made, one sums over all the possible elements of 
\beq
\mc{E}\pa{\bs{k}} = \Bigg\{ \paa{\eps_{\bs{t}}}_{\bs{t} \in \J{k}} \; :  \;  \eps_{\bs{t}} \in \paa{\pm 1, 0} \; \;  \e{and}  
 \; \;  \sul{j=1}{\bs{d}_1} \eps_{\bs{d},j} = 0  \quad \e{for} \; \e{all} \; 
\bs{d} \in \D{k}   \Bigg\} \;.
\enq

The sums and integrations in \eqref{definition fonctionnelle complexe HN pour determinant} are also ordered by the sets of triplets 
$\J{k}$ and doublets $\D{k}$. One first starts to sum up
over $r_{\bs{d}}$, where $\bs{d}$ runs through $\D{k}$.  There are $\# \D{k}$ such sums in total,
corresponding to $\bs{d}$ running through the set $\D{k}$.  Finally, for each $\bs{d} \in \D{k}$, there is one integral over the corresponding
$x_{\bs{d}}$, one over the corresponding $\la_{\bs{d}}$ and $\bs{d}_1$ integrals
over the subordinate set of $z$-variables $\paa{z_{\bs{d},j}}_{j=1}^{\bs{d}_1}$.
\end{lemme}

\Proof

We define an auxiliary function
\beq
A\pa{\ga} = \sul{N \geq 1}{} \ga^{N} \mc{F}_N\pac{\nu,u,g}\pa{x}
\enq
which is holomorphic in $\ga$ on the open disc of radius $m^{-1}\!\pa{x}$, with $m\pa{x}=\e{O}\pa{x^{-w}}$
and $w$ given by \eqref{definition constante w dans estimation de FN},
as follows from the estimates on the growth of $\mc{F}_N$ with $x$. This means that $\ex{A\pa{\ga}}$ is holomorphic on the same disc. The radius of convergence of its Taylor series
around $\ga=0$  has its lower bound given by  $m^{-1}\pa{x}$.

The series for the Fredholm determinant is obtained by using the Faa-di-Bruno formula so as to compute the $n^{\e{th}}$ derivative of $\ex{A\pa{\ga}}$
at $\ga=0$
\beq
\f{1}{n!}\left. \f{ \dd^n }{\dd \ga^n} \ex{A\pa{\ga}} \right|_{\ga=0} =
\sul{ \substack{\paa{k} \\ \Sg s k_s =n } }{} \ex{A\pa{0}} \pl{j=1}{n} \paa{\f{1}{k_j!} \paf{ A^{\pa{j}}\pa{0}}{ j! }^{k_j}  }   \; \Rightarrow
\abs{ \f{1}{n!}\left. \f{ \dd^n }{\dd \ga^n} \ex{A\pa{\ga}} \right|_{\ga=0}   } \leq 
 \f{\pac{4 m\pa{x} }^n  }{n!}\left. \f{ \dd^n }{\dd \ga^n} \exp\paa{\f{1}{2-\ga} } \right|_{\ga=0} \;.
\enq
Where we used that $\abs{\mc{F}_N\pac{\nu,u,g}}\leq \pa{m\pa{x}}^{N}$, with $m\pa{x}=\e{O}\pa{x^{-w}}$. The last estimates allow one 
to see explicitly that the Taylor series at $\ga=0$ for $\ex{A\pa{\la}}$ has a radius of convergence that scales as $m^{-1}\pa{x}$. It is in particular 
convergent at $\ga=1$ leading to 
\beq
\ddet{}{I+V}\pac{ \nu,u,g }= \ddet{}{I+V}^{\pa{0}}\pac{\nu,u,g}\paa{1 + \sul{n \geq 1}{}
\sul{ \Sg s k_s =n }{}   \pl{j=1}{n} \pa{k_j!}^{-1}  \cdot   \pl{j=1}{n} \pac{ \mc{F}_{j}\pac{\nu,u,g}\pa{x}  }^{k_j}     }  \; .
\label{equation Serie de Natte brutte pour le determinant}
\enq
In this language of doublets and triplets, the expression for the product in \eqref{equation Serie de Natte brutte pour le determinant} reads
\bem
 \pl{j=1}{n} \pa{ \mc{F}_{j}\pac{\nu,u,g}\pa{x} } ^{k_j}       =
 \sul{  \substack{r_{\bs{d}}=1 \\ \bs{d} \in D_{\paa{k}} }  }{  \bs{d}_1  } \!\!
\sul{ \substack{ \sul{j=1}{\bs{d}_1 } \!\! \eps_{\bs{d},j} =0  \\   \eps_{\bs{d},j} \in \paa{\pm 1,0}  }  }{} \!\!\!
\sul{\tau_{\bs{d}}=\ua/\da}{}
\pl{\bs{d} \in \D{k}}{}  \; \Int{+\infty}{ x } \dd x_{\bs{d}}  \Oint{ \ga^{\tau_{\bs{d}}} }{} \f{ \dd \la_{\bs{d}} }{ 4\pi }  
\Int{ \paa{\Sg_{\Pi}^{\tau_{\bs{d}}}}^{  \pa{ r_{\bs{d} , \bs{d}_1 } } }  }{ } \hspace{-2mm} \pl{j=1}{\bs{d}_1} \f{ \dd z_{\bs{d},j} }{ \pa{2i\pi}} \;  \\
\times \pl{\bs{d}\in \D{k}}{}  H_{\bs{d}_1,r_{\bs{d}}}
\Big( \la_{\bs{d}} , \{ z_{\bs{d},j} \}_{j=1}^{\bs{d}_1}, x_{\bs{d}} ; \{ \eps_{\bs{d},j} \}_{j=1}^{\bs{d}_1} \Big) \pac{ \nu ,u}
\; \exp\paa{\sul{\bs{t}\in J_{\paa{k}} }{} \eps_{\bs{t}}g\pa{z_{\bs{t}}} }
 \; .
\end{multline}
The result follows. \qed

The expression \eqref{definition fonctionnelle complexe HN pour determinant} for the functionals 
involved in the Natte series is more explicit then as it was given in theorem
\ref{theorem representation serie de Natte}.

\subsection{Proof of theorem \ref{theorem representation serie de Natte}}

The first part of theorem \ref{theorem representation serie de Natte}, \textit{ie} the very form of the expansion \eqref{ecriture serie de Natte intro}
is a consequence of lemma \ref{lemme form plus explicite serie Natte et fonctionnelle}. The latter provides moreover a more
explicit form for the functionals $\mc{H}_n\pac{\nu,\ex{g},u}$.

The well ordered asymptotic expansion in $x$ for each functional $\mc{H}_n\pac{\nu,\ex{g},u}$, as given in 
\eqref{ecriture serie Natte detaille pour chaque Hn Intro}, is a direct consequence of the existence of a similar 
representation for $\mc{F}_N\pac{\nu,u,g}$ together the  correspondence \eqref{equation Serie de Natte brutte pour le determinant} 
between $\mc{F}_N\pac{\nu,u,g}$ and $\ddet{}{I+V}$. Finally, the existence of a representation for 
$\mc{F}_N\pac{\nu,u,g}$ in the spirit of \eqref{ecriture serie Natte detaille pour chaque Hn Intro}
can be readily obtained by inserting the well-ordered  
series representation for $\Pi$ and $\Pi^{-1}$ (we remind that $\Pi^{-1}= \, ^t \e{Comat}\pa{\Pi} $ since $\ddet{}{\Pi}=1$)
given in proposition \ref{proposition DA Pi} into the integral representation for $\mc{F}_N$, 
\eqref{definition fN}, \eqref{ecriture rep int F1}, \eqref{ecriture rep int FN}. \qed

\subsection{Higher order Natte series}

The higher order Natte series is a generalization of the Natte series derived in the previous sub-sections. 
It gives a direct access to part of the asymptotic expansion without having to compute the effective form of the functionals
$D_{N,r}$ and then carry out the contour integrals. Indeed, even if it is possible in principle to compute explicitly, order-by-order 
the functionals $D_{N,r}$ and thus $H_n$, this task becomes  very quickly monstrously cumbersome. In 
order to get the corrections, it is more desirable to apply the procedure below (or its obvious extension to higher order asymptotics)
if one wants to access to the higher order correction then those contained in $\ddet{}{I+V}^{\pa{0}}\pac{\nu,u,g}$.

\begin{prop}
The Fredholm determinant of $I+V$ admits the below convergent Natte series representation:
\bem
\ddet{}{I+V}\pac{\nu,u,g}= \ddet{}{I+V}^{\pa{0}}\pac{\nu,u,g}
\exp\paa{\; \Int{+\infty}{ x } \dd  x^{\prime}  \; \pac{x^{\prime}}^{-\f{3}{2}}a_1\!\pa{x^{\prime}}  +  
 \pac{x^{\prime}}^{-2}\pac{ a^{\e{osc}}_2\!\pa{ x^{\prime} }+ a_{2}^{\e{no}}\!\pa{ x^{\prime} } }  }
\;  \\
\times \Bigg\{ 1 + \sul{n \geq 1}{}
\sul{ \bs{k} \in \mc{K}_n }{}  \sul{ \paa{\eps_{\bs{t}}} \in \mc{E}\pa{\bs{k}}  }{}   \; 
\wt{H}_n \pa{x, \paa{k}, \paa{\eps_{\bs{t}}} }\pac{\nu,u, \Pi_{ \bs{t} \in \J{k} }^{} \ex{  \eps_{\bs{t}}g\pa{z_{\bs{t}}}  }  
} \; \Bigg\}\; .
\end{multline}
There $\wt{H}_n$ is defined as in \eqref{definition fonctionnelle complexe HN pour determinant}, \eqref{definition des fonctionnelles H_N},
but with the minor difference that the functionals $D_N$ in \eqref{definition des fonctionnelles H_N} should be replaced by 
the functionals $\wt{D}_N$ as given in \eqref{definition des fonctionnelles DN tilde}. Also, $a_1$, $a_2^{\e{osc}}$, $a_2^{\e{nosc}}$
are given by \eqref{formule intermediare pour correction a2}, \eqref{formule pour a2 Osc}, \eqref{formule pour a1 time like},
\eqref{formule pour a1 space like}. Note that here we have explicitly insisted on their dependence on the large-parameter $x^{\prime}$. 
The fundamental difference between the higher order Natte series and the one discussed previously is that
for $x$ large enough and for an $n$-indepenent $\e{O}$:
\beq
\abs{ H_n \pa{ x, \paa{k}, \paa{\eps_{\bs{t}}}  }\bigg[ \nu,u, \pl{ \bs{t} \in \J{k} }{} \ex{  \eps_{\bs{t}}g\pa{z_{\bs{t}}}  } \bigg] } =
\e{min}\paa{ 
\e{O}\pa{ \f{1}{x^{nw}}} \; , \; \e{O}\pa{  \f{\log^2 x}{x^3} , \f{a_1 \log  x}{x^{\f{5}{2}}}, \f{a_2^{\e{osc}} \log x}{x^3}, 
 \f{a_1}{x^{\f{3}{2}+w}},  \f{a^{\e{osc}}_2}{x^{2+w}} }  } \;.
\nonumber
\enq
The constant $w$ is as defined in \eqref{definition constante w dans estimation de FN}. 
\end{prop}

\Proof
 One starts by performing the decomposition
\beq
\log\ddet{}{I+V}\pac{\nu,u,g}=  \log\ddet{}{I+V}^{\pa{0}}\pac{\nu,u,g} +
\Int{+\infty}{ x } \dd  y \pa{ \f{a_1\pa{y}}{ y^{\f{3}{2}} }  +  \f{a_2^{\e{osc}}\pa{y}+a_2^{\e{no}}\pa{y}}{ y^2 } }
+ \log\ddet{}{I+V}^{\pa{\e{sub2}}}\!\pac{\nu,u,g} \;.
\nonumber
\enq
There, $\log\ddet{}{I+V}^{\pa{\e{sub2}}}\!\pac{\nu,u,g} $ corresponds to that part of  $\log\ddet{}{I+V}^{\pa{\e{sub}}}\!\pac{\nu,u,g} $
 \eqref{formule explicite exacte pour logdet sub}  where all terms that give rise to the integral involving $a_1$ and $a_2^{\e{osc/no}}$
 have been substracted. Namely, 
\beqa
\log\ddet{}{I+V}^{\pa{\e{sub2}}}\pac{\nu,u,g} = \Int{+\infty}{ x}\!  \dd x^{\prime} \;  \Oint{ \ga }{} \f{\dd \la}{4\pi} \;  G\pa{\la}
\left\{
\e{tr}\pac{ \Dp{\la}\Pi\pa{\la} \, \sg\pa{\la} \, \Pi^{-1}\pa{\la} } - \f{{\bf 1}_{\ga^{\pa{0}}}\!\pa{\la} }{y}
\e{tr}\paa{ \pac{\Pi^{\pa{1}}}^{\prime}\!\pa{\la}\sg_3 }
\right.\hspace{1cm}   && \label{ecriture decomposition log det sub} \\
 \left.
-\f{{\bf 1}_{\ga^{\pa{0}}}\!\pa{\la} }{y^{\f{3}{2}}} \e{tr}\paa{\pac{\Pi^{\pa{2}}}^{\prime}\!\pa{\la}
- \Pi^{\pa{0}}\!\pa{\la}\pac{\Pi^{\pa{1}}}^{\prime}\!\pa{\la} - \Pi^{\pa{1}}\!\pa{\la}\pac{\Pi^{\pa{0}}}^{\prime}\!\pa{\la} }  \sg_3
- \f{{\bf 1}_{\ga^{\pa{0}}} \!\pa{\la} }{y^2} \e{tr}\paa{\pac{\Pi^{\pa{3}}}^{\prime}\!\pa{\la} - \Pi^{\pa{1}}\!\pa{\la}\pac{\Pi^{\pa{1}}}^{\prime}\!\pa{\la}}  \sg_3
 \right\} \; .&&  \nonumber 
\eeqa
We stress that the variable of integration $x^{\prime}$ corresponds to the large parameter (denoted by $x$ before) that enters in the formulation of
the RHP for $\Pi$ and on which $\Pi$ depends implicitly. 

The expansion of $\log\ddet{}{I+V}^{\pa{\e{sub}}}\pac{\nu,u,g}$ goes along the same lines as before, with the minor difference that
the functionals $D_{N,r}$ are defined slightly differently. Indeed one has to substract from the Neumann series like expansion for 
$\e{tr}\pac{ \Dp{\la}\Pi\pa{\la} \, \sg\pa{\la} \, \Pi^{-1}\!\pa{\la}}$  all the subleading contributions that appear in 
\eqref{ecriture decomposition log det sub}. 
For this purpose we define 
\beq
\hspace{-3mm}\tr\pac{\De\pa{z_r}\dots \De\pa{z_1} \sg\pa{\la} \nabla\pa{z_{r+1}} \dots \nabla\pa{z_N}  } - R_{N,r}\pa{\paa{z},x}=
\sul{ \substack{ \Sg \eps_k=0 \\ \eps_k \in \paa{\pm 1, 0}} }{}
\wt{D}_{N,r} \pa{\la,  \paa{z_j}_{1}^{N}, x ; \paa{\eps_j}_{j=1}^N }\pac{\nu,u} \; \ex{ \sul{p=1}{N} \eps_p g\pa{z_p} } \hspace{-2mm}.
\label{definition des fonctionnelles DN tilde}
\enq
In order to define $R_{N,r}\pa{ \paa{z},x}$ we represent the asymptotic expansion of 
$\De\pa{z}$ slightly differently then in \eqref{formule DA matrice Delta}. 
\beq
\De\pa{z} \simeq \sul{ p\geq 1 }{} x^{-\f{p}{2}} M^{\pa{p}}\pa{z,x} \qquad \e{with} \qquad  
\left\{ \ba{cccc}  M^{\pa{2p+1}}\!\pa{z;x}  &=&  \f{d^{\pa{p}}\!\pa{z}  }{ \pa{z-\la_0}^{2p+1} }  
\cdot \bs{1}_{\mc{D}_{\la_0,2 \de} \setminus \mc{D}_{\la_0,\de^{\prime}} } \!\! \pa{z} 
\vspace{2mm} \\
				M^{\pa{2p}}\!\pa{z;x}  &=& \sul{\eps= \pm}{} \f{ V^{ \pa{\eps; p-1}  }\!\pa{z}  }{ p! \pa{z -\eps q }^{p-1} } 	
				\cdot \bs{1}_{\mc{D}_{\eps q, 2\de} \setminus \mc{D}_{\eps q, \de^{\prime}} } \!\! \pa{z}  \; . \ea \right.
\enq
There we took $\de$ small enough and $\de>\de^{\prime}>0$. In terms of such matrices one has
\bem
R_{N,r}\pa{\paa{z},x}= \de_{N,1} \de_{r,N} \pac{ \sul{p=1}{4}  x^{ -\f{p}{2}}  \e{tr}\pac{M^{\pa{p}}\!\pa{z_1,x}  \sg_3 } } 
\; +\; \f{\de_{N,3} \de_{r,3}}{x^2}   \e{tr}\pac{ M^{\pa{1}}\!\pa{z_3,x}  M^{\pa{2}}\!\pa{z_2,x} M^{\pa{1}}\!\pa{z_1,x} \sg_3  }  \\
\;  + \;  \sul{r=1}{2}\de_{N,2} \pa{-1}^r \pac{ \sul{p,p^{\prime}=1}{2}  x^{ -\f{p+p^{\prime}}{2} } 
\e{tr}\pac{M^{\pa{p}}\!\pa{z_1,x}  \sg_3 M^{\pa{p^{\prime}}}\!\pa{z_2,x} } } \; .
%
%
%
\nonumber
\end{multline}
Finally, defining $\wt{H}_n$ as in \eqref{definition fonctionnelle complexe HN pour determinant}, \eqref{definition des fonctionnelles H_N},
but with the minor difference that the functionals $D_N$ in \eqref{definition des fonctionnelles H_N} should be replaced by 
the functionals $\wt{D}_N$ given in \eqref{definition des fonctionnelles DN tilde}. One gets the desired representation. 
  \qed

A similar Natte series can be obtained for other quantities that are also related with the correlation functions in integrable models. 

\begin{prop}
Let $F_1$ be as defined in \eqref{definition fction fplus-moins},  
then the below Fredholm minor admits a Natte series representation:
\bem
\paa{ \Int{\msc{C}_E}{} \f{\dd \la}{2\pi} e^{-2}\!\pa{\la} \; + \;  \Int{-q}{q} \f{\dd \la}{2\pi} 4\sin^{2}\pac{\pi\nu\pa{\la}} F_1\pa{\la}E\pa{\la} }
 \ddet{}{I+V}\pac{\nu,u,g}= \ddet{}{I+V}^{\pa{0}}\pac{\nu,u,g} \;  \\ 
\times \Bigg\{ \f{ \mc{S}_0^{-1} \bs{1}_{\intoo{q}{+\infty}\pa{\la_0} } }{ \sqrt{-2\pi u^{\prime \prime}\pa{\la_0}} } + \f{\nu\pa{q}}{x u^{\prime}\pa{q}}\mc{S}_+^{-1} 
-  \f{\nu\pa{-q}}{x u^{\prime}\pa{-q}}\mc{S}_-^{-1}   
 + \;    \sul{n \geq 1}{}
\sul{ \bs{k} \in \wt{\mc{K}}_n }{} \;  \sul{ \{  \eps_{\bs{t}} \}  \in \wt{\mc{E}}\pa{\bs{k}}  } { }   \; 
\hspace{-3mm}\wt{H}_n^{\pa{+}} \pa{x, \bs{k}, \paa{\eps_{\bs{t}}} }\pac{\nu,u, \Pi_{ \bs{t} \in \J{k} }^{} \ex{  \eps_{\bs{t}}g\pa{z_{\bs{t}}}  } } \Bigg\}
 \; .
\label{equation series natte mineur fred}
\end{multline}
There the summation runs through all the possible choice of integers $k_1,\dots k_{N+1}$ belonging to 
\beq
\mc{K}_n = \Bigg\{ \bs{k} =  \pa{k_1,\dots, k_{n+1}} \; : \; k_s \in \mathbb{N} \; ,\; s=1,\dots, n \quad \e{and} \quad k_{n+1}\in \mathbb{N}^* \quad 
k_{n+1} + \sul{s=1}{n} s k_s  = n  \Bigg\} \;.
\enq
The remaining summation run through sets that are labelled by doublets and triplets belonging to 
\beqa
\J{k} &=&\Big\{  \pa{s,p,j},  s \in \intn{1}{n+1} \, , \; p \in \intn{1}{k_s} \, , \; j \in \intn{1}{s-\de_{s,n+1}n} \Big\} \nonumber \hspace{3mm}\\
\D{k} &=& \Big\{  \pa{s,p,j},  s \in \intn{1}{n} \, , \; p \in \intn{1}{k_s} \Big\} \;.  \nonumber
\eeqa
Indeed, then 
\beq
\wt{\mc{E}}\pa{\bs{k}}=\Bigg\{  \paa{\eps_{\bs{t}}}_{ \bs{t} \in \J{k} } \; : \;  \eps_{\bs{t}} \in \paa{\pm 1, 0} \;   \; ,  \;    \sul{j=1}{ \bs{d}_1 } \!\! \eps_{\bs{d},j} =0  
\;\; \e{with} \;\; \bs{d} \in \D{k} \; \; \e{and} \; \; \sul{p=1}{k_{n+1}} \eps_{k_{n+1},p,1} =1 \Bigg\} \;.
\enq

Finally, the functionals $\wt{H}^{\pa{+}}_n\pa{x,\paa{k},\paa{\eps_{\bs{t}}}}$ read
\beq
\wt{H}_n^{\pa{+}} \pa{x, \paa{k}, \paa{\eps_{\bs{t}}} }\bigg[\nu,u, \pl{ \bs{t} \in \J{k} }{} \ex{  \eps_{\bs{t}}g\pa{z_{\bs{t}}}  }  \bigg]=
\Int{ \Sg_{\Pi}^{\pa{n}} }{} \pl{p=1}{n} \f{\dd z_{n+1,p,1}}{2i\pi} 
\wt{C}_n\Big( \{ z_{n+1,p,1} \}_{p=1}^{n}, \{\eps_{n+1,p,1}\}_{p=1}^{n}, x  \Big)
\pac{\nu, u} \pl{p=1}{n} \ex{\eps_{n+1,p,1} g\pa{z_{n+1,p,1}}} \;. \nonumber
\enq
when $k_{n+1}=n$ and in all other cases, 
\beqa
 \wt{H}_n^{\pa{+}}   
\pa{  x, \paa{k}, \paa{\eps_{\bs{t}}}_{\bs{t}\in \J{k}} }   \bigg[ \nu,u , \pl{ \bs{t}\in \J{k} }{} \ex{\eps_{\bs{t}} g\pa{z_{\bs{t}} } }  \bigg]   =
  \sul{  \substack{ r_{\bs{d}}=1   \\ \bs{d} \in \D{k} }    }{  \bs{d}_{1}  }  \hspace{1mm}
\sul{\tau_{\bs{d}}=\ua/\da }{}
\hspace{2mm} \pl{ \bs{d}\in \D{k} }{}  \; \Int{+\infty}{ x } \dd x_{\bs{d}}  \Oint{ \ga^{\tau_{\bs{d}}} }{}  \f{ \dd \la_{\bs{d}} }{ 4\pi }
\Int{  \paa{\Sg_{\Pi}^{\tau_{\bs{d}}}} ^{  \pa{  r_{\bs{d}} , \bs{d}_1  } }  }{ } \hspace{-2mm}\f{ \dd^{\bs{d}_1} z_{\bs{d},j} }{ \pa{2i\pi}^{ \bs{d}_1 }} 
\Int{\Sg_{\Pi}^{\pa{k_{n+1}}}}{} \pl{p=1}{k_{n+1}} \f{\dd z_{k_{n+1},p,1}}{2i\pi} \hspace{1cm} && \nonumber
\\
\times \pl{j=1}{n}\f{1}{k_j!} \; \times \; 
C_{k_{n+1}}\Big( \{ z_{k_{n+1},p,1} \}_{p=1}^{k_{n+1}}, x , \{ \eps_{k_{n+1},p,1} \}_{p=1}^{k_{n+1}}  \Big)\pac{\nu,u} \cdot 
\pl{\bs{d}\in \D{k}}{}  H_{\bs{d}_1,r_{\bs{d}}}
\Big( \la_{\bs{d}} , \{ z_{\bs{d},j} \}_{j=1}^{\bs{d}_1}, x_{\bs{d}} ; \{ \eps_{\bs{d},j} \}_{j=1}^{\bs{d}_1} \Big) \pac{ \nu ,u}
\cdot \pl{ \bs{t}\in \J{k} }{} \ex{\eps_{\bs{t}} g\pa{z_{\bs{t}} } } \;. \hspace{1cm} &&  \nonumber
\eeqa
Note that we have used above the notation introduced in lemma \ref{lemme form plus explicite serie Natte et fonctionnelle}. Also we have set 
\beqa
i \f{ \e{tr}\pac{ \De\pa{z_N} \dots \De\pa{z_1} \sg^- } - \de_{N,1} \sul{p=1}{2}x^{-\f{p}{2}} \e{tr}\pac{M^{\pa{p}}\!\pa{z_1,x} \sg^{-}} }
{ \pl{p=1}{N-1}\pa{z_p-z_{p-1}} } &=& 
 \sul{ \Sg \eps_s=1 }{} \wt{C}_N\big( \{z_j \}_1^N,x,\{ \eps_j \}_{1}^{N} \Big) \pac{\nu,u} \pl{p=1}{N} \ex{ \eps_p g\pa{z_p}}  \nonumber \\
i  \e{tr}\pac{ \De\pa{z_N} \dots \De\pa{z_1} \sg^- } \cdot  \pl{p=1}{N-1}\pa{z_p-z_{p-1}}^{-1}  &=& 
 \sul{ \Sg \eps_s=1 }{} C_N\big( \{ z_j \}_1^N,x,\{\eps_j \}_{1}^{N} \big) \pac{\nu,u} \pl{p=1}{N} \ex{ \eps_p g\pa{z_p}} \;.
\nonumber
\eeqa
The sums in the two equations above run over all choices of the variables $\eps_{s}$, $s=1,\dots,N$
with $\eps_{s} \in \paa{\pm 1, 0}$ and $\sum_{s=1}^{N}\eps_s=1$. 
\end{prop}

\Proof

By using the integral representation for $\chi$ \eqref{forme explicite de chi en terms f plus moins}, one readily gets that 
\beq
i\pac{\chi_{\infty}}_{12} \equiv \lim_{\la \tend +\infty} \e{tr}\pac{\la \chi\pa{\la} \sg^-}  
=  \Int{-q}{q} \f{\dd \la}{2\pi} 4\sin^{2}\pac{\pi\nu\pa{\la}} F_1\pa{\la}E\pa{\la}  \;.
\enq
Also, it is easy to convince oneself that 
\beq
i\pac{\Pi_{\infty}}_{12} = 
i\pac{\chi_{\infty}}_{12} + \Int{\msc{C}_E}{} e^{-2}\!\pa{\la} \f{\dd \la}{2\pi} \;.
\enq
The claim follows by expanding $\pac{\Pi_{\infty}}_{12}$ into a higher order Natte series (where the first few terms of the asymptotics 
have been taken into account) and then taking the product of this series with the Natte series for the Fredholm determinant. The details 
are left to the reader. \qed

It follows from the leading asymptotics given in \eqref{equation series natte mineur fred} that, in the case of the time-like regime, 
the saddle-point $\la_0$ does not contribute to the leading order.  It can however be checked that it does eventually contribute. Its contribution 
is a $\e{O}\Big( x^{-\tf{5}{2}} , x^{-\tf{5}{2}} \big(x^{-\nu\pa{q}}  + x^{-\nu\pa{-q}} \big)^2 \Big)$.


\section{Conclusion}

In this paper we have obtained the first few terms in the leading $x\tend +\infty$ asymptotics of the Fredholm determinant
of a class of integrable integral operator that provide a starting point for the analysis of the large-distance/long-time
asymptotic behavior in integrable models away from their free fermion point. Also, we have derived a new series representation
for the Fredholm determinant, the so-called
Natte series. This series is well adapted for an asymptotic analysis of the Fredholm determinant and can thus be thought of as
being an analogue of the Mellin-Barnes integral representation for hypergeometric functions.
In two subsequent paper, the Natte series will appear as a central tool in computing the large-distance/long-time asymptotic behavior of
the correlation functions in the non-linear Schr\"{o}dinger model away from its free fermion point 
\cite{KozKitMailTerNatteSeriesNLSECurrentCurrent,KozReducedDensityMatrixAsymptNLSE}.
As a byproduct of our analysis, we have been able to bring a little more order to the structure of the asymptotic expansion of Fredholm determinants
of operators belonging to the class of the generalized sine kernel.
It would be interesting to extend/push forward the form of the full asymptotic expansion of the determinant
given in theorem \ref{theorem representation serie de Natte}, in particular by providing a closed form 
(\textit{ie} the explicit values of coefficients/functionals), at least in the case of some particular kernel such as the sine kernel.



\section*{Acknowledgements}

I am supported by the EU Marie-Curie Excellence Grant MEXT-CT-2006-042695. I would like to thank
N. Kitanine, J.M. Maillet, N. Slavnov and V. Terras for stimulating discussion and their interest in this work.
I also thank the Theoretical group of DESY for hospitality, which maked this work possible.
I am also grateful to the organizers of the summer school 
"Finite-size technology in low-dimensional quantum systems (V)" held at the Pedro Pascual center for theoretical physics 
during which part of this work has been carried out.


\appendix

\section{Several Properties of CHF}
\label{Appendix Properties CHF and Barnes}

One can check that for $z \in \R^+$
\beq
\Psi\pa{1,\f{3}{2}; -\ex{i0^+} z} -\Psi\pa{1,\f{3}{2}; -\ex{-i0^+} z}= 2 i \sqrt{\f{\pi}{z}} \ex{-z} \; .
\label{relation de discontinuite pour CHF Erreur}
\enq

$\Psi\pa{a,c;z}$ has an asymptotic expansion at $z \tend \infty$ given by
\beq
 \Psi\pa{a,c;z}=\sum_{n=0}^{M} \pa{-1}^n \f{ \pa{a}_n \pa{a-c+1}_n }{ n! }z^{-a-n} \,+\e{O}\pa{z^{-M-a}}\;  ,\qquad -\frac{3\pi}2<\arg(z)<\frac{3\pi}2.
\label{asy-Psi}
\enq
It also satisfies to the monodromy properties
\begin{align}
 &\Psi(a,1;ze^{2i\pi})= \Psi(a,1;z)e^{-2i\pi a}+
 \frac{2\pi ie^{-i\pi a+z}}{\Gamma^2(a)}
 \Psi(1-a,1;-z),  &\Im (z)<0,
\label{cut-Psi-1}\\
 &\Psi(a,1;ze^{-2i\pi})= \Psi(a,1;z)e^{2i\pi a}-
 \frac{2\pi ie^{i\pi a+z}}{\Gamma^2(a)}
 \Psi(1-a,1;-z),  &\Im (z)>0.
\label{cut-Psi-2}
\end{align}

Tricomi's CHF can be expressed in terms of Humbert's CHF
\beq
\Psi\pa{a,c;z}= \Ga\pab{1-c}{a-c+1} \Phi\pa{a,c;z}  + \Ga\pab{c-1}{a} z^{1-c} \Phi\pa{a-c+1,2-c;z}  \; .
\label{Appendix Special Functions  Psi s'ecrit comme Phi}
\enq
There exists a similar formula expression Tricomi's CHF
$\Psi\pa{a,c;z}$ in terms of Humbert's one
\begin{equation}
\Phi\pa{a,c;z}
 =\f{\Gamma\pa{c}}{\Gamma\pa{c-a}} \ex{i\eps a \pi} \Psi\pa{a,c;z}
 +\f{\Gamma\pa{c}}{\Gamma\pa{a}}\ex{i\eps \pi \pa{a-c}+z} \Psi\pa{c-a,c;-z},
\label{Appendix Special Functions Phi s'ecrit comme Psi}
\end{equation}
where $\eps= \e{sgn}\pa{\Im (z)}$, and
\begin{equation}
 \Phi\pa{a,c;z}=\sum_{n=0}^\infty \frac{(a)_n}{(c)_n}\frac{z^n}{n!}.
\end{equation}

The Barnes' $G$ function satisfies to the reflection property
\beq
G\pa{1-z}=\f{G\pa{1+z}}{ \pa{2\pi}^z} \exp\paa{ \Int{0}{z} \pi x \cot\pac{\pi x}  \; \dd x} \; ,
\label{Appendix special functions Int rep Barnes}
\enq
which holds for $\Re\pa{z}<1$ in the usual sense (and also everywhere else by analytic continuation).


\section{Proof of the asymptotic expansion for $\Pi$}
\label{Appendix Proofs asymptotic expansion for Pi and determinant}

\subsection{Two lemmas}

We first need a technical lemma
\begin{lemme}
\label{Lemme developpement produit matrices}
Let the matrices $\De_{j}$ take the form
\beq
\De_{j}= \big[ \mf{e}_j \big]^{\f{\sg_3}{2}} \wt{\De}_j \big[ \mf{e}_j \big]^{-\f{\sg_3}{2}} = \pa{ \ba{cc}  a_j & b_j \mf{e}_j \\
																			c_j \mf{e}_j^{-1}& d_j \ea } \; ,
\enq
where the entries $a_j, b_j, c_j, d_j$ do not depend on $\mf{e}_j$. Then
\beq
\De_N\dots \De_1 = \sul{a=0}{ \pac{\f{N}{2}} } \sul{ \bs{j}^{\pa{a}}  \in \mc{B}_{a; N } }{}
\pa{\ba{cc}  A_{ \bs{j}^{\pa{a}} } &  \mf{e}_N B_{ \bs{j}^{\pa{a}} }  \\
		     \mf{e}_N^{-1} C_{ \bs{j}^{\pa{a}} }  &  D_{ \bs{j}^{\pa{a}} }   \ea} \, \cdot \,
\paf{ \mf{e}_{j_2} \dots \mf{e}_{j_{2a}} }{ \mf{e}_{j_1} \dots \mf{e}_{j_{2a-1}} }^{\sg_3}  \; .
\enq
Above, the sum runs through all choices of $2a$-uples of integers $\bs{j}^{\pa{a}}=\pa{j_1,\dots,j_{2a}}$
with $\bs{j}^{\pa{a}}$ belonging to
\beq
\mc{B}_{a ; N}=\paa{ \pa{j_1,\dots, j_{2a} }\in  \pac{\mathbb{N}^*}^{2a}  \; : \; 1\leq j_1<\dots < j_{2 a}\leq N } \;.
\enq
The entries of each matrix appearing in the sum are linear polynomials the entries of the matrices $\wt{\De}_{p}$.
\end{lemme}
This lemma allow us to trace back all the dependence on the fractional power of $x$ in the products
$\De\pa{z_N}\dots \De\pa{z_1}$ of the non-trivial parts of the jump matrices for $\Pi$.

\Proof

The result clearly holds for $N=1$ as then, the only possibility is to take $a=0$.
%
%
%
%
%
%
%
%
%
%
%
%

We prove the induction hypothesis for the $11$ entry. It goes similarly for all the others.
By applying the induction hypothesis to $\De_N\dots \De_1$ and then explicitly multiplying out with $\De_{N+1}$, we get that
\bem
\pac{\De_{N+1}\dots \De_1}_{11}
= \sul{a=0}{ \pac{\f{N}{2}} } \sul{ \bs{j}^{\pa{a}} \in \mc{B}_{a; N} }{}
\paf{ \mf{e}_{j_2} \dots \mf{e}_{j_{2a}} }{ \mf{e}_{j_1} \dots \mf{e}_{j_{2a-1}} }
\paa{ a_{N+1}A_{ \bs{j}^{\pa{a}} }  + \f{\mf{e}_{N+1}}{\mf{e}_{N}} b_{N+1} C_{ \bs{j}^{\pa{a}} } } \\
=  \sul{a=0}{ \pac{\f{N}{2}} } \sul{ \bs{j}^{\pa{a}} \in \mc{B}_{a; N}   }{}
\paf{ \mf{e}_{j_2} \dots \mf{e}_{j_{2a}} }{ \mf{e}_{j_1} \dots \mf{e}_{j_{2a-1}} }
a_{N+1}A_{ \bs{j}^{\pa{a}} }
+\sul{a=0}{ \pac{\f{N}{2}} } \sul{  \bs{j}^{\pa{a}} \in \mc{B}_{a; N-1} }{}
\paf{ \mf{e}_{j_2} \dots \mf{e}_{j_{2a}} \mf{e}_{N+1} }{ \mf{e}_{j_1} \dots \mf{e}_{j_{2a-1}} \mf{e}_{N} }
 b_{N+1} C_{ \bs{j}^{\pa{a}} }   \\
 + \sul{a=1}{ \pac{\f{N}{2}} } \sul{ \substack{ \bs{j}^{\pa{a}} \in \mc{B}_{a; N} \\ j_{2a}=N }  }{}
\paf{ \mf{e}_{j_2} \dots \mf{e}_{j_{2a-2}} \mf{e}_N \mf{e}_{N+1} }{ \mf{e}_{j_1} \dots \mf{e}_{j_{2a-1}} \mf{e}_{N} }
 b_{N+1} C_{ \bs{j}^{\pa{a}} }  \;.
\end{multline}
The result follows as the above sums can be seen organized in respect to the partition
\bem
\bigcup\limits_{a=1}^{\pac{\tf{N+1}{2}}} \mc{B}_{a ; N+1} = \Bigg\{ \bigcup\limits_{a=1}^{\pac{\tf{N}{2}}} \mc{B}_{a ; N} \Bigg\} \bigcup
\Bigg\{ \bigcup\limits_{a=1}^{\pac{\tf{N}{2}}}  \paa{ 1\leq j^{}_1<\dots< j_{2a}\leq N-1 \; , \; j_{2a+1}=N  \; , \; j_{2a+2}=N+1 } \Bigg\} \\
\bigcup \Bigg\{ \bigcup\limits_{a=1}^{\pac{\tf{N}{2}}}   \paa{ 1\leq j^{}_1<\dots< j_{2a-1}\leq N-1 \; , \; j_{2a}=N+1} \Bigg\} \;.
\end{multline}
 \qed

\begin{lemme}
\label{Lemme integrale Fleu et ses derivees}
Let $\mc{F}_N\pa{z_1,\dots,z_N}$ be a holomorphic function on $\mc{D}= \mc{D}_{v,\de_1} \times \dots \times \mc{D}_{v,\de_N}$,
where $0 < \de_N < \dots < \de_1$ and $v \in \Cx$. Let
$\Dp{}\mc{D}=\Dp{}\mc{D}_{v,\de_1}\times \dots \times \Dp{}\mc{D}_{v,\de_N}$ be the skeleton of $\mc{D}$ and
$n_p$ a set of positive integers.
Then, for $\la$ lying outside of $\ov{\mc{D}}_{v,\de_1}$, one has
\beq
\Oint{\Dp{}\mc{D} }{}  \f{\dd^N z }{\pa{2i\pi}^N}   \f{ \mc{F}_N\pa{z_1,\dots , z_N} }{ \pa{ \la - z_1 }\pl{k=1}{N-1} \pa{z_k-z_{k+1}}}
\; \cdot \; \pl{p=1}{N} \f{ 1 }{ \pa{z_p-v}^{n_p+1}  } =
\sul{k_N=0}{r_N}  \dots \sul{k_1=0}{r_1} \f{ 1 }{ \pa{\la-v}^{r_0} }  \cdot
 \Bigg\{ \f{1}{k_1 ! } \f{\Dp{}^{k_1}}{\Dp{}z_1^{k_1} } \dots    \f{1}{k_N ! } \f{\Dp{}^{k_N}}{\Dp{}z_N^{k_N} } \mc{F}_N \Bigg\}_{\mid z_p= v} \;.
\enq
where we agree upon
\beq
r_p = \sul{\ell = p}{N} n_{\ell} + N-p - \sul{\ell= p +1}{N} k_{\ell} \qquad \e{and} \quad n_0=0 \; .
\enq
\end{lemme}
The proof is a straightforward induction. Note that the total highest possible order of derivatives of $\mc{F}_N$ that is produced by the above contour
integral is  $\sum_{\ell=1}^{N} n_{\ell} +  N -1 $. It corresponds to no-derivation in respect to the variables $z_2,\dots,z_N$
and a derivative in respect to $z_1$ of order $\sum_{\ell=1}^{N} n_{\ell} +  N -1 $. All other choices of the integers $\paa{k_a}$
lead to a total order of the partial derivatives that is strictly smaller.

\subsection{Proof of proposition \ref{proposition DA Pi}}
We are now in position to prove proposition \ref{proposition DA Pi}.

Here, we will only discuss the case of the time-like regime. The proof in the case of the space-like regime goes very similarly, so that we omit it here.

We have already established that, for $x$-large enough, $\Pi\pa{\la}$ is given in terms of a uniformly convergent Neumann series 
\eqref{serie de Neumann Pi+}:
\beq
\Pi\pa{\la}=
I_2 + \sul{N\geq 1}{} \f{\Pi_N \pa{\la}}{x^{N}}
\qquad \e{with} \qquad
\Pi_N \pa{\la} = x^{N} \Int{ \Sg_{\Pi}^{\pa{N}} }{} \f{ \dd ^N z }{ \pa{2i\pi}^N}
\f{ \De\pa{z_N} \dots \De\pa{z_1}  }
{\pa{\la-z_1} \pl{p=1}{N-1} \pa{z_p-z_{p+1}}   } \; .
\label{equation donnat Pi et PiN preuve Prop DA}
\enq
Above each N-fold integral runs across the inslotted contour $\Sg_{\Pi}^{\pa{N}}$ as defined in Fig.~\ref{contour integration encastree for
finale a deux N} and the equality holds for $\la$ uniformly away from the boundary $\Sg_{\Pi}$. We remind that in this Neumann series 
the matrices $\De\pa{z_k}$  are subordinate to the jump matrix $I_2+\De\pa{\la}$ for $\Pi\pa{\la}$ 
solving the $\Pi$-type RHP associated  with the jump contour
$\Sg_{\Pi}\pac{z_k}$, \textit{cf} definition \ref{Definition Matrices Delta Nabla Sur contour encastre}. 

To prove the claim of proposition \ref{proposition DA Pi}, we build on \eqref{equation donnat Pi et PiN preuve Prop DA} so as to
obtain a more precise form of the asymptotic expansion of $\Pi_N\pa{\la}$.

Recall that each contour $\Sg_{\Pi}\pac{z_k}$ entering in the definition of the inlsotted contour $\Sg_{\Pi}^{\pa{N}}$ can be
divided into its exterior part $\wt{\Ga}\pac{z_k}$ and three circles $\Dp{}\mc{D}_{q,\de_k} \cup \Dp{}\mc{D}_{-q,\de_k} \cup \Dp{}\mc{D}_{\la_0,\de_k}$.
There $0<\de_N<\dots<\de_1$ and $\de_1$ is small enough, in particular it is such that $\de_1< \tf{\abs{\la_0 \pm q}}{2}$.
However, the very choice of the contour $\wt{\Ga}$ implies that
\beq
\norm{ \De }_{L^{\infty}\pa{ \wt{\Ga}\pac{z_k} }} + \norm{ \De}_{L^{2}\pa{ \wt{\Ga}\pac{z_k} }} +\norm{ \De }_{L^{1}\pa{ \wt{\Ga}\pac{z_k} }} =
\e{O}\pa{x^{-\infty}} \; .
\label{ecriture estimation matrice De sur Ga tilde}
\enq
Hence, from the point of view of the asymptotic expansion, one can drop all contributions to $\Pi_N\pa{\la}$ stemming from those parts of
the multiple integral in \eqref{equation donnat Pi et PiN preuve Prop DA}, where at least one variable is integrated along
$\wt{\Ga}$. Indeed, due to the estimates \eqref{ecriture estimation matrice De sur Ga tilde}, such an integration would only produce 
$\e{O}\pa{x^{-\infty}}$ terms.

The matrix $A_N\!\pa{\la}$ appearing in \eqref{ecriture form detaillee AE} contains exactly these contributions,
and hence $A_N\!\pa{\la}=\e{O}\pa{x^{-\infty}}$ uniformly away from $\Sg_{\Pi}$.

It thus now remains to focus on the effect of the integration on the boundary of the three disks centered at $\pm q$ and $\la_0$.
In other words,
\beq
\Pi_N\pa{\la} = A_N\pa{\la} +  x^{N} \sul{ \bs{\eps} \in \mc{E}_N }{} \;   
\Oint{\Dp{}\mc{D}_{\bs{\eps}} }{} \hspace{-1mm} \f{\dd^N z}{\pa{2i\pi}^N} \cdot
 \f{ \De\pa{z_N} \dots \De\pa{z_1}  }
{ \pl{k=1}{N} \pa{z_{k-1}-z_k}    }  \qquad \e{where} \quad z_0=\la \; .
\label{ecriture decomposition PiN en partie exp petite et reste}
\enq
The above sum corresponds to summing up over all the possible choices of the integration contour
$\Dp{}\mc{D}_{v_{\bs{\eps}_k},\de_k}$ for each variables $z_k$.
More precisely, one sums over all the $N$-dimensional vectors $\bs{\eps}$ belonging to 
\beq
\mc{E}_N=\paa{\bs{\eps}=\pa{\bs{\eps}_1,\dots,\bs{\eps}_N}  \; : \; \bs{\eps}_k \in \paa{\pm 1,0} } \; .
\enq
We also agree upon the shorthand notation $v_{+}=q$, $v_{-}=-q$ and $v_0=\la_0$.
Finally, the integration contour  $\Dp{}\mc{D}_{ \bs{\eps} }$ in each summand corresponds to the Cartesian product of N-circles
$\Dp{}\mc{D}_{\bs{\eps}} = \Dp{} \mc{D}_{v_{\bs{\eps}_1},\de_1}\times \dots \times \Dp{}\mc{D}_{v_{\bs{\eps}_N},\de_N}$
of decreasing radii  $0<\de_N<\dots<\de_1$, with $\de_1$ small enough.

We stress that there exists natural constraints on the possible choices of the $\eps_k$.
Indeed, if $z_j$ and $z_{j+1}$ both belong to a sufficiently small neighborhood of $\la_0$, then 
$\De\pa{z_j} \De\pa{z_{j+1}} = 0$. Hence, choices of $N$-dimensional vectors $\bs{\eps}$
having two neighboring coordinates ($\eps_j$ and $\eps_{j+1}$ for some $j$) equal to zero  
do not contribute to the sum in  \eqref{ecriture decomposition PiN en partie exp petite et reste}. 
 
\vspace{2mm}

The asymptotic expansions of $\De\pa{z}$ on each of the three disks all take the generic form:
\beq
\De\pa{z} \simeq  \sul{n \geq 0}{}
\f{  \pac{\mf{e}\pa{z;\eps}}^{\f{\sg_3}{2}} \cdot \wt{\De}^{\pa{n}}\!\pa{z} \cdot \pac{\mf{e}\pa{z;\eps} }^{-\f{\sg_3}{2}}   }{x^{n+1} \pa{z-v_{\eps}}^{n\pa{2-\abs{\eps}}+1} }  \qquad
\e{uniformly} \; \e{in} \; z \in  \mc{D}_{v_{\eps},2 \de} \setminus \mc{D}_{v_{\eps},\de^{\prime} }  \qquad \eps \in \paa{\pm 1, 0} \;.
\label{ecriture DA De en terme De tilde}
\enq
The radii are such that $\de>\de^{\prime}>0$ and $\de$ is taken sufficiently small, but are arbitrary otherwise. 
\eqref{ecriture DA De en terme De tilde} is to be understood in the sense of an asymptotc expansion, \textit{ie} up to a truncation 
to any given order in $x$. 
The detailed expression for the matrices $\wt{\De}^{\pa{n}}\pa{z}$ and $\mf{e}\pa{z;\eps}$ differ on each of the disks 
(\textit{ie} for $\eps=\pm 1$ or $0$). However, $\mf{e}\pa{z;\eps}$ are holomorphic on any sufficently small neighborhood of $\pm q$ or $\la_0$. 
Also, the matrix $\wt{\De}^{\pa{n}}\!\pa{z}$ does not depend on $x$ anymore. 
The function $\mf{e}\pa{z;\eps}$ contains a fractional power of $x$ and also an oscillating term:
\beq
\mf{e}\pa{z;\eps} = \left\{  \ba{lcc}  \ex{ix u\pa{q}} x^{-2 \nu\pa{z}}   & \e{for} & \eps=  1 \\
									\ex{ix u\pa{-q}} x^{ 2 \nu\pa{z}}   & \e{for} & \eps=  -1 \\
									\ex{ix u\pa{\la_0}} x^{-\f{1}{2}}  & \e{for} & \eps=  0 \ea \right. 	\;.
\enq

We are now in position to establish the asymptotic expansion of the second term in \eqref{ecriture decomposition PiN en partie exp petite et reste}. 

Expanding each matrix $\De\pa{z_n}$ into its asymptotic series \eqref{ecriture DA De en terme De tilde}, using that the latter is uniform 
on the compact domain of integration we obtain the asymptotic expansion of $\Pi_{N}$:
\beq
\Pi_N\pa{\la} \simeq   \sul{r \geq 0}{}  \f{1}{x^r} \sul{ \bs{\eps} \in \mc{E}_N }{} 
\sul{ \bs{n} \in \mc{N}_N^{\pa{r}} }{}
\; \Int{ \Dp{}\mc{D}_{\bs{\eps}}   }{} \hspace{-2mm} \f{\dd^N z}{\pa{2i\pi}^N}
\;\cdot   \f{ \pac{\mf{e}_N}^{\f{\sg_3}{2}} \wt{\De}^{\pa{\bs{n}_N}}\!\pa{z_N} \pac{\tf{\mf{e}_{N-1}}{\mf{e}_N}}^{\f{\sg_3}{2}} \hspace{-3mm} \dots
\pac{\tf{\mf{e}_{1}}{\mf{e}_2}}^{\f{\sg_3}{2}}  \wt{\De}^{\pa{\bs{n}_1}}\!\pa{z_1}   \pac{\mf{e}_1}^{-\f{\sg_3}{2}}  }
{ \pa{\la-z_1} \cdot \pl{k=2}{N} \pa{z_{k-1}-z_k}     \cdot   \pl{p=1}{N} \pa{z_p-v_{ \bs{\eps}_p}}^{ \bs{n}_p\pa{2-\abs{ \bs{\eps}_p}}+1} } \; .
\label{equation DA Pi comme somme integrale produits DA De}
\enq
There appears a summation over $N$-dimensional integer valued vectors $\bs{n}$ belonging to 
\beq
\mc{N}^{\pa{r}}_N=\paa{\bs{n}=\pa{\bs{n_1},\dots,\bs{n}_N} \; : \;  n_k \in \mathbb{N} \; , \; \Sg_{p=1}^{N} \bs{n}_p = r } \; .
\enq
Note that in order to lighten the notations slightly, we have set $\mf{e}_k \equiv \mf{e}\pa{z_k;\bs{\eps}_k}$.
Also, just as in \eqref{ecriture DA De en terme De tilde}, we did not make the remainder explicit.

Lemma \ref{Lemme developpement produit matrices} ensures the existence of holomorphic functions 
$A^{\pa{\bs{n}}}_{\bs{j}^{\pa{a}}}\pa{ \paa{z_k}} $, \dots, $D^{\pa{\bs{n}}}_{ \bs{j}^{\pa{a}} }\pa{ \paa{z_k}}$ of $z_1,\dots, z_N$ such that
\beq
\pac{\mf{e}_N}^{\f{\sg_3}{2}}\!\! \wt{\De}^{\pa{\bs{n}_N}}\!\pa{z_N} \pac{\f{\mf{e}_{N-1}}{\mf{e}_N}}^{\f{\sg_3}{2}}\hspace{-2mm}\dots
\pac{\f{\mf{e}_{1}}{\mf{e}_2}}^{\f{\sg_3}{2}} \!\!\! \wt{\De}^{\pa{\bs{n}_1}}\!\pa{z_1} \pac{\mf{e}_1}^{- \f{\sg_3}{2}} =
\sul{ a =0}{ \pac{\f{N}{2}} } \sul{ \bs{j}^{\pa{a}} \in \mc{B}_{a;N} }{}
\pa{\hspace{-2mm}\ba{cc}  A^{\pa{\bs{n}}}_{ \bs{j}^{\pa{a}} } \pa{ \paa{z_k}}  &  \mf{e}_N B^{\pa{\bs{n}}}_{ \bs{j}^{\pa{a}} } \pa{ \paa{z_k}}  \\
		     \mf{e}_N^{-1} C^{\pa{\bs{n}}}_{\bs{j}^{\pa{a}} } \pa{ \paa{z_k}}  &  D^{\pa{\bs{n}}}_{\bs{j}^{\pa{a}} } \pa{ \paa{z_k}}   \ea \hspace{-2mm}}  
\paf{ \mf{e}_{j_2} \dots \mf{e}_{j_{2 a }} }{ \mf{e}_{j_1} \dots \mf{e}_{j_{2 a-1}} }^{\sg_3}  \hspace{-3mm}.
\label{ecriture developpement explicite puissance x produit matrices}
\enq

Due to the form taken by the matrices $\De^{\pa{n}}\!\pa{z}$, not all configurations of the $2a$-uples $\bs{j}^{\pa{a}}$ appear in 
\eqref{ecriture developpement explicite puissance x produit matrices}.
Indeed, when $z_k \in \Dp{}\mc{D}_{\la_0,\de_k}$ (\textit{ie} $\bs{\eps}_k=0$) the matrix $\wt{\De}^{\pa{n}}\!\pa{z_{k}}$ is proportional to $\sg^-$
(\textit{cf} \eqref{ecriture generique DA diverse parametrices time-like}). It appears in \eqref{equation DA Pi comme somme integrale produits DA De}
with a pre-factor $\mf{e}_k^{-1}$. 
 Therefore, for $z_k \in \Dp{}\mc{D}_{\la_0,\de_k}$ the only non-vanishing terms in the sum over $\bs{j}^{\pa{a}}\in \mc{B}_{a;n}$
are those corresponding to choices of $2a$-uples $\bs{j}^{\pa{a}}=\pa{ j_1,\dots,j_{2a} }$ such that $j_{p}=k$ for some $p$.
In other words, each time an integration variable belongs to $\Dp{}\mc{D}_{\la_0,\de_k}$ for some $k$, 
the associated oscillating exponent $\mf{e}^{-1}\!\pa{\la_0;0}$ is always present.
Moreover, all matrix entries in the expansion 
\eqref{ecriture developpement explicite puissance x produit matrices} that appear (after taking the matrix products) 
in front of the monomials $\tf{  \pa{\mf{e}_{j_2}\dots \mf{e}_{j_{2a}}}^{\pm 1}  }{  \pa{\mf{e}_{j_1}\dots \mf{e}_{j_{2a-1}}}^{\pm 1}  }$
vanish whenever a function $\mf{e}_{j_p}\equiv \mf{e}\pa{z_{j_p};\bs{\eps}_{j_p}}$ corresponding to  $z_{j_p} \in \Dp{}\mc{D}_{\la_0,\de_{j_p}}$ 
would appear in the numerator.
More precisely, if there exists a $p$ such that $j_p=k$ then 
\begin{itemize}
\item for $p \in 2\mathbb{N}+1$ (\textit{ie} $\bs{\eps}_{j_{2\ell+1}}=0$ for some $\ell$), one has
$B^{\pa{\bs{n}}}_{ \bs{j}^{\pa{a}} }=D^{\pa{\bs{n}}}_{ \bs{j}^{\pa{a}} }=0$ ;
\item for $p \in 2\mathbb{N}$ (\textit{ie} $\bs{\eps}_{j_{2\ell}}=0$ for some $\ell$), one has
$A^{\pa{\bs{n}}}_{ \bs{j}^{\pa{a}}  }=C^{\pa{\bs{n}}}_{ \bs{j}^{\pa{a}}  }=0$ .
\end{itemize}

Putting together \eqref{equation DA Pi comme somme integrale produits DA De} and \eqref{ecriture developpement explicite puissance x produit matrices}
leads to the below form of the asymptotic expansion for $\Pi_N$
\beq
\Pi_N\pa{\la} \simeq    \sul{r \geq 0}{} \f{1}{x^r}
\sul{ \bs{n} \in \mc{N}^{\pa{r}}_N }{}  \; \sul{ \bs{j}^{\pa{a}}  \in \mc{B}_{a;N} }{} \;   \sul{ \bs{\eps} \in \mc{E}_N }{}
I_{\bs{\eps} ; \bs{n} } \pac{ M_{\bs{j}^{\pa{a}} } } \;.
%
%
%
%
%
\label{ecriture PiN comme somme integrale fnelle matrices}
\enq
There $M_{\bs{j}^{\pa{a}}} $ stands for the matrix appearing in the expansion \eqref{ecriture developpement explicite puissance x produit matrices}
and $I_{\bs{\eps}; \bs{n}}$ is a functional depending on the choices of the entries of the $N$-dimensional vectors $\bs{\eps}$ and $\bs{n}$.
It acts on holomorphic functions (or matrices in the sense of entrywise action)
$\mc{D}_{\bs{\eps}}=\Dp{}\mc{D}_{v_{\bs{\eps}_1,\de_1}} \times \dots \times \Dp{}\mc{D}_{v_{\bs{\eps}_N,\de_N}}$ according to:

\beq
I_{ \bs{\eps}; \bs{n} }\pac{ \mc{F}_N } =   \Oint{ \Dp{}\mc{D}_{ \bs{\eps}} }{} \! \f{\dd^N y}{ \pa{2i\pi}^N } \cdot 
\f{ \mc{F}_N\pa{y_1,\dots, y_N} }{\pa{\la-y_1 }  \cdot \pl{s=2}{N} \pa{y_{s-1}-y_s} \cdot
\pl{s=1}{N}\pa{y_s-v_{\bs{\eps}_s}}^{\pa{2-\abs{\bs{\eps}_s}} \bs{n}_s+1} }  \;.
\label{definition fonctionnelle conduisant aux dérivées}
\enq
The functional $I_{\bs{n};\bs{\eps}}$ can be estimated by computing the residues at $v_{\bs{\eps}_s}$. This produces partial derivatives of $\mc{F}_N$ 
at the points $y_s=v_{\bs{\eps}_s}$. From now on, we focus on the analysis  of the action of $I_{\bs{n};\bs{\eps}}$
on the $11$ entry of the matrix $M_{\bs{j}^{\pa{a}}}$. The case of all the other entries can be treated similarly.

Depending on the choice of the components of the N-dimensional vector $\bs{\eps}$ and hence of the evaluation points $v_{\bs{\eps}_p}$,
after performing the integration induced by $I_{\bs{n};\bs{\eps}}$ (and having computed the eventual derivatives) 
the ratio $\tf{ \pa{\mf{e}_{j_2}\dots \mf{e}_{j_{2a}}} }
{ \pa{\mf{e}_{j_1}\dots \mf{e}_{j_{2a-1}} } }$ present in the 11 entry of \eqref{ecriture PiN comme somme integrale fnelle matrices} reduces to:
\begin{itemize}
\item $ \f{\mf{e}^m\!\pa{q;+}}{\mf{e}^m\!\pa{-q;-}} $ for some  $-a \leq m \leq a$ ;
%
%
%
%
\item or $\f{ \mf{e}^p\!\pa{q;+}  \mf{e}^{b-p}\!\pa{-q;-}  }{ \mf{e}^{b}\!\pa{\la_0;0}} \cdot \paf{\mf{e}\!\pa{q;+}}{\mf{e}\!\pa{-q;-}}^m$
for some  $1\leq b \leq m$, \; $0\leq p \leq b$\;,   $-\pa{a-b}\leq m \leq a -b$ ;
\end{itemize}

Hence, we get that there exists two sets of constant $c_{\bs{j}_{\ell}}^{\pa{m}}$, and
$c_{\bs{j}_{\ell}}^{\pa{m,\, b,\,p}}$,
\beq
\sul{ \bs{n} \in \mc{N}^{\pa{r}}_N  }{} \sul{  \bs{\eps}\in\mc{E}_N } {}
I_{ \bs{n} ; \bs{\eps} } \pac{ \pac{M_{\bs{j}^{\pa{a}}}}_{11} }
= \sul{m=-a}{a} \paf{ \mf{e}\pa{q;+} }{ \mf{e}\pa{-q;-}}^m c^{\pa{m}}_{  \bs{j}^{\pa{a}} }
%
%
+  \sul{  b=1  }{ \ell } \sul{p=0}{b} \sul{  m=b-a  }{a-b}  \f{ \mf{e}^p\!\pa{q;+} \mf{e}^{b-p}\!\pa{-q;-} }{ \mf{e}^b\!\pa{\la_0;0} }
\cdot \paf{ \mf{e}\!\pa{q;+} }{ \mf{e}\!\pa{-q;-} }^{m}
c^{\pa{m,\, b,\, p}}_{ \bs{j}^{\pa{a}} } \;.
\enq

Each derivative of the factor $\mf{e}_k$ in respect to $z_k$, when $z_k$ is in a neighborhood of $\pm q$, produces one power of $\log x$.
This $\log x$ term appears due to a differentiation of the exponent $x^{-2 \bs{\eps}_k \nu\pa{z_k}}$. It thus follows that the coefficients
$c^{\pa{m}}_{  \bs{j}^{\pa{a}} }$ and $c^{\pa{m,\, b,\, p}}_{  \bs{j}^{\pa{a}} }$ are polynomials in $\log x$. 
In the following, we determine the degree of these polynomials. This will allow us to show that 
\beq
\max{ \e{deg}\pa{ c^{\pa{m}}_{\bs{j}^{\pa{a}}} }} = r + N-2m-\de_{m,0} \qquad \e{and} \qquad
\max{ \e{deg}\pa{ c^{\pa{m,b,p}}_{\bs{j}^{\pa{a}}} }} = r + N -2\pa{b+m}
\label{ecriture resultat final sur les degree des plys en lnx}
\enq
where the sup is taken over all possible choices of $\bs{n}$, $\bs{\eps}$, $\bs{j}^{\pa{a}}$. 
Once that \eqref{ecriture resultat final sur les degree des plys en lnx} is established we get the claim.

\subsubsection*{The degree of $c^{\pa{m}}_{  \bs{j}^{\pa{a}} }$}

\vspace{2mm}

As already argued, when  $\bs{\eps}_k=0$, there necessarily appears $\mf{e}\pa{z_k;0}$ in the denominator of $\pac{M_{\bs{j}^{\pa{a}}}}_{11}$.  
As no function $\mf{e}\pa{z_{\ell};0}$, $\ell\not=k$, can appear in the numerator, this implies that 
in such a situation $\mf{e}\pa{\la_0,0}$ appears with some strictly postive exponent after computing the integrals. 
Therefore, one ends-up with a term that does not corresponds to the 
coefficient $c_{\bs{j}^{\pa{a}}}^{\pa{m}}$. Hence, contribution to the coefficients $c^{\pa{m}}_{  \bs{j}^{\pa{a}} }$ can only stem from 
these choice of N-dimensional vectors $\bs{\eps}$ whose entries are in $\paa{\pm 1}$. 
This means that when focusing on $c^{\pa{m}}_{\bs{j}^{\pa{a}}}$, all "admissible" choices of the $N$-dimensional vector $\bs{\eps}$
can be parameterized as
\beq
\bs{\eps}= ( \; \underbrace{\eps_1,\dots , \eps_1}_{\ell_1}\; , \underbrace{\eps_{2}, \dots , \eps_2 }_{\ell_2}, \dots , \;
\underbrace{\eps_p, \dots, \eps_p}_{\ell_p} \; ) \; \qquad \e{with} \qquad \eps_s =\pa{-1}^{s-1} \eps_1 \; \eps_1 \in \paa{\pm 1} 
\quad \e{for} \; \e{some} \; p \leq N\;.
\label{definition suite eps type bord pur}
\enq
We now compute explicitly the action of the functional $I_{\bs{\eps};\bs{n}}$ corresponding to some $\bs{\eps}$ given by 
\eqref{definition suite eps type bord pur}. For this purpose,
it is convenient to relabel the integration variables $y_i$ appearing in \eqref{definition fonctionnelle conduisant aux dérivées}
in a form that is subordinate to such a representation of the vector $\bs{\eps}$. Namely,
\beq
\pa{y_1,\dots, y_N} = \paa{ z_{1,1}, \dots, z_{1,\ell_1},\,  z_{2,1}, \dots , z_{s,t}, \dots, z_{p,\ell_p} } \; , \qquad ie \quad
z_{s,t}=y_{\, \ov{\ell}_s+t} \; ,  \quad \e{where}  \quad \ov{\ell}_s=\sul{r=1}{s-1}\ell_r \; .
\label{definition du mapping entre variable y et z}
\enq
We relabel the entries of the vector $\bs{n}$ in a similar way, \textit{ie}  $n_{s,t}=\bs{n}_{\,\ov{\ell}_s+t}$.
Then, the functional $I_{\bs{\eps} ; \bs{n}}$ reads
\beq
I_{ \bs{\eps} ; \bs{n} } \pac{ \pac{M_{\bs{j}^{\pa{a}}}}_{11}  }=    \Oint{ \Dp{}\mc{D}_{ \bs{\eps}}  }{}  \! \f{\dd^N z}{ \pa{2i\pi}^N } \cdot 
\f{ \pac{M_{\bs{j}^{\pa{a}}}}_{11}\!\pa{ \paa{z}}   } { \pl{s=1}{p} \paa{  \pa{z_{s-1, \ell_{s-1}}-z_{s,1} }  \cdot\pl{t=1}{ \ell_s-1 }\pa{z_{s,t} -z_{s,t+1} }} }
 \cdot  \pl{s=1}{p} \pl{t=1}{ \ell_s }\f{1}{ \pa{z_{s,t}-v_{\bs{\eps}_s}}^{ n_{s,t}+1} } \; .
\enq
Here we agree upon $\ell_0=0$ and $z_{0,0}=\la$. The above integral is directly computed by an inductive application of lemma \ref{Lemme integrale Fleu et ses derivees}:
\beq
I_{ \bs{\eps} ; \bs{n} } \pac{ \pac{M_{\bs{j}^{\pa{a}}}}_{11}  }  =
\sul{k_{s,t}=0}{r_{s,t}} \pl{s=1}{p} \pl{t=1}{\ell_{s}} \paa{ \f{1}{ k_{s,t}! } \f{\Dp{}^{k_{s,t}} }{ \Dp{}z_{s,t}^{k_{s,t}}} }
\cdot \pac{    \pac{M_{\bs{j}^{\pa{a}}}}_{11}\! \pa{\paa{z}}  \pl{s=1}{p} \pa{z_{s-1, \ell_{s-1}} - v_{\eps_{s}}}^{-r_{s,0}}    }_{z_{s,t}=v_{\eps_s}}
\hspace{-8mm} \;.
\label{equation calcul explicit I eps n sans la0}
\enq
In \eqref{equation calcul explicit I eps n sans la0} one sums over integers $k_{s,t}$  with $s=1,\dots,p$ and $t=1,\dots, \ell_s$
where each $k_{s,t}$ is summed from $0$ to
\beq
r_{s,t}= \sul{j=t}{\ell_s} n_{s,j}  \; + \;  \ell_s - t  \; -\; \sul{j=t+1}{ \ell_p} k_{s,t}  \qquad \e{with} \quad n_{s,0} = 0\;.
\enq

It follows that each block of variables  $\pa{z_{s,1}, \dots, z_{s,\ell_s}}$ associated to the same  $\eps_s$,
is subject to partial derivatives of total order $\sum_{t=1}^{\ell_s} k_{s,t}$. Hence, the maximal total order of all the derivatives
that may act on this block of variables is
$r_{s,1}^{\e{max}}=\sum_{j=1}^{\ell_s} n_{s,j} + \ell_s -1$.  The unique way of realizing this maximal order
is through a single derivative of order $r_{s,1}^{\e{max}}$ with respect to the variable $z_{s,1}$. 
We stress that in this case,
all the other variables of the block are simply set equal to $v_{\eps_s}$.
Very similarly, the maximal total order of all the partial derivatives that may act on a sub-block of variables
$\pa{z_{s,t}, \dots, z_{s,\ell_s}}$ associated to the same $\eps_s$ is $r_{s,t}^{\e{max}}=\sum_{j=t}^{\ell_s} n_{s,j} + \ell_s -t$.
The unique way of realizing this maximal order is by a derivative of order $r_{s,t}^{\e{max}}$ with respect to the variable $z_{s,t}$.
Then $z_{s,t+1}, \dots, z_{s,\ell_s}$ should be set equal to $v_{\eps_s}$.

As we have already mentioned, the function $\mf{e}_{j_k}\equiv\mf{e}\pa{y_{j_k};\eps_k}$ depends on $x$. Hence, its derivative in respect to $y_{j_k}$
generates powers of $\log x$. Therefore, the highest degree in $\log x$ appearing in $I_{\bs{\eps};\bs{n}}\big[M_{\bs{j}^{\pa{a}}} \big]$
will be generated by a derivative of the highest order possible  in respect to the variables $y_{j_k}$, with $k=1,\dots, 2a$.

\vspace{2mm}
Hence, to be able to determine this maximal degree in $\log x$, we first have to order the
indices $j_k$ according to the block to which they belong. For this purpose, we set
\beq
\mc{A}_s= \paa{k \; : \; j_k \in \intn{\ov{\ell}_s +1 }{ \ov{\ell}_{s+1} } } \qquad \e{and} \qquad a_s = \min\paa{k \; : \; k \in \mc{A}_s} \;.
\enq
Suppose that one deals with a block of variables $\pa{z_{s,1},\dots, z_{s,\ell_s}}$ such that $\# \mc{A}_s\not= 0$. Then the highest possible power of 
$\ln x$ that an integration over the variables of this block can produce will be issued by the action of a derivative of the highest order possible
on the variable $z_{s,j_{a_s}-\ov{\ell}_s}$. Thence, an integration over this block of variables generates a polynomial in $\ln x$ of degree
$r_{s, j_{a_s}-\ov{\ell}_s}^{\e{max}}$. Clearly, if $\# \mc{A}_s = 0 $, its associated set of variables and functions 
cannot generate, once upon being integrated, any power of $\log x$.

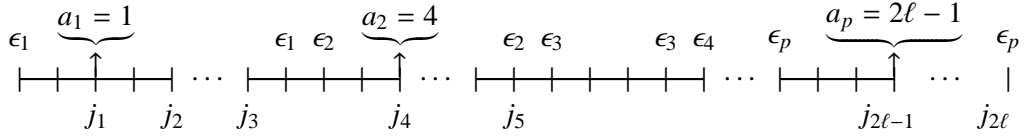
\begin{figure}[h]
\begin{center}

\begin{pspicture}(13.5,2)

\psline(0.5,1)(2.5,1)

\psline(3.5,1)(5.5,1)

\psline(6.5,1)(9.5,1)

\psline(10.5,1)(12,1)


\rput(1.5,1.7){$\underbrace{a_1=1}$}

\rput(5.5,1.7){$\underbrace{a_2=4}$}


\rput(12,1.7){$\underbrace{a_p= 2\ell-1}$}


\rput(0.5,1){$|$}
\rput(1,1){$|$}
\rput(1.5,1){$|$}
\rput(2,1){$|$}
\rput(2.5,1){$|$}
\rput(3,1){$\dots$}
\rput(3.5,1){$|$}
\rput(4,1){$|$}
\rput(4.5,1){$|$}
\rput(5,1){$|$}
\rput(5.5,1){$|$}
\rput(6,1){$\dots$}
\rput(6.5,1){$|$}
\rput(7,1){$|$}
\rput(7.5,1){$|$}
\rput(8,1){$|$}
\rput(8.5,1){$|$}
\rput(9,1){$|$}
\rput(9.5,1){$|$}
\rput(10,1){$\dots$}
\rput(10.5,1){$|$}
\rput(11,1){$|$}
\rput(11.5,1){$|$}
\rput(12,1){$|$}

\rput(12.7,1){$\dots$}
\rput(13.5,1){$|$}

\rput(0.5,1.5){$\eps_1$}
\rput(1.5,0.5){$j_1$}
\rput(1.5,1.2){$\ua$}
\rput(2.5,0.5){$j_2$}

\rput(3.5,0.5){$j_3$}
\rput(4,1.5){$\eps_1$}
\rput(4.5,1.5){$\eps_2$}
\rput(5.5,0.5){$j_4$}
\rput(5.5,1.2){$\ua$}

\rput(7,0.5){$j_5$}
\rput(7,1.5){$\eps_2$}
\rput(7.5,1.5){$\eps_3$}
\rput(9,1.5){$\eps_3$}

\rput(9.5,1.5){$\eps_4$}

\rput(10.5,1.5){$\eps_p$}
\rput(11.9,0.5){$j_{2\ell-1}$}
\rput(12,1.2){$\ua$}
\rput(13.3,0.5){$j_{2\ell}$}
\rput(13.5,1.5){$\eps_p$}

\end{pspicture}

\caption{Definition of the sets $\mc{A}_s$ and of its minimal element $a_s$.
One has $\mc{A}_1=\paa{1,2,3}$, $\mc{A}_2=\paa{4,5}$, \newline $\mc{A}_3=\emptyset$, \dots, $\mc{A}_p=\paa{2\ell-1, 2\ell}$.
The $\eps_k$ delimit the block of variables of length $\ell_k$ associated to $\eps_k$.
\label{contour pour representer ensemble As et son min cas bord purs}}
\end{center}
\end{figure}

We now characterize the oscillating term appearing in $I_{\bs{\eps};\bs{n}}\pac{ M_{\bs{j}^{\pa{a}}}}$.
If a given block $\pa{z_{s,1}, \dots, z_{s,\ell_s}}$ corresponds to a set $\mc{A}_s$
having an even number of elements ($\# \mc{A}_s \in 2\mathbb{N}$), then after taking the derivatives and once upon
evaluating at $z_{s,t}=v_{\eps_s}$,  the associated ratio of the functions  $\mf{e}_{j}$ cancels out. Indeed, there are as many identical factors in the
denominator that in the numerator. For instance, when $a_s$ is even one has 
\beq
\left. \f{ \mf{e}_{ j_{a_s} } \dots \mf{e}_{ j_{a_s + \# \mc{A}_s } } }
{ \mf{e}_{ j_{a_{s}+1} } \dots \mf{e}_{ j_{a_s+ \# \mc{A}_{s} -1}} }
\right|_{ z_{s,t}=v_{\eps_s}} = 1 \;.
\enq

However,  if a given block $\pa{z_{s,1}, \dots, z_{s,\ell_s}}$ corresponds to a set $\mc{A}_s$
with an odd number of elements ($\# \mc{A}_s \in 2\mathbb{N}+1$), then after taking the derivatives,
 the associated  ratio of the functions $\mf{e}_{j}$ reduces to $\pac{\mf{e}\pa{v_{\eps_s};\eps_s}}^{\pa{-1}^{a_s}}$\!. Indeed
\beq
a_s \in 2\mathbb{N} \Rightarrow
\left. \f{ \mf{e}_{j_{a_s}} \dots \mf{e}_{j_{a_s+ \# \mc{A}_{s}-1 }} }
				{ \mf{e}_{j_{a_s+1}} \dots \mf{e}_{j_{a_s+ \# \mc{A}_{s}-2 } } }  \right|_{z_{s,t}=v_{\eps_s}}
\hspace{-5mm}= \mf{e}\pa{v_{\eps_s}; \eps_s}  \quad  \e{and}\quad  \;a_s \in 2\mathbb{N}+1 \Rightarrow
\left. \f{ \mf{e}_{j_{a_s+1}} \dots \mf{e}_{j_{a_s+ \# \mc{A}_{s} -2 }} }
				{ \mf{e}_{j_{a_s}} \dots \mf{e}_{j_{a_s+ \# \mc{A}_{s} -1 } } }  \right|_{z_{s,t}=v_{\eps_s}}
\hspace{-5mm}=   \f{1}{\mf{e}\pa{v_{\eps_s}; \eps_s}} \;.
\enq
Therefore, we obtain that
\beq
I_{\bs{\eps}; \bs{n}} \pac{ \pac{M_{\bs{j}^{\pa{a}}}}_{11} } = P_{ \bs{\eps}; \bs{n} }^{\pa{\bs{j}^{\pa{a}}}}\! \pa{\log x}  \cdot \hspace{-2mm}  
\pl{ \substack{ s=1  \\  \# \mc{A}_s\in 2\mathbb{N}+1  } } { p }\hspace{-2mm}
\pac{\mf{e}\pa{v_{\eps_s};\eps_{s}} }^{\pa{-1}^{a_s}}  \quad \e{where} \quad
\deg\Big( P_{ \bs{\eps}; \bs{n} }^{\pa{\bs{j}^{\pa{a}}}} \Big) = \!\! \sul{ \substack{ s=1  \\ \# \mc{A}_s \not= 0} }{p}
\paa{ \sul{k=j_{a_s}-\ov{\ell}_s}{\ell_s} \!\! n_{s,k} + \ell_s - \pa{j_{a_s}-\ov{\ell}_s} } \;.
\label{equation pour forme general contribution oscillatoire de I}
\enq

Now, in order to obtain the coefficient $c^{\pa{m}}_{  \bs{j}^{\pa{a}} }$ we should sum up
\eqref{equation pour forme general contribution oscillatoire de I} over $\bs{n}\in\mc{N}_N^{\pa{r}}$ and also over 
 all the possible configurations of vectors $\bs{\eps}$ parameterized as in \eqref{definition suite eps type bord pur} and such that
we eventually generate the power $\pa{ \tf{\mf{e}\pa{q;+}}{ \mf{e}\pa{-q;-} } }^m$. Then, among such configurations, we should
look for those that correspond to a polynomial $P_{\bs{\eps};\bs{n}}\pa{\log x}$ of highest degree. 

Given a fixed number of flips $p$ in \eqref{definition suite eps type bord pur}, one maximizes the degree 
in \eqref{equation pour forme general contribution oscillatoire de I} by choosing the lengths $\ell_s$ is such a way that $\# \mc{A}_s \not=0$
for any $s$ and such that $j_{a_s}= \ov{\ell}_s+1$. One can do so for all $s$, but $s=1$. Indeed, in the latter case one necessarily has
$j_{a_1}=j_1\geq 1$. Therefore, for such a choice of lengths $\ell_s$, once upon choosing $n_{1,t}=0$ for $t=0,\dots,j_{1}-1$ one obtains that 
this maximal degree of is $ r+N-p -\pa{j_1-1}$. Note that, we have used $\sum_{s,t} n_{s,t}=r$  and $\sum_s \ell_s=N$. 

\vspace{2mm}
There is also a condition on the number of flips that are necessary to generate the oscillatory factors
\newline$\pa{ \tf{\mf{e}\pa{q;+}}{ \mf{e}\pa{-q;-} } }^m$. Due to the form of the oscillatory factor in
\eqref{equation pour forme general contribution oscillatoire de I}, we get that one sequence $\pa{\eps_s,\dots,\eps_s}$
generates at most one factor $\pac{ \mf{e}\pa{\eps_s q;\eps_s} }^{\tau} $, $\tau=\pm 1$. Hence, if $m\not=0$
there are at least $2m$ flips necessary to generate the factors $\pa{ \tf{\mf{e}\pa{q;+}}{ \mf{e}\pa{-q;-} } }^m$.
If $m=0$, then one still has one sequence $\pa{\eps_1,\dots,\eps_1}$ of length $\ell_1=N$. Therefore, one has $p\geq \max\pa{2m,1}$.
Taking the lowest possible value, \textit{ie} $p=\max\pa{2m,1}$, we get that
\beq
\max_{\bs{\eps} \,; \,  \bs{n}} \;  \deg\pa{ P_{ \bs{\eps}; \bs{n} }^{\pa{\bs{j}^{\pa{a}}}} }    = n+N-2m - \de_{m,0} +\pa{j_1-1} \;.
\enq

In order to obtain estimates for $\pac{\Pi_N}_{11}$ one should still sum up over all the possible configurations of $2a$-uples $\bs{j}^{\pa{a}}$. 
Therefore the highest degree in $\log x$ of $\big[ \Pi_{N,r}^{\pa{m}} \big]_{11}$ is obtained by setting $j_1=1$. 
That is, we reproduce \eqref{ecriture resultat final sur les degree des plys en lnx}.

\subsubsection*{The degree of $c^{\pa{m,\, b, \, p}}_{  \bs{j}^{\pa{a}} }$}

It follows from the previous discussions that each time an integration over $\Dp{}\mc{D}_{\la_0,\de}$ occurs in 
\eqref{ecriture PiN comme somme integrale fnelle matrices}-\eqref{definition fonctionnelle conduisant aux dérivées},
there appears, once upon integrating, a factor $\mf{e}^{-1}\!\pa{\la_0,0}$. Hence, the oscillating factor in front of $c^{\pa{m,\, b, \, p}}_{  
\bs{j}^{\pa{a}} }$ is necessarily generated by these choices of $N$-dimensional vectors $\bs{\eps}$ where precisely $b$ entries are equal to zero
(\textit{ie} there are exactly $b$ integrations over $\Dp{}\mc{D}_{\la_0,\de}$).

Taking into account the fact that, as previously argued, two neighboring entries of the vector $\bs{\eps}$ cannot simultaneously vanish, 
we get that such vectors $\bs{\eps}$ can be parameterized as 
\beq
\bs{\eps}= ( \; \underbrace{\eps_1,\dots , \eps_1}_{\ell_1}\; , \dots , \underbrace{\eps_{\tau_1}, \dots , \eps_{\tau_1} }_{\ell_{\tau_1}}, 0,
\eps_{\tau_1+1}, \dots, \eps_{\tau_b}, 0 ,  \dots , \;
\underbrace{\eps_p, \dots, \eps_p}_{\ell_p} \; )  \; ,  \;\;\e{where} \quad  \sul{ r=1}{p} \ell_r= N-b \;.
\label{definition suite eps mixe plus moins et zeros}
\enq
We relabel the integration variables $y_i$ appearing in \eqref{definition fonctionnelle conduisant aux dérivées}
in a form subordinate to \eqref{definition suite eps mixe plus moins et zeros}:
\beq
\pa{y_1,\dots, y_N} = \paa{ z_{1,1}, \dots z_{\tau_a,\ell_{\tau_a}}, \om_a, z_{\tau_a+1,1} \dots, z_{p,\ell_p} } \; , \qquad ie \quad
z_{s,t}=y_{\,\ov{\ell}_s+t} \; , \qquad \om_{a}= y_{\,\ov{\ell}_{\tau_a}+1}
\label{definition du mapping entre variable y et z omega}
\enq
where we agree upon
\beq
 \ov{\ell}_s=\sul{r=1}{s-1}\ell_r \; + \;  \#\paa{\;  k  \; : \;  \tau_k <s \; } \; .
\enq

We also relabel the entries of the vector $\bs{n}$ in a similar way, \textit{ie}  $n_{s,t}=\bs{n}_{\,\ov{\ell}_s+t}$
and $n_a^{\pa{0}}= \bs{n}_{\, \ov{\ell}_{\tau_a}+1}$.
The action of the associated functional $I_{\bs{\eps} ; \bs{n}}$ takes the form
\bem
I_{ \bs{\eps} ; \bs{n} } \pac{ \pac{M_{\bs{j}^{\pa{a}}}}_{11}  }=    \Oint{ \Dp{}\mc{D}_{ \bs{\eps}}  }{}  \f{\dd^{N-b} z}{ \pa{2i\pi}^{N-b} } \cdot \f{\dd^b \om }{ \pa{2i\pi}^b }
  \pl{ \substack{ s=1  \\ s \not= \tau_a + 1} }{p}   \f{ 1 } {\pa{z_{s-1, \ell_{s-1}}-z_{s,1} }}
  \cdot \pl{s=1}{p}\pl{t=1}{ \ell_s -1 }  \f{ 1 } {\pa{z_{s,t} -z_{s,t+1} }} 
 \cdot  \f{ \pac{M_{\bs{j}^{\pa{a}}}}_{11} \! \pa{\paa{z}; \paa{\om}}  }{ \pl{s=1}{p} \pl{t=1}{ \ell_s } \pa{z_{s,t}-v_{\eps_s}}^{ n_{s,t}+1} }  \\
\times \pl{a=1}{b} \f{ 1 }{ \pa{z_{\tau_a, \ell_{\tau_a}}-\om_a  }  \pa{\om_a - z_{\tau_a+1, 1}  }  } \; \cdot \;
\pl{a=1}{b}  \f{1}{ \pa{\om_a - \la_0 }^{2n_a^{\pa{0}}+1}  }
 \; .
\end{multline}
The integrals over $\om_a$ are readily computed. We set
\beq
\mc{G}_{N-b}\pa{\paa{z}} = \pl{a=1}{b} \paa{ \f{1}{  \pa{2n_a^{\pa{0}}}!  }  \f{ \Dp{}^{2n_a^{\pa{0}}}  }{ \Dp{}\om_a^{2n_a^{\pa{0}}} } }   \cdot
\pac{   \f{ \pac{M_{\bs{j}^{\pa{a}}}}_{11} \! \pa{\paa{z}; \paa{\om}}  }{ \pa{z_{\tau_a, \ell_{\tau_a}}-\om_a  }  \pa{\om_a - z_{\tau_a+1, 1}  }   }    }_{\om_a=\la_0} \;.
\enq
Then, the analysis boils down to the case previously studied:
\beq
I_{ \bs{\eps} ; \bs{n} } \pac{ \pac{M_{\bs{j}^{\pa{a}}}}_{11} }=    \Oint{ \Dp{}\mc{D}^{\e{red}}_{ \bs{\eps}}  }{}  \f{\dd^{N-b} z}{ \pa{2i\pi}^{N-b} }
 \mc{G}_{N-b}\pa{ \paa{z}}
\pl{ \substack{ s=1  \\ s \not= \tau_a + 1} }{p} \f{1} { \pa{z_{s-1, \ell_{s-1}}-z_{s,1} } }
\pl{s=1}{p} \pl{t=1}{ \ell_s -1 } \f{1}{\pa{z_{s,t} -z_{s,t+1} }} 
 \cdot  \pl{s=1}{p} \pl{t=1}{ \ell_s }\f{1}{ \pa{z_{s,t}-v_{\eps_s}}^{ n_{s,t}+1} }  \; .
\enq
The integrals runs over the contour $\Dp{}\mc{D}^{\e{red}}_{\bs{\eps}}$ which corresponds to that part of the initial contour $\Dp{}\mc{D}_{\bs{\eps}}$
where the integrals over the variable $\om_a$ have been suppressed. Therefore
\beq
I_{ \bs{\eps} ; \bs{n} } \pac{ \pac{M_{\bs{j}^{\pa{a}}}}_{11} }=
\sul{k_{s,t}=0}{r_{s,t}} \pl{s=1}{p} \pl{t=1}{\ell_{s}} \paa{ \f{1}{ \pa{k_{s,t}}! } \f{\Dp{}^{k_{s,t}} }{ \Dp{}z_{s,t}^{k_{s,t}}} }
\cdot \Bigg[   \f{\mc{G}_{N-m}\pa{\paa{z}} }
{  \pl{ \substack{s=1 \\ s \not= \tau_a +1 } }{p} \pa{z_{s-1, \ell_{s-1}} - v_{\eps_{s}}}^{r_{s,0}} }   \Bigg]_{\mid z_{s,t}=v_{\eps_s}} \hspace{-5mm} .
\enq
The sum over the integers $k_{s,t}$ runs from $0$ to
$r_{s,t}= \sul{j=t}{\ell_s} n_{s,j}  \; + \;  \ell_s - t  \; -\; \sul{j=t+1}{ \ell_s} k_{s,t}$ \;.

Similarly to the previous analysis, we set
\beq
\mc{A}_s= \paa{\;  k \; : \; j_{k}^{} \in \intn{ \ov{\ell}_s+1 }{ \ov{\ell}_s + \ell_s  } \; } \qquad \e{and} \qquad
 a_s = \min\paa{\; k \; : \; k \in \mc{A}_k  \; }
\;.
\enq
It is then easy to see by using similar arguments to those invoked for $c_{\bs{j}^{\pa{a}}}^{\pa{m}}$ that
\beq
I_{\bs{\eps}; \bs{n}} \pac{ \pac{M_{\bs{j}^{\pa{a}}}}_{11} } = \f{ P^{\pa{\bs{j}^{\pa{a}}}}_{ \bs{\eps}; \bs{n} } \pa{\log x}  }{ \mf{e}^{b}\pa{\la_0; 0}  }
 \cdot  \hspace{-2mm}\pl{ \substack{ s=1  \\  \# \mc{A}_s\in 2\mathbb{N}+1  } } { p }  \hspace{-2mm}
\pac{\mf{e}\pa{v_{\eps_s};\eps_{s}} }^{\pa{-1}^{a_s}}  \quad \e{where} \quad
 \deg\Big( P_{ \bs{\eps}; \bs{n} }^{ \pa{\bs{j}^{\pa{a}}} } \Big) = \sul{ \substack{ s=1  \\ \# \mc{A}_s \not= 0} }{p}
\paa{ \sul{k=j_{a_s}-\ov{\ell}_s}{\ell_s} n_{s,k} + \ell_s - \pa{j_{a_s}-\ov{\ell}_s} } \;.
\label{equation pour forme general contribution oscillatoire de fonctionnelle 0 et plus moins}
\enq

In order to obtain the maximal degree in $\log x$ associated to the oscillating term 
\beq
\f{ \mf{e}^p\!\pa{q,+} \mf{e}^{b-p}\!\pa{-q,-} }{  \mf{e}^b\!\pa{\la_0; 0} }  \cdot \paf{ \mf{e}\! \pa{q,+}  }{  \mf{e}\!\pa{ -q; -} }^m
\label{forme exposant}
\enq
present in $\pac{\Pi_N}_{11}$, we should maximize the degree of the previous polynomial 
in \eqref{equation pour forme general contribution oscillatoire de fonctionnelle 0 et plus moins} 
under the constraint that the sequence $\eps_a$ in \eqref{definition suite eps mixe plus moins et zeros} ought to change 
its value at least $b+m$ times (this in order to produce the sought form of the oscillatory term with its associated power-law
behavior) and that these changes are such that eventually \eqref{forme exposant} is generated. 

We should also maximize this degree in respect to all the possible choices 
of $2a$-uples  $\bs{j}^{\pa{a}}$ of various lengths $2a$ and over the allowed vectors $\bs{n} \in \mc{N}_N^{\pa{r}}$. 
In order to obtain this maximal degree, one should choose a minimal number of flips $(m+b)$, choose the lengths $\ell_k$ and the $j_k$
in such a way that $j_{a_s}=\ov{\ell}_s+1$. Finally, one should also take $n_a^{\pa{0}}=0$ for all $a$. This leads to the conclusion that the maximal 
degree in $\ln x$ is $r+N-2\pa{m+b}$.  \qed

\section{Fine bounds on $\Pi_N$}
\label{appendix Final bounds for PiN}

In this appendix we provide bounds for the matrices $\Pi_{N}^{\pa{m,b,p}}$ entering in the decomposition for $\Pi_N$ given in proposition 
\ref{proposition DA Pi}. 

\begin{prop}
\label{proposition fine bounds for PiN}

Let $\Sg_{\Pi}$ be a contour appearing in the RHP for $\Pi$ and $U$ any open set such that $\e{d}\pa{U,\Sg_{\Pi}}>0$. 
Let $\Pi_N$ be as defined by \eqref{ecriture Pi comme serie asymptotique detaille} and, agreeing upon 
$\bs{\eta}=1$ in the space-like regime and $\bs{\eta}=-1$ in the time-like regime, let
\beq
\mf{e}\pa{z;\eps} = \left\{  \ba{lcc}  \ex{ix u\pa{q}} x^{-2 \nu\pa{z}}   & \e{for} & \eps=  1 \\
									\ex{ix u\pa{-q}} x^{ 2 \nu\pa{z}}   & \e{for} & \eps=  -1 \\
									\ex{ix u\pa{\la_0}} x^{- \bs{\eta} \f{1}{2}}  & \e{for} & \eps=  0 \ea \right. 	\;.
\enq
Then the matrix $\Pi_N\!\pa{\la}$  admits the representation
\beq
\Pi_N\pa{\la} = A_N\pa{\la} \; + \; \sul{b=0}{ \pac{\tf{N}{2}} } \sul{ p=0 }{ b }  \sul{m= b - \pac{\tf{N}{2}} }{ \pac{\tf{N}{2}} -b }
\paf{ \mf{e}\pa{q;+} }{ \mf{e}\pa{-q;-} }^{m-\bs{\eta} p}  \paf{ \mf{e}\pa{\la_0;0} }{ \mf{e}\pa{-q;-} }^{\bs{\eta} b}
\Pi_N^{\pa{m,b,p}}\pa{\la} \;.
\label{equation developpement PiN explicite appendix final}
\enq
For $x$-large enough, the matrices $\Pi_N^{\pa{m,b,p}}\!\pa{\la}$ and $A_N\!\pa{\la}$ depend smoothly on $x$ and holomorphically on $\la \in U$. 
One has the decomposition 
\beq
\Pi_N^{\pa{m,b,p}}\pa{\la} \; = \; \sul{ \eps \in \paa{\pm 1, 0} }{} 
\pac{\mf{e}\pa{v_{\eps};\eps}}^{\f{\sg_3}{2}} \Pi_{N;\eps}^{\pa{m,b,p}}\pa{\la} \pac{\mf{e}\pa{v_{\eps};\eps}}^{-\f{\sg_3}{2}} \qquad 
\e{where} \qquad v_{\pm }= \pm q  \quad  \e{and}  \; v_0=\la_0 \;.
\label{equation decomposition Pimbp}
\enq
The matrix $\Pi_{N;\eps}^{\pa{m,b,p}}\pa{\la}$ is such that it does not contain any oscillating factor in its entries. Moreover, for all 
$k\in \mathbb{N}$ there exists an N-independent constant $C>0$ such that 
\beq
 \norm{A_N}_{L^{\infty}\pa{U}} \leq \f{C^N}{x^{k}} \qquad \e{and}  \qquad 
\norm{ \Pi_{N;\eps}^{\pa{m,b,p}} }_{ L^{\infty}\pa{U} } \leq C^N x^{ \wt{w} N } \quad \e{with} \quad 
\wt{w}= 2 \max_{\eps=\pm} \paa{ \sup_{\Dp{}\mc{D}_{\eps q , \de } }\abs{ \Re\pa{\nu-\nu\pa{\eps q}} } }\;. 
\label{ecriture estimation norme AN et PIN et tout le bla bla}
\enq
These estimates also hold for the first order partial derivatives (in respect to $x$ or $\la$). 
\end{prop}

\Proof 

Recall that the matrix $\Pi_N$ can be represented in terms of Cauchy transforms (or their $+$ boundary values) on $\Sg_{\Pi}$:
\beq
\Pi_N\pa{\la} = \mc{C}^{\De}_{\Sg_{\Pi}} \circ \dots \circ  \mc{C}^{\De}_{\Sg_{\Pi}} \pac{I_2}\pa{\la}
= \paa{\mc{C}^{\De}_{\Sg_{\Pi}} }^N\hspace{-2mm}\pac{I_2}\pa{\la} \;.
\enq
Above and in the following, $\mc{C}^{\De}_{\msc{C}}\pac{M}\pa{\la}$ for $\la \not\in \msc{C} $ corresponds to the case where in the 
integral representation \eqref{definition operateur Cauchy + cas matrice L2} for this operator we substitute the $+$ boundary value with $\la\not\in 
\msc{C} $. This is clearly a well defined expression. 
We decompose the jump contour for $\Pi$ according to $\Sg_{\Pi}=\Dp{}\mc{D} \cup \wt{\Sg}_{\Pi}$ with $\Dp{}\mc{D}=
\Dp{}\mc{D}_{q,\de} \cup \Dp{}\mc{D}_{-q,\de} \cup \Dp{}\mc{D}_{\la_0,\de}$. 

The exponentially small in $x$ terms gathered in $A_N$ can be written as 
\beq
A_N\pa{\la} = \sul{k=0}{N-1} \paa{\mc{C}^{\De}_{\Dp{}\mc{D}}}^{N-1-k}\circ  \paa{ \mc{C}^{\De}_{\wt{\Sg}_{\Pi}} }   \circ   
\paa{\mc{C}^{\De}_{\Sg_{\Pi}}}^{k} \!\! \pac{I_2}\pa{\la} \; ,
\enq
whereas
\beq
\Pi_{N}\pa{\la}-A_N\pa{\la} = \sul{\bs{\eps} \in \mc{E}_N }{} \mc{C}^{\De}_{\Dp{}\mc{D}_{v_{ \bs{\eps}_1,\de}}}\circ  \dots \circ 
\mc{C}^{\De}_{\Dp{}\mc{D}_{v_{\bs{\eps}_N,\de}}} \;.
\enq
There, the sum runs through $\bs{\eps} \in \mc{E}_N=\paa{\bs{\eps}=\pa{\bs{\eps}_1,\dots, \bs{\eps}_N} \; : \; \bs{\eps}_s \in \paa{\pm 1 , 0}}$. 
One can readily convince oneself that for any matrix function $M$ such that $\De M \in L^{2}\pa{\wt{\Sg}_{\Pi}}$ there exists a constant $c^{\prime}$ such that 
\beq
\norm{ \mc{C}^{\De}_{\wt{\Sg}_{\Pi}}\pac{M} }_{L^{2}\pa{\Dp{}\mc{D}}} \leq c^{\prime} \norm{ \De M  }_{L^{2}\pa{ \wt{\Sg}_{\Pi}}  } \;.
\enq
Thence setting $c=\max\paa{ c^{\prime}, c\pa{\Sg_{\Pi}}, c\pa{\Dp{}\mc{D}}}$ (we recall that for a curve $\Ga$, $c\pa{\Ga}$ stands for 
the norm of the $+$ boundary value of the Cauchy operator on $L^{2}\pa{\Ga}$), one gets
\beq
\norm{A_N}_{ L^{\infty}\pa{U} } \leq N \f{ \pac{2c \mc{N}_{\Sg_{\Pi}}\!\pa{\De}}^{N-1} }{\pi^ 2\e{d}\pa{ U,\Sg_{\Pi} } } 
\mc{N}_{\wt{\Sg}_{\Pi}}\!\pa{\De}
\qquad \e{with} \qquad 
\mc{N}_{\msc{C}}\pa{\De}= \norm{\De}_{L^{2}\!\pa{\msc{C}}}\; + \;  \norm{\De}_{L^{\infty}\!\pa{\msc{C}}} \;.
\nonumber
\enq
Thus, the claim follows for $A_N$ as, by construction, $\mc{N}_{\wt{\Sg}_{\Pi} }\pa{\De}  = \e{O}\pa{x^{-\infty}}$. 

It remains to obtain estimates for the remaining, algebraically small in $x$, part.  
For this we set 
\beq
\wh{\De}\pa{z} = \pac{ \mf{e}\pa{\vp\pa{z} ; \eps\pa{z}} }^{-\f{\sg_3}{2}} \De\pa{z}  \pac{ \mf{e}\pa{\vp\pa{z} ; \eps\pa{z}} }^{\f{\sg_3}{2}}
\quad \e{with} \quad \left\{ \ba{cc c } \vp\pa{z} &= & 
					q \bs{1}_{\Dp{}\mc{D}_{q,\de}} -  q \bs{1}_{\Dp{}\mc{D}_{-q,\de}}  + \la_0 \bs{1}_{\Dp{}\mc{D}_{\la_0,\de}} \\
			\eps\pa{z}   &=&   \bs{1}_{\Dp{}\mc{D}_{q,\de}} -   \bs{1}_{\Dp{}\mc{D}_{-q,\de}}   \ea \right.  \;.
\nonumber
\enq

Then, by carrying out similar expansions to \eqref{equation DA Pi comme somme integrale produits DA De} and \eqref{ecriture developpement explicite puissance x produit matrices} it is easy to convince oneself that 
\beq
I_2 \pac{ \Pi_{N;\eps}^{\pa{m,b,p}} }_{k_1,k_{N+1} } \! \pa{\la}  =  \sul{\bs{\eps}\in \mc{E}_N}{}\! ^{\prime} \; \sul{k_a=1}{2} \!^{\prime}
\;\; \mc{C}^{\wh{\De}_{k_2k_1}}_{\Dp{}\mc{D}_{v_{\eps_1,\de}}}\circ  \dots \circ 
\mc{C}^{\wh{\De}_{k_{N+1},k_N}}_{\Dp{}\mc{D}_{v_{\eps_N,\de}}}\!\pac{I_2} \;.
\enq
Above $\wh{\De}_{ ab }$ stands for the $ab$ entry of $\wh{\De}$. Also, there appears $\wh{\De}$ instead of $\De$ as the oscillating factors 
have been already pulled-out, as in \eqref{equation developpement PiN explicite appendix final}-\eqref{equation decomposition Pimbp}. 
Also the primes $^{\prime}$ in front of the two sums are there to indicate that these are constrained. Namely, 
one should sum-up only over thoses choices of $\bs{\eps}\in \mc{E}_N$  and $k_a\in \paa{1,2}$, $a=2,\dots, N$ which, upon the replacement $\wh{\De} \mapsto \De$ would 
give rise to the oscillating factor associated with $\pac{ \Pi_{N;\eps}^{\pa{m,b,p}} }_{k_1,k_{N+1} } $. 
By using the continuity of the $+$ boundary value Cauchy operator on $L^{2}\pa{\Dp{}\mc{D}}$, one shows that 
there exists a constant $c$ such that for any $\eps, \tau \in \paa{\pm 1, 0}$
and any $f \in  L^{2}\pa{\Dp{}\mc{D}}$:
\beq
\norm{ \mc{C}^{I_2}_{\Dp{}\mc{D}_{v_{\eps},\de}}\pac{f I_2} }_{L^{2}\pa{\Dp{}\mc{D}_{\tau,\de}}} \leq 
 c \norm{ f }_{L^2\!\pa{\Dp{}\mc{D}_{v_{\eps},\de}} } \;.
\enq
Then, 
\bem
\norm{ \Pi_{N;\eps}^{\pa{m,b,p}} }_{L^{\infty}\pa{U}} \hspace{-3mm} 
\leq \f{ c^{N-1} }{ \e{d}\pa{U,\Sg_{\Pi}} }
\sul{\bs{\eps}\in \mc{E}_N}{}\! ^{\prime}  \sul{k_a=1}{2} \!^{\prime}
\norm{ \wh{\De}_{k_2k_1} }_{L^{2}\big( \Dp{}\mc{D}_{v_{\eps_1,\de}} \big) } \pl{a=2}{N-1} \norm{ \wh{\De}_{k_{a+1}k_a} }_{L^{\infty}\big( \Dp{}\mc{D}_{v_{\eps_a,\de}} \big) }
\norm{ \wh{\De}_{k_{N+1}k_N} }_{L^{2}\big( \Dp{}\mc{D}_{v_{\eps_N,\de}} \big) } \\
\leq 
\f{ \pa{2c}^{N-1}  }{ \e{d}\pa{U,\Sg_{\Pi}} }  \mc{N}_{\Dp{}\mc{D}}\!\pa{\wh{\De}} \;.
\end{multline}
Since there exists $c^{\prime}>0$ such that  $ \mid\mid \wh{\De} \mid\mid _{L^{2}\pa{\Dp{}\mc{D}}}  \leq c^{\prime} x^{\wt{w}} $, 
the claim follows. Also, we stress that, by construction, 
$\Pi_{N;\eps}^{\pa{m,b,p}}$ does not contain any oscillating terms in $x$ in its asymptotic expansion when $x\tend +\infty$. 
\qed

\end{document}